\documentclass[12pt]{article}
\pdfoutput=1
\usepackage{amsmath,amssymb,amsfonts,color,graphicx,color,cite}
\usepackage{mciteplus}
\usepackage{slashed}
\usepackage[bottom]{footmisc}
\usepackage{float}
\usepackage[plainpages=false,colorlinks,linkcolor=blue,citecolor=blue,urlcolor=blue]{hyperref}
\usepackage[normalem]{ulem}

\usepackage{booktabs}

\hypersetup{pdfauthor={Bastian Feigl, Heidi Rzehak and Dieter Zeppenfeld},
            pdftitle={SUSY Background to Neutral MSSM Higgs Boson Searches}}

\mciteSetMidEndSepPunct{;\newline}{.}{\relax}

\evensidemargin \oddsidemargin
\usepackage{sectsty}
\sectionfont{\Large}
\subsectionfont{\large}
\subsubsectionfont{\normalsize}

\allowdisplaybreaks

\begin{document}
\thispagestyle{empty}

\def\thefootnote{\fnsymbol{footnote}}
\def\pslash#1{{\setbox0=\hbox{$#1$}
  \rlap{\ifdim\wd0>.18em\kern.18\wd0\else\kern.18\wd0\fi /}#1}}

\begin{flushright}
KA--TP--20--2011\\
FR-PHENO-2011-014\\
\end{flushright}

\vspace{0.5cm}

\begin{center}

{\large\sc {\bf SUSY Background to Neutral MSSM Higgs Boson Searches}}

\vspace{1cm}

{\sc
B.~Feigl$^{1}$%
\footnote{email: bastian.feigl @ kit.edu}%
, H.~Rzehak$^{2}$%
\footnote{email: heidi.rzehak @ physik.uni-freiburg.de}%
~and D.~Zeppenfeld$^{1}$ %
\footnote{email: dieter.zeppenfeld @ kit.edu}
}

\vspace*{.7cm}

{$^1$Institut f\"ur Theoretische Physik, Karlsruher Institut f\"ur Technologie \\
D--76128 Karlsruhe, Germany

\vspace*{0.1cm}

$^2$ Physikalisches Institut, Albert-Ludwigs-Universit\"at Freiburg, D--79104
Freiburg, Germany 
}

\end{center}

\vspace*{0.1cm}

\begin{abstract}
\noindent
Within the Minimal Supersymmetric Standard Model (MSSM) the production and decay 
of superpartners can give rise to backgrounds for Higgs boson searches. Here MSSM 
background processes to the vector boson fusion channel with the Higgs boson 
decaying into two tau leptons or two W-bosons are investigated, giving rise to 
dilepton plus missing transverse momentum signals of the Higgs boson. Starting 
from a scenario with relatively small masses of the supersymmetric (SUSY) 
particles, with concomitant large cross section of the background processes, 
one obtains a first conservative estimate of the background. Light chargino 
pair production plus two jets, lightest and next-to-lightest neutralino 
production plus two jets as well as slepton pair production plus two jets are 
identified as important contributions to the irreducible SUSY background. 
Light chargino and next-to-lightest neutralino production plus two jets and 
next-to-lightest neutralino pair production plus two jets give rise to 
reducible backgrounds, which can be larger than the irreducible ones in 
some scenarios. The relevant distributions are shown and additional cuts 
for MSSM background reduction are discussed. Extrapolation to larger squark 
masses is performed and shows that MSSM backgrounds are quite small for 
squark masses at the current exclusion limits.
\end{abstract}


\def\thefootnote{\arabic{footnote}}
\setcounter{page}{0}
\setcounter{footnote}{0}

\newpage

\section{Introduction}

One of the major tasks at the LHC is the search for Higgs bosons. In the
Minimal Supersymmetric Standard Model (MSSM), for a wide range of parameters,
one of the neutral Higgs bosons couples to vector bosons like a
Standard Model Higgs boson i.e.~with nearly no suppression of its coupling 
to gauge bosons with respect to the one in the Standard Model. For this Higgs boson, 
the search topologies of a Standard Model Higgs boson can be applied. 
The discovery potential for a MSSM Higgs boson at the LHC has been 
discussed in Refs.~\cite{CMSTDR, ATLASTDR}. 

\begin{figure}[b]
  \begin{center}
    \includegraphics[height=0.2\textwidth]{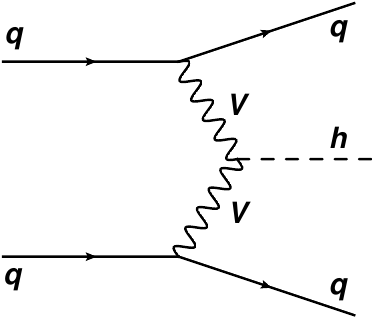}
  \end{center}
  \caption{$h$ production in weak boson fusion, with $V={W,Z}$.}
  \label{vbfdiag}
\end{figure}

A particularly interesting Higgs boson production channel is
weak boson fusion (VBF, Fig.~\ref{vbfdiag}) which has been studied
in~\cite{MSSMHiggsinVBF} for the MSSM. The two
tagging jets accompanying the VBF channel are an important characteristic
allowing for sensitivity also at lower Higgs boson mass values. For $H\to WW$
and $H\to \tau\tau$ decay, yielding dilepton plus missing transverse momentum
signatures, the major
Standard Model background is due to processes like top quark pair
production (plus jets),
 $W$~boson pair production plus two jets,
tau lepton pair production plus two jets,
or $Z$~boson
production plus two jets
.

In the MSSM, additional contributions to the background may arise from
production of superpartners 
of the Standard Model type particles. In previous studies these background
contributions have been taken into account in specific MSSM search channels
where the Higgs boson is produced in supersymmetric (SUSY) cascade 
decays~\cite{CMSTDR,Higgscascade,HiggscascadeATLAS}. For
Standard Model type processes SUSY background contributions have been discussed in the
context of calibration processes~\cite{Baer}. 
However, first MSSM Higgs boson searches at the LHC did not include SUSY
backgrounds \cite{MSSM_Higgs_ATLAS, MSSM_Higgs_CMS1, *MSSM_Higgs_CMS2} (in the corresponding
data analyses only the SM Higgs boson search topologies were used).

In this paper 
contributions from processes with SUSY particles to the background of the VBF
Higgs boson production channel with subsequent Higgs boson decay into
tau leptons or $W$~bosons are investigated. For the produced tau leptons and
$W$~bosons the fully leptonic decay modes into electrons and muons are taken
into account:
\begin{eqnarray}
  h \rightarrow & \tau^+\tau^- &\rightarrow \ell^+\ell^- \;
  \nu_\ell\bar{\nu}_\ell \nu_\tau\bar{\nu}_\tau\\ 
  h \rightarrow & W^+W^- &\rightarrow \ell^+\ell^- \; \nu_\ell\bar{\nu}_\ell
  \nonumber 
\end{eqnarray}
For searches of a relatively light MSSM Higgs boson (with mass around 120~GeV)
the VBF production channel with a
subsequent decay into  tau leptons is one promising discovery
channel~\cite{MSSMHiggsinVBF} whereas the channel with a subsequent decay into
$W$~bosons is less important as a discovery channel but might be useful
for measuring Higgs boson couplings \cite{HinWW_cpl}.

The final state of the signal processes is characterized by two charged leptons and 
missing energy from the decay of the Higgs boson, plus two jets from the scattered 
quarks. Background processes with SUSY particles can lead to the same final state 
topology and therefore contribute to the irreducible background: 
In the discussed scenarios, with the lightest neutralino as stable lightest supersymmetric 
particle (LSP), these background processes are chargino pair 
production plus two jets with subsequent decays of the charginos into a charged 
lepton, a lightest neutralino and neutrinos, lightest and next-to lightest 
neutralino production plus two jets where the next-to lightest neutralino decays 
into a charged lepton pair and a lightest neutralino as well as slepton pair 
production plus two jets with the sleptons decaying into charged leptons, 
lightest neutralinos and possibly neutrinos. 
Processes with other chargino and neutralino combinations, like 
next-to lightest neutralino pair production
plus two jets or production of a chargino and a next-to lightest neutralino plus two jets, 
contribute to the reducible background. These processes give rise to additional jets
or leptons which might be missed by the detector.
The goal of this paper is an 
estimate whether the SUSY background is already sufficiently suppressed and 
therefore is of no concern in the Higgs boson analyses or whether further cuts 
have to be applied for a sufficient reduction of potential SUSY backgrounds. To reach 
this goal a conservative estimate of the SUSY background is helpful. Therefore, 
our starting point is a scenario with relatively small masses of the SUSY particles 
(though these masses are disfavoured by the current LHC exclusion limits 
\cite{atlassusy1, *atlassusy2, *atlassusy3,
 cmssusy1, *cmssusy2, *cmssusy3}). Since SUSY cross sections generally 
decrease with increasing squark mass, the low mass scenario provides an upper 
bound for SUSY backgrounds in a much larger range of scenarios. In a second step, 
larger, more realistic squark and gluino masses are considered and the resulting 
further reduction of the SUSY backgrounds is investigated. 

While concerning the background only
the Standard Model processes have been 
discussed in great detail~\cite{CMSTDR, ATLASTDR, MSSMHiggsinVBF, HinWW_cpl},
the theoretical prediction of the VBF Higgs boson 
production  cross section 
is known with a high accuracy in the Standard Model as well as in the MSSM:
The next-to leading order (NLO) QCD 
contributions amount to about 10~\% of the leading  order cross
section~\cite{VBFNLOQCD1, *VBFNLOQCD2, *VBFNLOQCD3, *VBFNLOQCD4, VBFEW1,
  *VBFEW2}. The NLO electroweak corrections are of a 
similar size, about 5 -- 10~\% of the leading order cross
section~\cite{VBFEW1, *VBFEW2, VBFSUSY2}. Gluon induced contributions to the VBF channel
have been investigated~\cite{VBFgluonind} and the interference effects of
gluon and weak boson fusion have been studied~\cite{VBFGFint1, *VBFGFint2};
both types of contributions have been found to be very small. Recently,
approximate NNLO QCD corrections to VBF have been calculated and 
the theoretical uncertainty on the total cross section from higher order
corrections and the uncertainties due to the parton distribution functions was 
estimated to be at the 2~\% level each~\cite{VBFNNLOQCD}.
The NLO SUSY QCD corrections have found to be 
negligible~\cite{VBFSUSYQCD, VBFSUSY1,VBFSUSY2}  but also the NLO SUSY
electroweak corrections are rather small, for a large part of the MSSM
parameter space they are of the order of or even below 1~\% of the leading order
cross section~\cite{VBFSUSY1, VBFSUSY2}. In special regions of the parameter
space, especially for small SUSY masses~\cite{VBFSUSY1} and in the
non-decoupling region of the Higgs bosons~\cite{VBFSUSY2}, the corrections can
be larger; in the latter case they can result in contributions up to 10~\% of
the leading order cross section. We will focus on a light MSSM Higgs boson in
the decoupling region which means that it couples like a Standard Model Higgs
boson to the gauge bosons.

Our paper is organized as follows:
In Section~\ref{scenarios} we present the different MSSM scenarios which we
use in our analysis and list the processes which give rise to potentially significant SUSY
backgrounds to VBF Higgs production in these scenarios.
We discuss the tools used in the analysis as well as parameters and
approximations in 
Section~\ref{procedure}. 
In Sections \ref{irreducible} and \ref{reducible} we show our results for 
the irreducible and reducible SUSY backgrounds in a specific mSUGRA inspired scenario. 
The dependence on squark and gluino mass values is presented in Section~\ref{massdep}.
In Section~\ref{otherscenarios} we focus on three other scenarios, one with
light sleptons, another one with a very light LSP and a third one with small
mass differences of the sparticles in the chargino decay chain.
The situation for the 7 TeV LHC is discussed in Section~\ref{7tev}.
We conclude in Section~\ref{conclusion}.

\section{The scenarios}
\label{scenarios}

We choose MSSM scenarios with the lightest
neutralino $\chi^0_1$ being the LSP, which is
stable as we assume R-parity conservation.
One widely considered parameter set within this class of MSSM scenarios is the
SPS1a point \cite{SPSpoints}, where the parameters are given at the GUT-scale
in an mSUGRA type model and then evolved down to lower scales. 
The squark masses in the SPS1a scenario are relatively light which leads to 
higher production cross sections when squarks are involved. Therefore, this 
scenario is suited for finding an estimate of 
the upper bound of the SUSY contributions to the VBF background (note,
however, that this is a very conservative estimate  as the new measurements
from ATLAS and CMS \cite{atlassusy1, *atlassusy2, *atlassusy3, cmssusy1,
*cmssusy2, *cmssusy3} especially constrain scenarios
with light squarks and 
gluinos with the exclusion bounds $m_{\tilde g} \approx m_{\tilde{q}}
\gtrsim 800$~GeV).
In the scenario that we choose as a starting point for our investigations,  the
values of the
low scale parameters of the SPS1a scenario are used except for the one of the
top 
trilinear coupling $A_t$. Instead, the low scale value of $A_t$ is assumed as
$A_t = -733$~GeV 
according to the $m_h^{\text{max}}$ scenario~\cite{mhmax}. The new
modified SPS1a scenario (SPS1amod) then has a lightest Higgs boson with a mass
value of $m_h \approx 118$~GeV,
which is above the LEP exclusion bound (the corresponding Higgs mass in the
SPS1a scenario is slightly below the LEP exclusion bounds, $m_h \approx
112$~GeV). According to previous studies~\cite{SMHiggsVBFtotau,
  MSSMHiggsinVBF, CMSTDR, ATLASTDR} a lightest MSSM Higgs boson with the
SPS1amod mass value is accessible via the VBF Higgs production channel with a 
subsequent decay into tau leptons. This holds also for a Standard Model like
Higgs boson in the $h \rightarrow WW$ channel~\cite{VBFHinWW}.

One further characteristic of the SPS1a point -- which is preserved in our 
modified scenario -- is that the branching ratio for the decay channel of a 
light tau slepton, $\widetilde \tau_1$, into a tau lepton, $\tau$, and a 
lightest neutralino is $BR(\widetilde{\tau} \rightarrow \tau \chi^0_1) = 1$. 
The light charginos, $\chi_1^{\pm}$, decay dominantly
into a tau slepton and a corresponding neutrino, $\chi_1^{\pm} \rightarrow
\widetilde{\tau}_1^{\pm} \stackrel{\text{\tiny(}-\text{\tiny)}}{\nu_{\tau}}$. 
Furthermore, the second lightest neutralino decays only 
into two opposite charged leptons and a lightest neutralino, $BR(\chi^0_2
\rightarrow \ell^+ \ell^- \chi^0_1)= 1$, where the leptons $\ell$ include electrons,
muons and tau leptons, $\ell = \{e, \mu, \tau\}$. The most important channels
contributing to the irreducible background are 
\begin{eqnarray}
p p \rightarrow j j \,\, \chi_1^+ \,\, \chi_1^-
&\rightarrow& j j \,\, \ell^+ \, \chi_1^0 \, \nu_\ell \,\, \ell^- \chi_1^0 \, \bar{\nu}_\ell \label{eq_chargino}\\
p p \rightarrow j j \,\, \chi_1^0 \,\, \chi_2^0
&\rightarrow& j j \,\, \chi_1^0 \,\, \ell^+ \, \ell^- \, \chi_1^0 \label{eq_neutralino}
\end{eqnarray}
with $l = \{e, \mu, \tau\}$, see Sections~\ref{sps1amod-tautau} and \ref{sps1amod-WW}.

Subleading contributions to the irreducible background in the SPS1amod scenario come from processes involving a
direct slepton pair production accompanied by two jets with subsequent decays of 
the sleptons into corresponding leptons and invisible particles.
\begin{eqnarray}
p p \rightarrow j j \,\, \widetilde{e}_{L,R}^+ \,\, \widetilde{e}_{L,R}^-
\rightarrow j j \,\, \ell^+ \chi_1^0 \,\, \ell^- \chi_1^0 \,\, + \slashed{p}_T \\
p p \rightarrow j j \,\, \widetilde{\mu}_{L,R}^+ \,\, \widetilde{\mu}_{L,R}^-
\rightarrow j j \,\, \ell^+ \chi_1^0 \,\, \ell^- \chi_1^0 \,\, + \slashed{p}_T \\
p p \rightarrow j j \,\, \widetilde{\tau}_{1,2}^+ \,\, \widetilde{\tau}_{1,2}^-
\rightarrow j j \,\, \ell^+ \chi_1^0 \,\, \ell^- \chi_1^0 \,\, + \slashed{p}_T
\end{eqnarray}
While the $\widetilde{e}_R$, $\widetilde{\mu}_R$ and $\widetilde{\tau}_1$ decay 
directly into the corresponding lepton and a neutralino, roughly a quarter of all 
$\widetilde{e}_L$, $\widetilde{\mu}_L$ and $\widetilde{\tau}_2$ decay via a chargino which further decays,
leading to further missing transverse momentum $\slashed{p}_T$. 
On the production level the $\widetilde{\tau}_1$ pair production
channel gives cross sections about 2 orders of magnitude smaller than
the chargino and next-to-lightest neutralino production channels 
(14 fb with only very basic cuts (see jet cuts from Eq.~\eqref{cuts_tau_min})
compared to 2.6 pb  
for (\ref{eq_chargino}) and 1.4 pb for (\ref{eq_neutralino})). $\widetilde{e}_{L,R}$ and 
$\widetilde{\mu}_{L,R}$ pair production cross sections are of the same order
as for the 
$\widetilde{\tau}_1$ leptons within the SPS1amod scenario.
However including the additional cuts from the Higgs boson analysis they
increase the background by roughly 10\% of the chargino / next-to-lightest neutralino background
cross section.
For higher squark and gluino masses the slepton contributions can become larger
than the chargino and next-to-lightest neutralino channels (see Sections~\ref{sleptons-sps1amod}~and~\ref{massdep}).

Several SUSY processes lead to a $\ell^+\ell^- j j + \slashed{p}_T$ signature
plus additional jets and leptons. As these additional particles
might escape undetected, those processes contribute to the reducible background of
VBF Higgs boson production. 
Details of our modeling of visible particles and our selection criteria
are given in Section~\ref{reducible}. 
In the SPS1amod scenario the dominant channels which contribute to the reducible background are:
\begin{eqnarray}
p p \rightarrow j j \,\, \chi_1^+ \,\, \chi_2^0
&\rightarrow& j j \,\, \ell^+ \, \chi_1^0 \, \nu_\ell \,\, \ell^+ \, \ell^- \, \chi_1^0    \label{eq_CH1pN2}\\
p p \rightarrow j j \,\, \chi_1^- \,\, \chi_2^0
&\rightarrow& j j \,\, \ell^- \, \chi_1^0 \, \bar{\nu}_\ell \,\, \ell^+ \, \ell^- \, \chi_1^0   \label{eq_CH1mN2}\\
p p \rightarrow j j \,\, \chi_2^0 \,\, \chi_2^0
&\rightarrow& j j \,\, \ell^+ \, \ell^- \, \chi_1^0 \,\, \ell^+ \, \ell^- \, \chi_1^0   \label{eq_N2N2}
\end{eqnarray}
More than two jets can occur from hadronic tau lepton or W boson decays.
These processes contributing to the reducible background are discussed in detail in Section~\ref{reducible}.
We also look at several other processes that turn out to give only small contributions to the background.
They are listed in Sections~\ref{procedure} and \ref{sps1asummary}.

 We use the program
{\tt SUSYHIT 1.3}~\cite{SUSYHIT,SDECAY,SuSpect,HDECAY} for the evolution to lower scales with the default
spectrum calculator {\tt SuSpect}~\cite{SuSpect} (changing to the spectrum
calculator {\tt SPheno 3.0.beta}~\cite{SPheno} gives comparable results with only
slight differences as expected). The resulting values are fed into the program 
{\tt FeynHiggs 2.6.5}~\cite{FeynHiggs1, *FeynHiggs2, *FeynHiggs3, *FeynHiggs4} for  precise 
values of the Higgs boson masses. The corresponding SLHA output
file~\cite{SLHA1, *SLHA2} is used as an input for the further calculations. 
As Standard Model input parameters in the spectrum calculator we take
\begin{equation}
\begin{array}{rclcrcl}
  \alpha_{em}^{-1}(M_Z) &=& 127.934 & \quad &
  G_F &=& 1.16639 \cdot 10^{-5} \;\text{GeV}^{-2} \\
  \alpha_s(M_Z) &=& 0.1172 & \quad &
   M_Z &=& 91.187 \;\text{GeV} \\
   M_b(M_b) &=& 4.25 \;\text{GeV} & \quad &
   M_t &=& 172.5 \;\text{GeV} 
\end{array}
\end{equation}
with the top quark mass value from \cite{topmass}.

For our analysis of the SUSY background to the VBF Higgs production channel
with a subsequent decay into two W bosons, we also consider a further modified
SPS1a  
scenario (SPS1amod2), where we aim for a higher Higgs mass. Therefore we
increase the 
 soft SUSY breaking parameter values in the top squark sector, 
$M_{q_3L}=881 \;\textrm{GeV}$, $M_{tR}=808 \;\textrm{GeV}$ and  
$A_t=-1833 \;\textrm{GeV}$, at the low scale. This leads to a Higgs boson mass
 of 
$m_h \approx 124 \;\text{GeV}$. As in the SPS1amod case, the chargino 
background channel is not changed when taking the modified scenarios instead
of the original  
SPS1a scenario. However, the neutralino channel cross section is a bit reduced
for 
SPS1amod2: The particle masses calculated
with 
SUSYHIT~\cite{SUSYHIT} are higher order corrected and therefore the branching 
ratios of the next-to-lightest neutralino decay are changed towards a
preference  
of a decay into tau leptons which leads to a decrease of the cross section due
to  
the further decay of the tau leptons into muons and electrons.

As the first LHC results become public and start to put constraints on low squark
and gluino
masses \cite{atlassusy1, *atlassusy2, *atlassusy3, cmssusy1,
*cmssusy2, *cmssusy3}, it becomes even more important to study the mass 
dependence of the SUSY backgrounds. Therefore we look into a series of scenarios, where 
we modify the squark and gluino mass related soft SUSY breaking terms:
\begin{itemize}
 \item For final states containing only quarks of the first and second generation,
       we modify $M_{q_1L}$, $M_{q_2L}$, $M_{uR}$, $M_{dR}$, $M_{cR}$ and $M_{sR}$
       by a factor $(1+\xi)$ with $0 \leq \xi \leq 2$. We also modify $M_3$ to
       ensure that the gluino is always heavier than the squarks. This leads
       to average 
       squark masses from 553 to 1581 GeV.
 \item When there is at least one b-quark in the final state, the mass of the
   stop 
       has a large impact on the cross section. So for this final state, we
       increase 
       the parameters that influence the third generation squark masses
       $M_{q_3L}$, 
       $M_{tR}$, $M_{bR}$, $A_{t}$, $A_{b}$ and $M_3$ by a factor $(1+\rho)$ with 
       $0 \leq \rho \leq 1$.
\end{itemize}

We also check a scenario where all SUSY particle masses are raised by increasing
$M_0$. The other values are changed according to the SPS1a slope~\cite{SPSpoints},
\begin{align}
M_0 = - A_0 = 130 \text{ GeV}, \qquad\ M_{\frac{1}{2}} = 2.5 M_0\,.
\end{align}
The parameters $A_t$, $M_L^{\widetilde{t}}$ and $M_R^{\widetilde{t}}$ are 
changed analogous to the procedure in SPS1amod2 in order to stay within the 
$m_h^{\text{max}}$ scenario~\cite{mhmax}. The SUSY mass values are then 
about 30\% higher than in the SPS1a scenario. 

Moving further away from the mSUGRA restrictions, we also study scenarios with different 
slepton masses, where the stau is heavier than the selectron and the smuon 
($m_{\widetilde{\tau}_1}=334.8 \;\text{GeV}$, 
 $m_{\widetilde{e}_L}=m_{\widetilde{\mu}_L}=141.9 \;\text{GeV}$). This is achieved by
changing the low-energy SUSY breaking parameters $M_{eL}$, $M_{\mu L}$, $M_{\tau L}$ and $M_{\tau R}$. 
For these slepton masses,
we look at two scenarios with different squark masses corresponding to $\xi=0$ and $\xi=1.0$.
These scenarios lead to an increased number of leptons in the final state, as the dominant
decay chain no longer involves tau leptons, but directly first and second generation (s)leptons.
So one decay with $BR(\tau \rightarrow \nu_\tau \,l \, \bar{\nu}_l)
\approx 0.35$ with $l = \{e,\,\mu\}$ is
avoided, which in addition leads to harder leptons.

Afterwards we check the influence of an almost massless LSP on our first scenario SPS1a-mod, where
the stau mass is set to about half the value of the chargino mass. This is no longer a genuine MSSM
scenario, but can illustrate the effect of the lepton hardness on our SUSY background 
processes.

Finally we illustrate the effect of two particles in the decay chain with a small mass difference.
This is a mSUGRA scenario, where we again start with SPS1a and increase the trilinear
coupling $A_0$ from -100 GeV to -750 GeV. This results in a small stau mass of 108 GeV, close
to the mass of the LSP which is at 99.3 GeV.

Some relevant SUSY particle masses and branching ratios in these scenarios are summarized in
Tables \ref{masstable} and \ref{brtable}.
\tabcolsep1.9mm
\begin{table}[t]
  \begin{center}
  \begin{tabular}{|c||c|c|c|c|c|}
  \hline
   & SPS1amod & $\xi= 1.0 $ & $\rho= 1.0$ & SPS1a-slope & light sleptons\\
  \hline
  \hline
  $m_{\widetilde{u}_L}=m_{\widetilde{c}_L}$ & 560.2 GeV & 1090.9 GeV & 576.1 GeV & 713.7 GeV & 560.1 GeV \\
  $m_{\widetilde{d}_L}=m_{\widetilde{s}_L}$ & 565.6 GeV & 1093.7 GeV & 581.6 GeV & 717.9 GeV & 565.6 GeV \\
  $m_{\widetilde{u}_R}=m_{\widetilde{c}_R}$ & 543.9 GeV & 1055.4 GeV & 559.6 GeV & 691.4 GeV & 543.9 GeV \\
  $m_{\widetilde{d}_R}=m_{\widetilde{s}_R}$ & 543.6 GeV & 1052.0 GeV & 559.3 GeV & 690.0 GeV & 543.6 GeV \\
  $m_{\widetilde{b}_1}$                     & 518.2 GeV & 533.5 GeV & 1003.6 GeV & 685.8 GeV & 518.2 GeV \\
  $m_{\widetilde{b}_2}$                     & 544.9 GeV & 560.4 GeV & 1047.2 GeV & 852.5 GeV & 544.9 GeV \\
  $m_{\widetilde{t}_1}$                     & 346.3 GeV & 365.6 GeV &  768.3 GeV & 665.4 GeV & 346.2 GeV \\
  $m_{\widetilde{t}_2}$                     & 608.4 GeV & 620.3 GeV & 1060.0 GeV & 976.1 GeV & 608.4 GeV \\
  \hline
  $m_{\widetilde{g}}$                       & 607.6 GeV & 1165.8 GeV & 1140.7 GeV & 783.6 GeV & 607.6 GeV \\
  \hline
  $m_{\chi_1^+}$                            & 181.1 GeV & 183.8 GeV & 182.0 GeV & 244.6 GeV & 180.2 GeV \\
  $m_{\chi_2^0}$                            & 181.6 GeV & 184.4 GeV & 182.5 GeV & 244.9 GeV & 180.8 GeV \\
  $m_{\chi_1^0}$                            &  97.5 GeV &  98.1 GeV &  97.6 GeV & 129.9 GeV &  97.1 GeV \\
  \hline
  $m_{\widetilde{e}_L}=m_{\widetilde{\mu}_L}$ & 199.7 GeV & 199.6 GeV & 199.6 GeV & 256.1 GeV & 141.9 GeV \\
  $m_{\widetilde{e}_R}=m_{\widetilde{\mu}_R}$ & 142.6 GeV & 142.6 GeV & 142.7 GeV & 181.7 GeV & 142.6 GeV \\
  $m_{\widetilde{\tau}_1}$                  & 133.0 GeV & 133.1 GeV & 132.8 GeV & 172.9 GeV & 334.8 GeV \\
  $m_{\widetilde{\tau}_2}$                  & 203.9 GeV & 203.8 GeV & 204.1 GeV & 259.1 GeV & 397.4 GeV \\
  \hline
  $m_{h}$                                   & 118.2 GeV & 118.4 GeV & 122.7 GeV & 124.0 GeV & 118.2 GeV \\
  \hline
  \end{tabular}
  \end{center}
  \caption{Particle masses in the scenarios SPS1amod,
    with higher first and second generation squark masses ($\xi=1.0$), with
    higher 
    stop masses ($\rho=1.0$), SPS1a-slope and with light selectrons and
    smuons.}  
  \label{masstable}
\end{table}

\tabcolsep1.9mm
\begin{table}[p]
  \begin{center}
  \begin{tabular}{|c||c|c|c|c|c|}
  \hline
   & SPS1amod & $\xi= 1.0 $ & $\rho= 1.0$ & SPS1a-slope & light sleptons\\
  \hline
  \hline
  $BR(\widetilde{u}_L\rightarrow\chi_1^+\,d)$ & 65.0 \% & 62.9 \% & 65.1 \% & 65.1 \% & 65.0 \% \\
  $BR(\widetilde{u}_L\rightarrow\chi_2^0\,u)$ & 31.8 \% & 30.7 \% & 31.9 \% & 32.1 \% & 31.8 \% \\
  $BR(\widetilde{u}_L\rightarrow\chi_1^0\,u)$ &  0.7 \% &  0.6 \% &  0.7 \% &  1.0 \% &  0.7 \% \\
  \hline
  $BR(\widetilde{u}_R\rightarrow\chi_1^0\,u)$ & 98.6 \% & 97.7 \% & 98.7 \% & 99.3 \% & 98.6 \% \\
  \hline
  $BR(\widetilde{t}_1\rightarrow\chi_1^+\,b)$ & 80.7 \% & 74.9 \% & 21.1 \% & 44.1 \% & 80.7 \% \\
  \hline
  $BR(\widetilde{b}_1\rightarrow\chi_1^0\,b)$ &  3.1 \% &  3.1 \% &  1.0 \% & 70.2 \% &  3.1 \% \\
  $BR(\widetilde{b}_1\rightarrow\chi_2^0\,b)$ & 21.4 \% & 22.1 \% & 12.0 \% &  6.0 \% & 21.4 \% \\
  \hline
  $BR(\widetilde{b}_2\rightarrow\chi_1^0\,b)$ & 14.0 \% & 13.0 \% & 29.5 \% &  0.7 \% & 14.0 \% \\
  $BR(\widetilde{b}_2\rightarrow\chi_2^0\,b)$ & 11.1 \% &  9.9 \% &  1.2 \% & 12.2 \% & 11.1 \% \\
  \hline
  $BR(\chi_1^+\rightarrow \widetilde{\tau}^+_1\,\nu_\tau)$   & 95.8 \% & 93.3 \% & 95.6 \% & 74.4 \% &  0.0 \% \\
  $BR(\chi_1^+\rightarrow \ell^+\,\widetilde{\nu}_\ell)$  &  0.0 \% &  0.0 \% &  0.0 \% &  0.1 \% & 71.4 \% \\
  $BR(\chi_1^+\rightarrow \widetilde{\ell}^+_L\,\nu_\ell)$   &  0.0 \% &  0.0 \% &  0.0 \% &  0.0 \% & 28.5 \% \\
  $BR(\chi_1^+\rightarrow W^+\,\chi_1^0)$                 &  4.2 \% &  6.7 \% &  4.4 \% & 25.5 \% &  0.1 \% \\
  \hline
  $BR(\chi_2^0\rightarrow \widetilde{\ell}_R^\pm\,\ell^\mp)$       & 11.9 \% & 12.3 \% & 10.6 \% &  8.1 \% &  0.4 \% \\
  $BR(\chi_2^0\rightarrow \widetilde{\ell}_L^\pm\,\ell^\mp)$       &  0.0 \% &  0.0 \% &  0.0 \% &  0.0 \% & 36.4 \% \\
  $BR(\chi_2^0\rightarrow \widetilde{\tau}_1^\pm\,\tau^\mp)$       & 88.1 \% & 87.0 \% & 89.4 \% & 87.3 \% &  0.0 \% \\
  $BR(\chi_2^0\rightarrow \widetilde{\nu}_\ell\,\bar{\nu}_\ell)$   &  0.0 \% &  0.2 \% &  0.0 \% &  0.3 \% & 63.2 \% \\
  $BR(\chi_2^0\rightarrow \widetilde{\nu}_\tau\,\bar{\nu}_\tau)$   &  0.0 \% &  0.5 \% &  0.0 \% &  0.8 \% &  0.0 \% \\
  $BR(\chi_2^0\rightarrow \chi_1^0 \,Z)$                           &  0.0 \% &  0.0 \% &  0.0 \% &  3.5 \% &  0.0 \% \\
  \hline
  $BR(\widetilde{\ell}_{L}^\pm\rightarrow \chi_1^0 \, \ell^\pm)$   & 63.7\%  & 70.7\%  &  66.1\% &  87.8\% & 100.0\% \\
  $BR(\widetilde{\ell}_{L}^\pm\rightarrow \chi_1^\pm \, \nu_\ell)$ & 23.3\%  & 18.8\%  &  21.8\% &   8.0\% &   0.0\% \\
  \hline
  $BR(\widetilde{\ell}_{R}^\pm\rightarrow \chi_1^0 \, \ell^\pm)$   & 100.0\% & 100.0\% & 100.0\% & 100.0\% & 100.0\% \\
  \hline
  $BR(\widetilde{\tau}_{1}\rightarrow \chi_1^0 \, \tau^\pm)$   & 100.0\% & 100.0\% & 100.0\% & 100.0\% & 94.0\% \\
  \hline
  $BR(\widetilde{\tau}_{2}\rightarrow \chi_1^0 \, \tau^\pm)$   & 64.8\%  &  70.4\% &  66.6\% &  85.8\% & 12.6\% \\
  $BR(\widetilde{\tau}_{2}\rightarrow \chi_1^\pm \, \nu_\tau)$ & 22.5\%  &  19.0\% &  21.4\% &   9.3\% & 54.7\% \\
  \hline
  $BR(h \rightarrow W^+ \, W^-)$                               &  8.2\%  &   8.3\% &  12.6\% &  15.1\% &  8.2\% \\
  $BR(h \rightarrow \tau^+ \tau^-)$                            &  7.4\%  &   7.4\% &   7.0\% &   6.7\% &  7.4\% \\
  \hline
  \end{tabular}
  \end{center}
  \caption{Branching ratios in the scenarios SPS1amod,
    with higher first and second generation squark masses ($\xi=1.0$), with
    higher 
    stop masses ($\rho=1.0$), SPS1a-slope and with light selectrons and smuons.
    Here, $\stackrel{\text{\tiny(}\sim\text{\tiny)}}{\ell}$ means the combined
    (s)electron and (s)muon channel. The branching  
    ratios of the $\widetilde{d}$ decays are comparable to the ones
    listed for $\widetilde{u}$ decays.} 
  \label{brtable}
\end{table}

\section{The procedure}
\label{procedure}

For our investigation we divide the processes in a production part with unstable SUSY particles in the final state 
and in a decay part which includes the decay chain of these SUSY particles.  
In the following we will describe the procedure and discuss its validity for the dominant irreducible background 
processes. The procedure is only slightly modified for the reducible background processes, these modifications
are illustrated at the end of this section.

We use the program {\tt MadGraph/MadEvent 4.4}~\cite{Madgraph} 
to generate parton level events with two jets and a pair of charginos or a 
lightest and next-to-lightest neutralino in the final state, respectively,
\begin{align}\label{charginoprod}
p p &\rightarrow j j \, \chi_1^+ \, \chi_1^- \,, \\\label{neutralinoprod}
p p &\rightarrow j j \, \chi_1^0 \, \chi_2^0 \,
\end{align}
where $p$ denotes the incoming protons with quarks of the first and second
generation and gluons, $p = \{u,\,d,\,c,\,s,\,
\bar{u},\,\bar{d},\,\bar{c},\,\bar{s},\,g\}$, and $j$ is an outgoing jet at
parton level, i.e. $j = \{u,\,d,\,c,\,s,\,
\bar{u},\,\bar{d},\,\bar{c},\,\bar{s},\,g\}$. The slepton pair production
processes
\begin{align}
p p &\rightarrow j j \, \widetilde{e}_{L,R}^+ \, \widetilde{e}_{L,R}^- \,, \nonumber\\
p p &\rightarrow j j \, \widetilde{\mu}_{L,R}^+ \, \widetilde{\mu}_{L,R}^- \,, \nonumber \\
p p &\rightarrow j j \, \widetilde{\tau}_{1,2}^+ \, \widetilde{\tau}_{1,1}^- 
\label{sleptonproc}
\end{align}
are generated with {\tt MadGraph 5}, version~1.3~\cite{Madgraph5}.
We abbreviate this process class by $\widetilde{\ell}^+\,\widetilde{\ell}^-\,jj$
in the following.
Final states containing at least
one b-quark are generated separately. In this case we also allow b-quarks in the
initial state. Contributions with at least one b-quark in the initial state,
but none in the final state are very small and can be neglected.
For all our calculations we use the parton distribution functions 
{\tt CTEQ6l1} \cite{cteq}. Most parts of the analysis are done for a center of
mass energy of 14 TeV at the LHC,
but in Sect.~\ref{7tev} results for 7 TeV are shown.

\begin{figure}[bt]
  \begin{center}
    \includegraphics[height=0.30\textwidth]{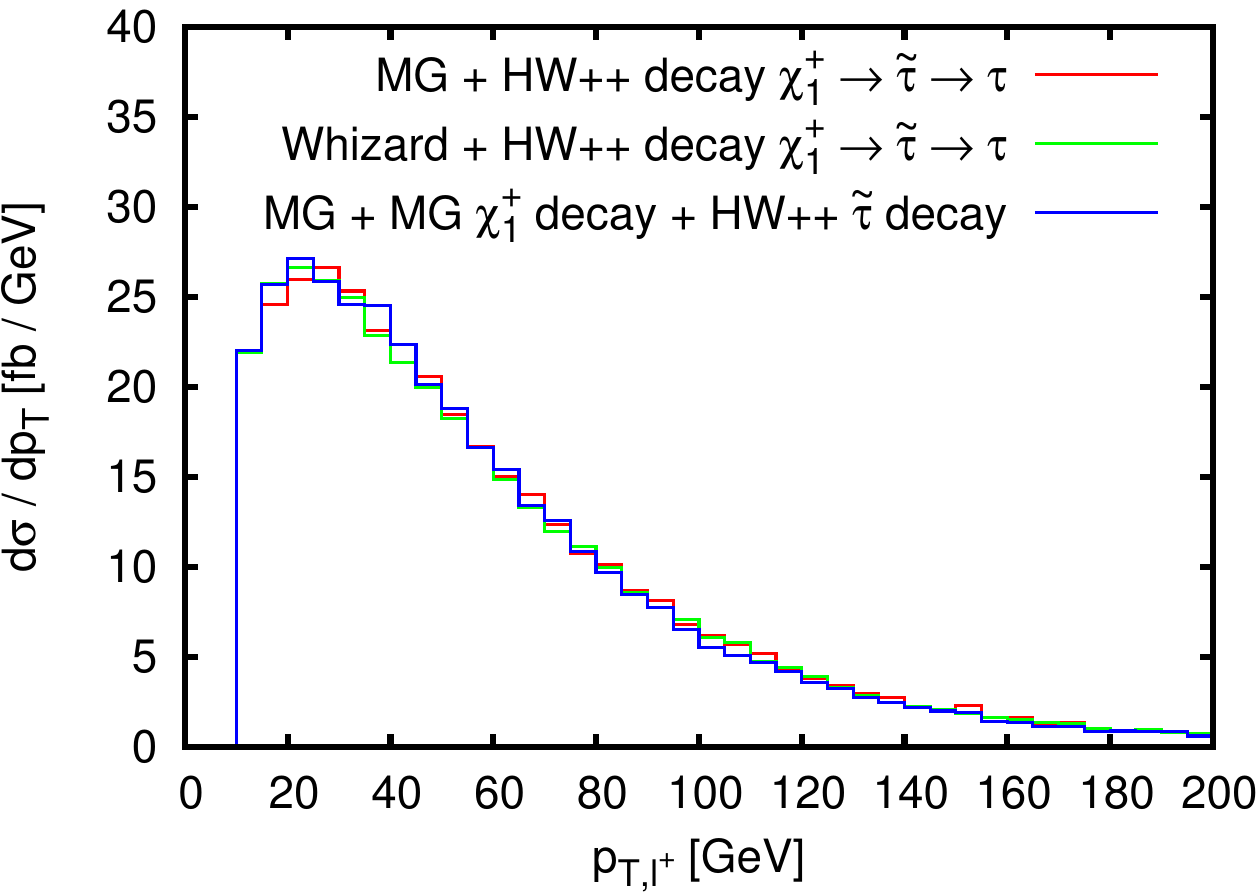}
    \hskip20pt
    \includegraphics[height=0.30\textwidth]{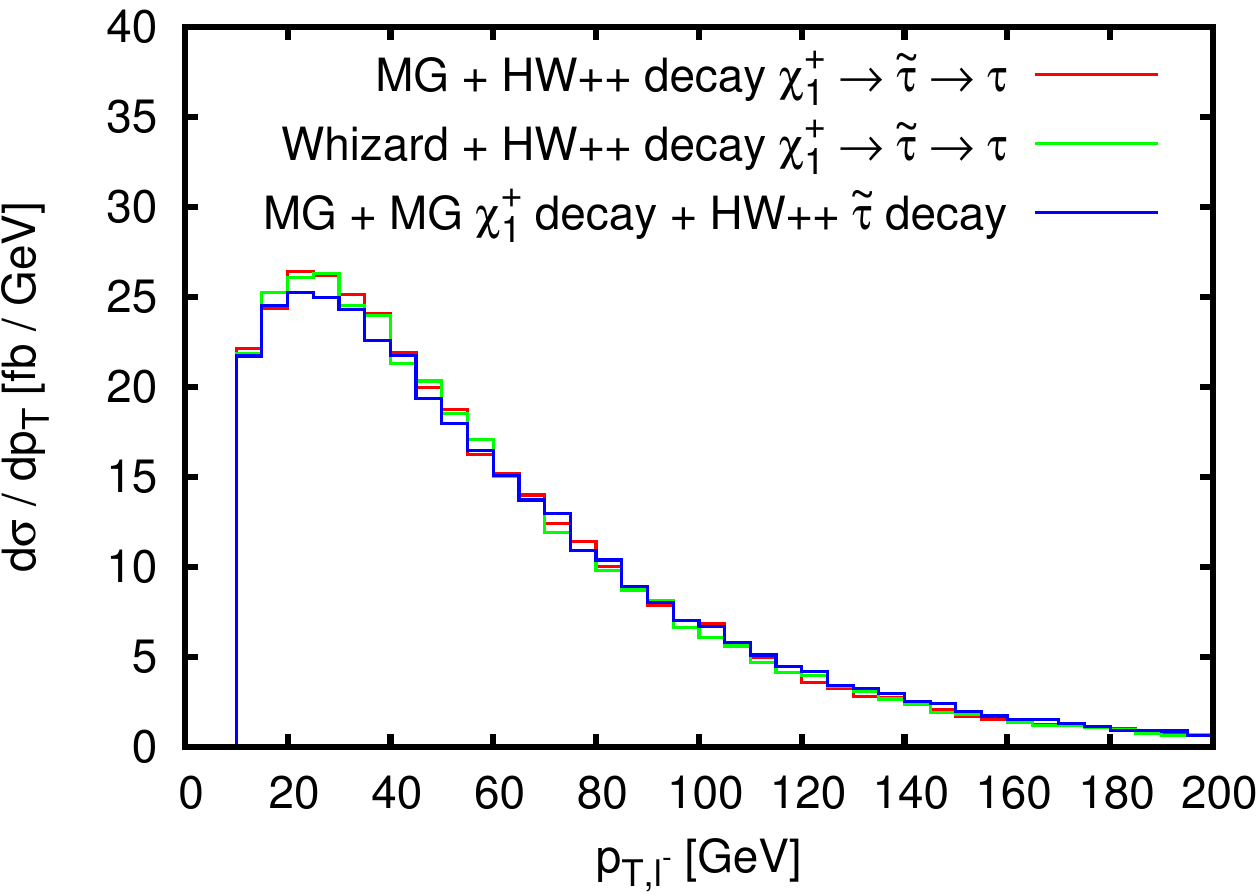}
    \vskip20pt
    \includegraphics[height=0.30\textwidth]{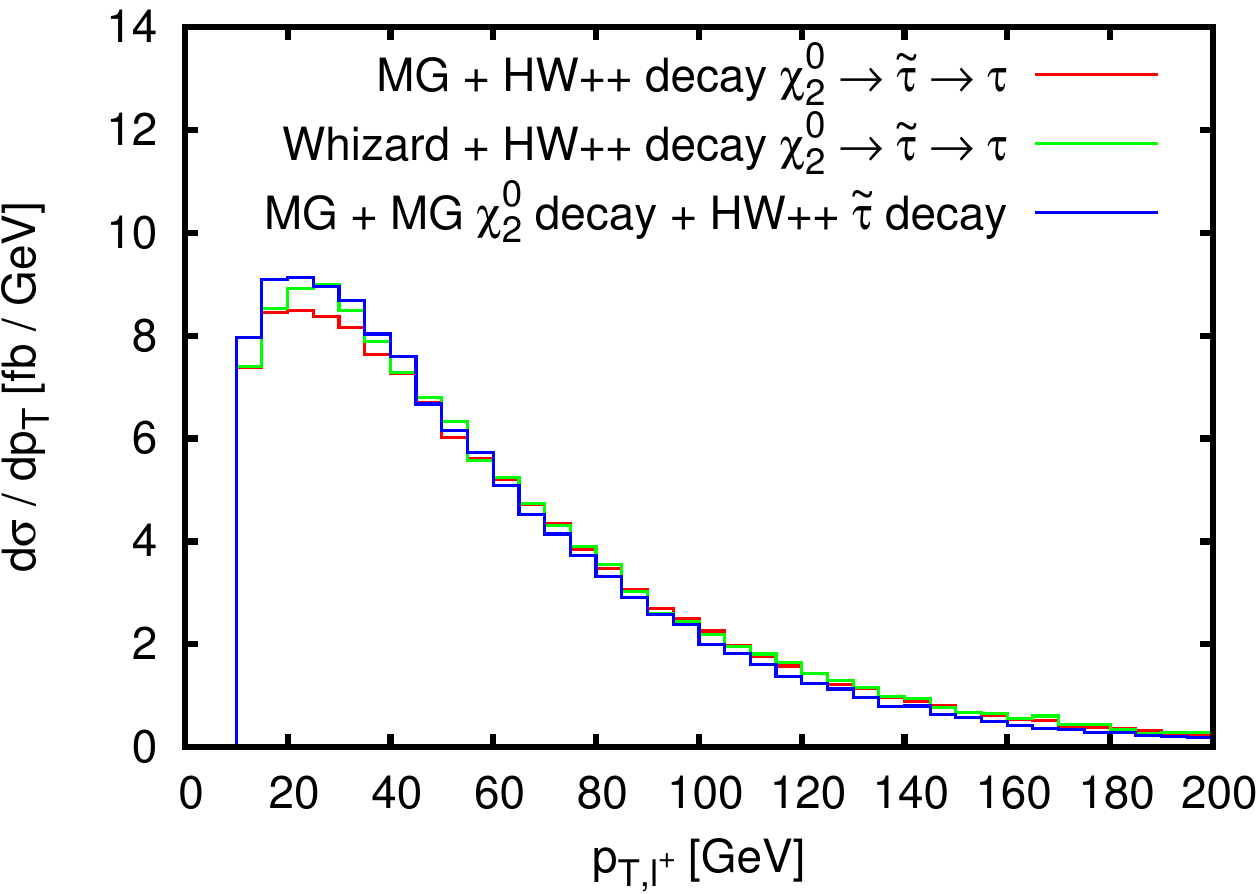}
    \hskip20pt
    \includegraphics[height=0.30\textwidth]{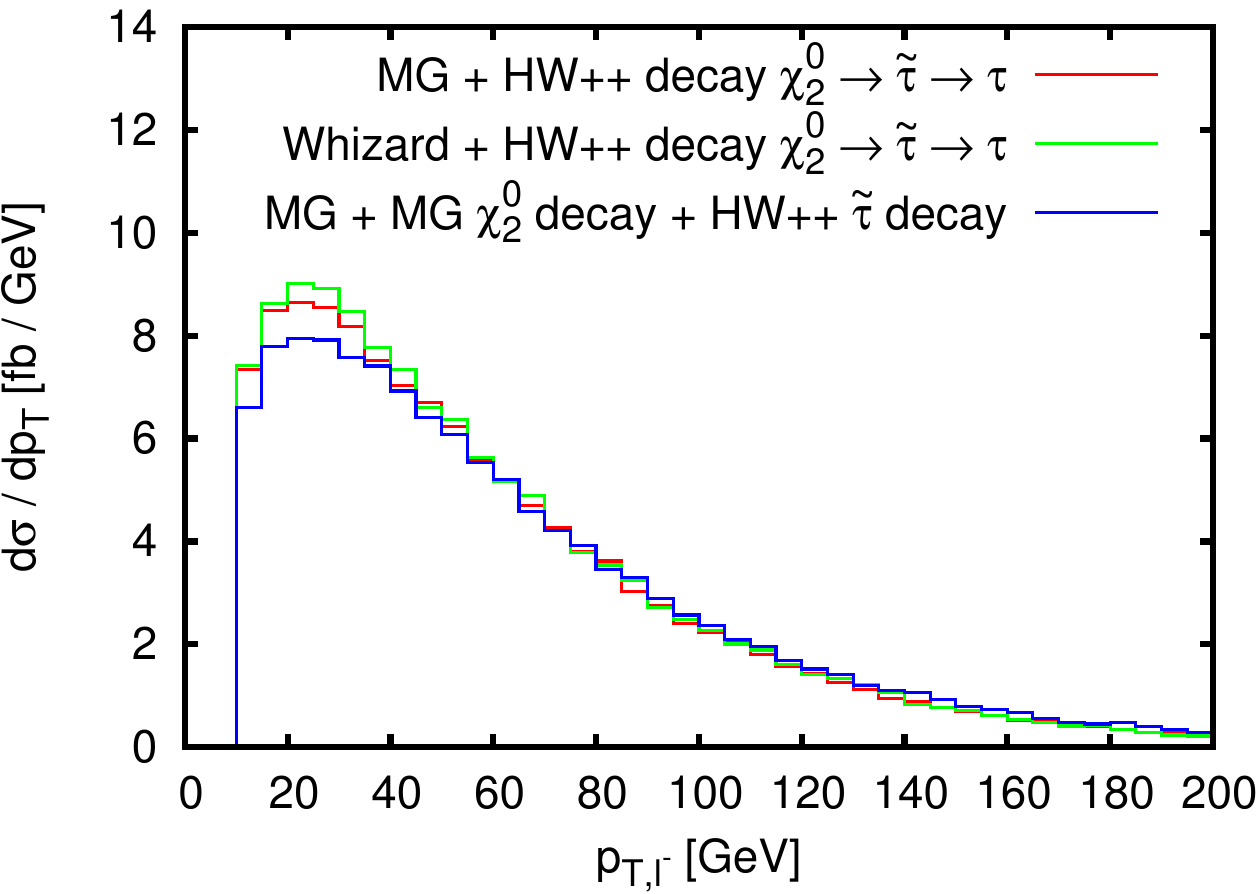}
  \end{center}
  \caption{Comparison of {\tt MadGraph} and {\tt Herwig++} performing the
           $\chi_1^\pm$ and $\chi_2^0$ decay in the process from Eq.~\eqref{charginostau} and Eq.~\eqref{neutralinostau}, 
           respectively, with scales $\mu_F=\mu_R=M_Z$. 
           The result of {\tt Whizard} with the complete decay calculated by {\tt Herwig++} is also shown.
           The $p_T$ distribution of the $\tau^+$ is shown on the left, while
           the right plot shows the same for $\tau^-$.}
  \label{MGHerwig}
\end{figure}

We compare the $\chi_1^+\,\chi_1^-\,jj$ and $\chi_2^0\,\chi_1^0\,jj$ cross sections 
to the ones generated with the program {\tt Whizard 2.0}~\cite{Whizard1, *Whizard2} and find 
agreement
(see Fig.~\ref{MGHerwig} for the comparison of the decay channels 
into stau leptons).

\begin{figure}[t]
  \begin{center}
    \includegraphics[height=0.30\textwidth]{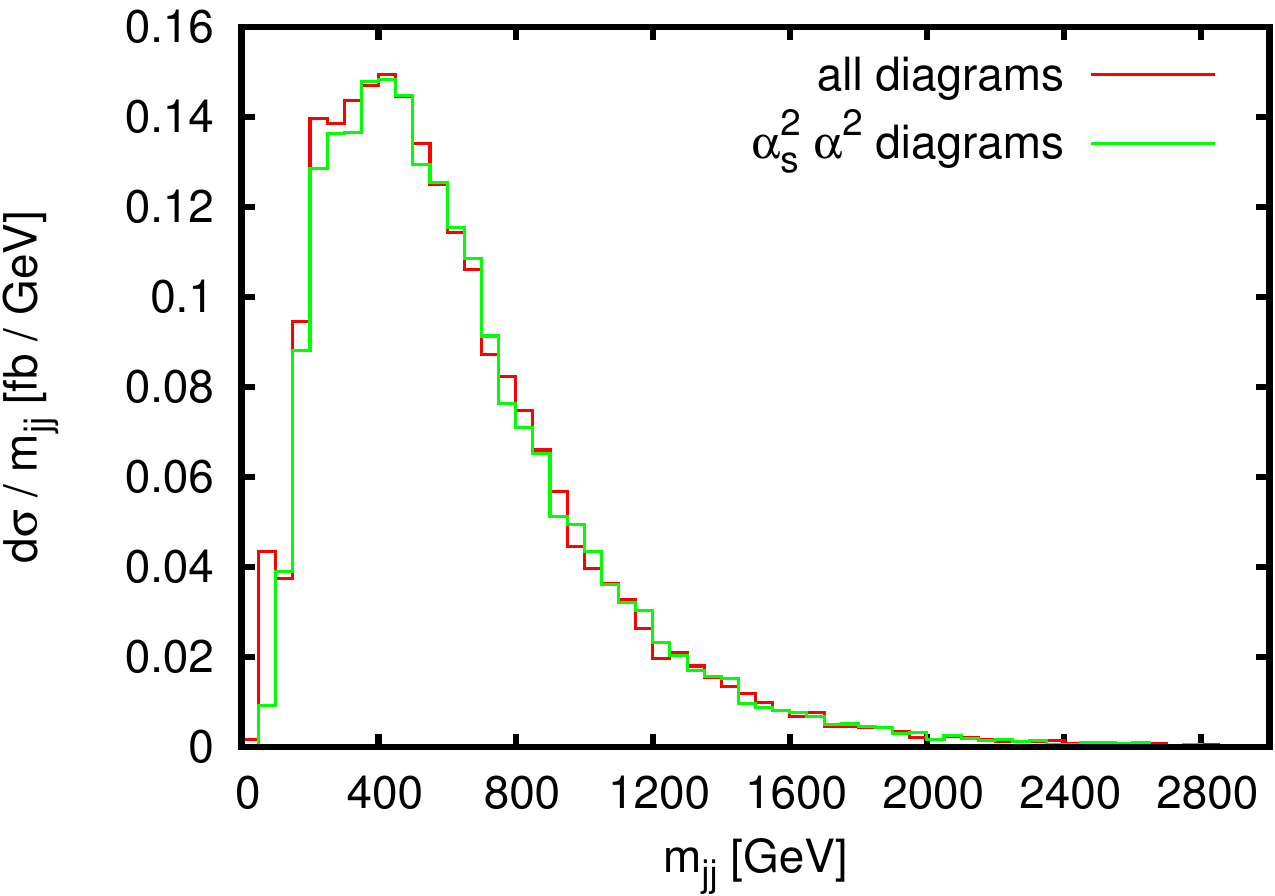}
    \hskip20pt
    \includegraphics[height=0.30\textwidth]{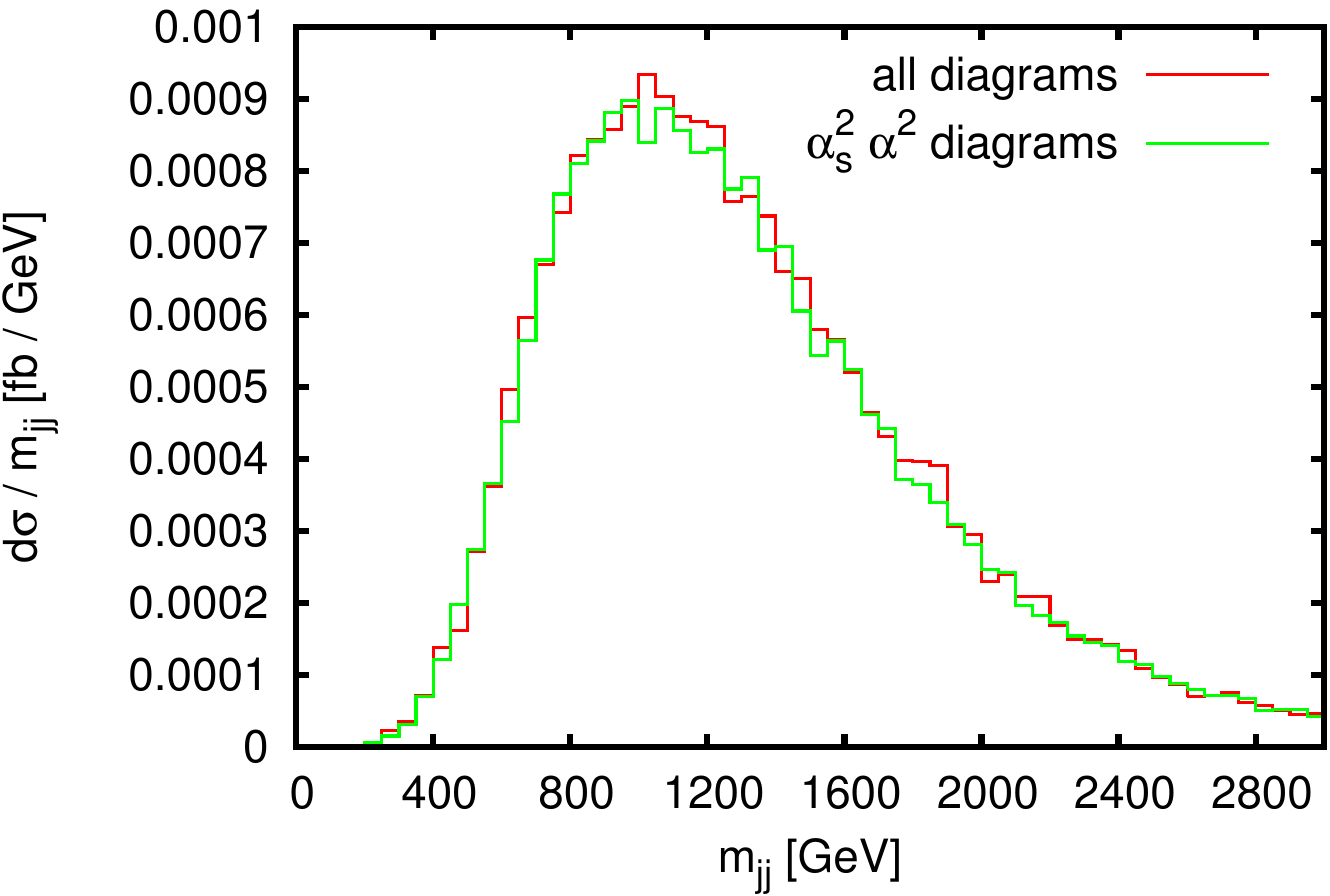}
  \end{center}
  \caption{Comparison of full process vs. approximation for $\chi_2^0\chi_1^0jj$ 
           production with minimal jet cuts (left) and an additional rapidity 
           separation cut of $\Delta\eta \geq 4.2$ (right) in the dijet mass.}
  \label{neutralinoapprox}
\end{figure}

For the neutralino production we take only the contributions of 
$\mathcal O(\alpha_s^2 \alpha^2)$ into account. We checked that this is 
a reasonable approximation as the difference compared to the full process
is very small.
Furthermore events that we discard via this approximation show a very small 
rapidity separation between the tagging jets and have a small invariant jj mass.
They would have been cut away by the requirement of a large rapidity 
separation for our signal processes (Fig.~\ref{neutralinoapprox}).

\begin{figure}[t]
  \begin{center}
    \includegraphics[height=0.2\textwidth]{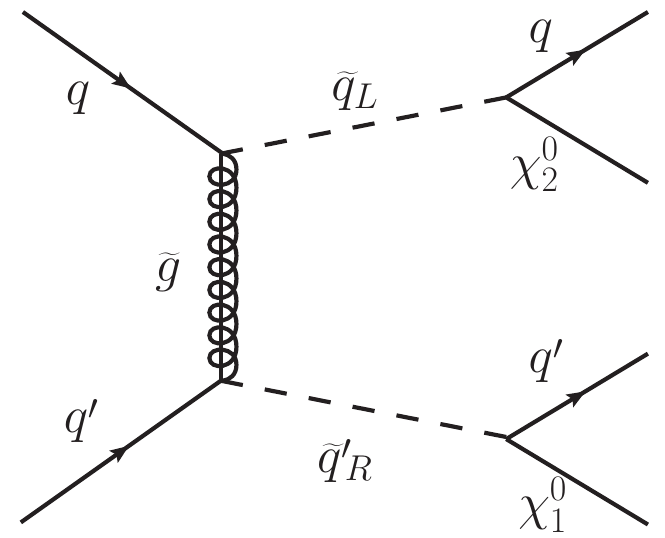}
  \end{center}
  \caption{Dominant Feynman Graph for $\chi_1^0\chi_2^0jj$ production in SPS1amod scenario.}
  \label{neutrprod}
\end{figure}

The  $\mathcal O(\alpha_s^2 \alpha^2)$ contributions of the $\chi_1^0\chi_2^0jj$ 
production correspond to the squark production diagrams with a subsequent
decay into neutralinos shown in Fig.~\ref{neutrprod}. As those diagrams give the
dominant contribution, the renormalization scale $\mu_R$ and the factorization
scale $\mu_F$ are chosen of the order of the squark masses, which is 
$\mu_R = \mu_F = 550$~GeV for the SPS1amod and SPS1amod2 scenario. 

\begin{figure}[t]
  \begin{center}
    \includegraphics[height=0.2\textwidth]{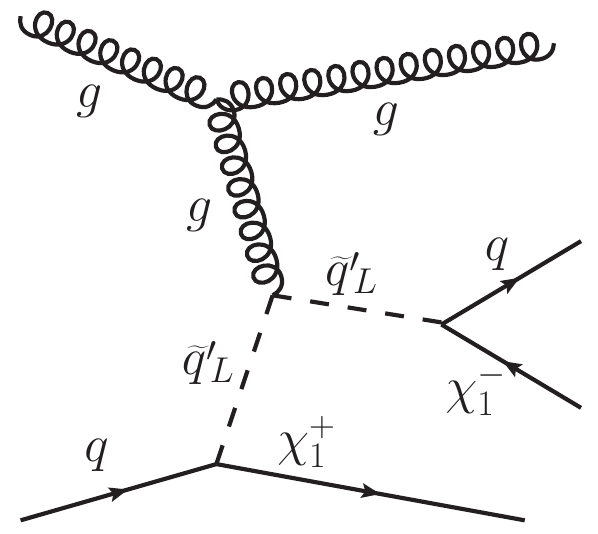}
    \hskip20pt
    \includegraphics[height=0.2\textwidth]{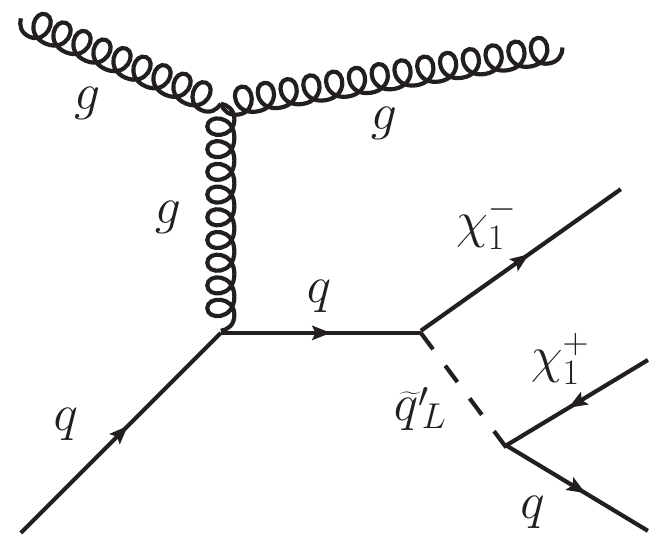}
    \hskip20pt
    \includegraphics[height=0.2\textwidth]{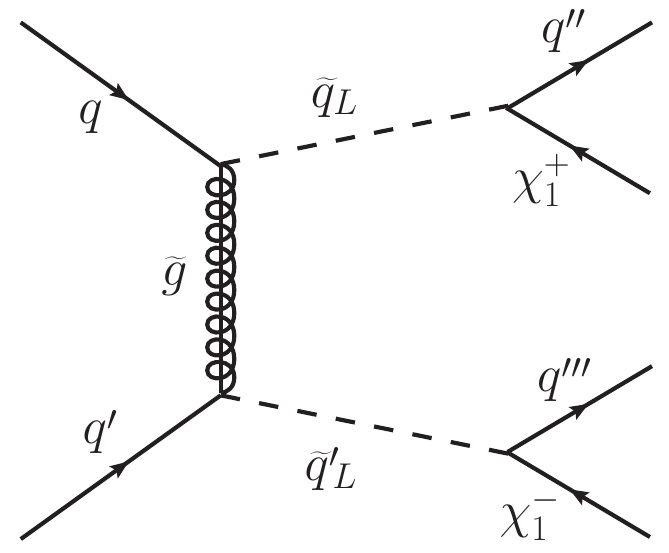}
  \end{center}
  \caption{Dominant Feynman graphs for $\chi_1^+\chi_1^-jj$ production in SPS1amod scenario.}
  \label{charprod}
\end{figure}

\begin{figure}[t]
  \begin{center}
    \includegraphics[height=0.2\textwidth]{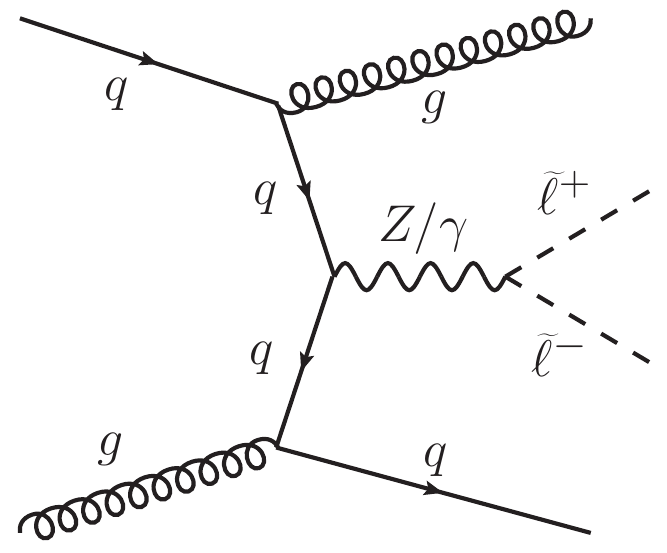}
    \hskip20pt
    \includegraphics[height=0.2\textwidth]{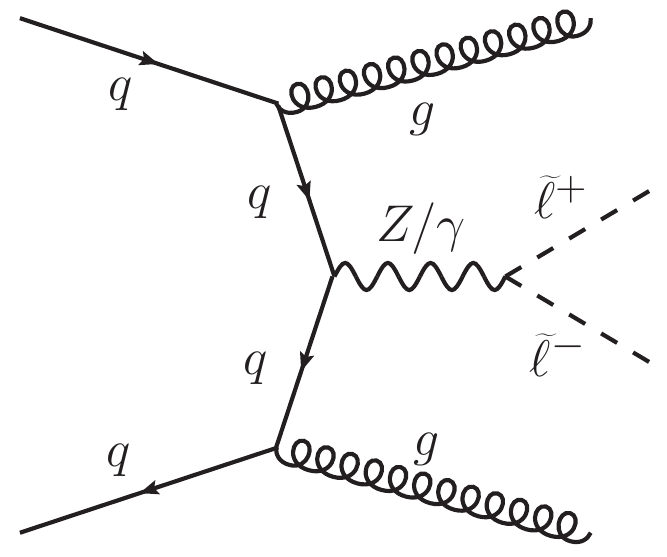}
    \hskip20pt
    \includegraphics[height=0.2\textwidth]{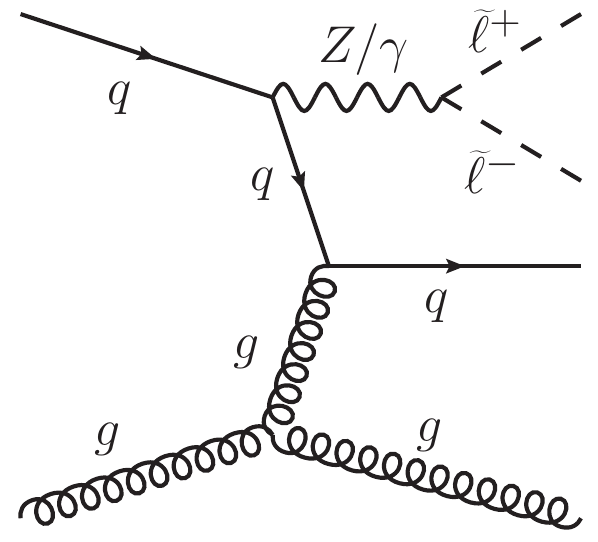}
  \end{center}
  \caption{Dominant Feynman graphs for $\widetilde{\ell}^+\, \widetilde{\ell}^-\, jj$ production in SPS1amod scenario.}
  \label{sleptonprod}
\end{figure}

In the case of the chargino
production, however, not only the squark pair production diagrams contribute
but also the other diagrams shown in Fig.~\ref{charprod}. Relatively, these
additional diagrams become even 
more important in the part of phase space which is interesting in
the context of VBF 
backgrounds. For that reason, for the chargino production channel we choose
the factorization scale as 
\begin{equation}
  \mu_F^2 = |p_T^{\text{jet}_1}| |p_T^{\text{jet}_2}| \label{facscale}
\end{equation}
and 
\begin{equation}
  \alpha_s^2 = \alpha_s(|p_T^{\text{jet}_1}|)\alpha_s(|p_T^{\text{jet}_2}|) \,, \label{alphas}
\end{equation}
  where $|p_T^{\text{jet}_i}|$ is the absolute value for the i-th jet ($p_T$
  ordered). With  
this choice we take into account the different scales of the event: for
t-channel processes the jet  $p_T$ is the scale governing the momentum
transfer between the QCD vertices, while for squark decays high $p_T$ jets
dominate and  hence $p_T \approx m_{\tilde{q}}/2$ becomes the effective scale.

The relevant Feynman diagrams for the slepton pair production are
shown in Fig.~\ref{sleptonprod}. Here no squarks are involved.
Like in the chargino case, the transverse momentum of the two
jets is the most relevant scale concerning the QCD part,
so we use the same scale in the slepton production processes.

Concerning
the scale uncertainty, we see the usual leading order $\alpha_s^2$ scale
dependence  
(listed in Table \ref{scaletable}), which means that for a very precise
analysis next-to leading order corrections would have to be
taken into account. For some of the topologies in the relevant SUSY processes
(e.g.~the last diagram in Fig.~\ref{charprod}), the
K-factor on the total cross section is known \cite{prospinosquarks,
  prospinosleptons}. For SPS1amod and $\mu_R=\mu_F=m_{\widetilde{q}}$, Prospino2 gives  
K-factors ranging from 1.25 for squark pair production
up to 1.5 for squark-antisquark production. However, for some of the relevant
topologies in our SUSY 
processes (e.g. the first two diagrams in Fig.~\ref{charprod}), the NLO QCD
corrections are not known, and therefore a scaling of the leading order
results is not possible.
The leading order scale variations of Table~\ref{scaletable} cover the change due to 
known NLO corrections and we therefore do not attempt to correct for higher order effects.
A full next-to leading order analysis is beyond the scope of this paper.
\tabcolsep3mm
\begin{table}[b]
  \begin{center}
  \begin{tabular}{|c||c|c|c|}
  \hline
   & $\zeta=0.5$ & $\zeta=1.0$ & $\zeta=2.0$\\
  \hline
  \hline
  $\chi_1^+\,\chi_1^-\,jj$ & $1.71 \;\text{fb}$  & $1.21 \;\text{fb}$ & $0.88 \;\text{fb}$\\
  $\chi_2^0\,\chi_1^0\,jj$ & $1.37 \;\text{fb}$  & $1.04 \;\text{fb}$ & $0.82 \;\text{fb}$\\
  $\widetilde{\ell}^+\, \widetilde{\ell}^-\, jj$ & $1.74 \;\text{fb}$  & $1.23 \;\text{fb}$ & $0.91 \;\text{fb}$\\
  \hline
  \end{tabular}
  \end{center}
  \caption{Scale dependence for $\chi_1^+\chi_1^-jj$, $\chi_2^0\chi_1^0jj$ and 
           $\widetilde{\ell}^+\,\widetilde{\ell}^-\,jj$ cross sections
           with $\mu_R=\mu_F=\zeta \cdot \mu_0$ (basic
           jet and lepton cuts plus $\Delta\eta_{jj}\geq4.2$ and 
           $\eta_{j,min} \leq \eta_{\ell} \leq \eta_{j,max}$ are applied, see
           Eqs.~\eqref{cuts_W_min}, \eqref{cuts_W_deltaeta}).} 
  \label{scaletable}
\end{table}

Due to the considered many particle final states and the large number of 
Feynman diagrams involved within the MSSM, 
a complete calculation of the cross sections including the full decay into leptons
on matrix element level with programs like 
{\tt MadGraph/MadEvent}~\cite{Madgraph} or {\tt Whizard}~\cite{Whizard1, *Whizard2}
did not seem to be feasible. Instead, we feed the unweighted events produced
with 
{\tt MadGraph/MadEvent} into the Monte Carlo program {\tt
Herwig++ 2.4.2}~\cite{Herwig} using Les Houches event files \cite{lhe}.
With {\tt Herwig++} we simulate the subsequent $\chi_1^\pm$, $\chi_2^0$ and
$\widetilde{l}$ decays at parton level according to the branching ratios 
calculated by the spectrum generator (which is SUSYHIT in our case). 
Within the decay chains, {\tt Herwig++} includes the spin correlations 
as described in \cite{Richardson:2001df}, but we lose the spin
information of the charginos and neutralinos at the interface between the two programs.
To check the accuracy of this approximation, especially concerning
spin effects, we consider the processes with the specific chargino decay chain
\begin{align}
p p
\rightarrow j j \, \chi^+_1 \, \chi^-_1 \rightarrow j j \, \widetilde{\tau}_1^+ \,
\nu_{\tau} \, \widetilde{\tau}_1^- \, \bar{\nu}_{\tau} \rightarrow j j \,
\chi^0_1 \, \chi^0_1 \, \tau^- \, \bar{\nu}_{\tau} \, \tau^+ \, \nu_{\tau} 
\label{charginostau}
\end{align}
and neutralino decay chain
\begin{align}
p p
\rightarrow j j \, \chi^0_1 \, \chi^0_2 \rightarrow j j \, \chi^0_1
\widetilde{\tau}_1^\pm  \, \tau^\mp 
 \rightarrow j j \,
\chi^0_1 \, \chi^0_1 \, \tau^\pm \, \tau^\mp \;,
\label{neutralinostau}
\end{align}
respectively, and perform the computation down to the $\widetilde{\tau}$-level
using {\tt MadGraph}.
When the $\chi_1^\pm \rightarrow \widetilde{\tau}^\pm
\stackrel{\text{\tiny(}-\text{\tiny)}}{\nu_{\tau}}$ decay is included in the 
{\tt MadGraph} calculation, the final state SUSY particles have
spin-0. Therefore  
comparing the {\tt MadGraph} result at this stage with the {\tt MadGraph}
result at  
the chargino level ({\tt Herwig++} always computes the rest of the decay chain
down to tau-lepton level) we get a good estimate on the impact of the chargino
spin and of the spin of the second lightest neutralino 
on the distributions.
We find that, for the charginos, the two different treatments of the 
$\chi_1^\pm \rightarrow \widetilde{\tau}^\pm
\stackrel{\text{\tiny(}-\text{\tiny)}}{\nu_{\tau}}$ decay agree very well.
There are only small differences in the tau lepton $p_T$-distributions: While
the $p_T$-distribution of the two tau 
leptons has the same shape for the chargino decay computed in {\tt Herwig++},
there is a small 
asymmetry in the {\tt MadGraph} case as can be seen in
Fig.~\ref{MGHerwig}. For the neutralinos, the results are similar; only the
asymmetry in the $p_T$-distribution of the tau leptons, when the $\chi_2^0$
decay is performed within the {\tt MadGraph} calculation, is enhanced (see
Fig.~\ref{MGHerwig}). As we 
use the same lepton cuts independent of the charge of the lepton and as we
take into account all channels, particularly the $\widetilde{\tau}^+$ and the
$\widetilde{\tau}^-$ channel in the neutralino case, neglecting 
the asymmetry still gives a reasonable result.
Overall, these small effects in the chargino as well as in the neutralino case
are negligible with respect to the accuracy of this leading order analysis.

For the SUSY processes which contribute to the reducible background of VBF Higgs boson production
we again generate parton level events using {\tt MadGraph/MadEvent 4.4}. The complete process list
is:
\vskip-6pt
\begin{minipage}{0.45\textwidth}
  \begin{align}
  p p &\rightarrow j (j) \, \chi_1^\pm \, \chi_2^0 \,, \label{ch1neu2} \\
  p p &\rightarrow (j) (j) \, \chi_2^0 \, \chi_2^0 \,, \label{neu2neu2} \\
  p p &\rightarrow j j \, \chi_3^0 \, \chi_1^0 \,, \\
  p p &\rightarrow j j \, \chi_4^0 \, \chi_1^0 \,, \\
  p p &\rightarrow j j \, \chi_4^0 \, \chi_2^0 \,,
  \end{align}
\end{minipage}\hfill
\begin{minipage}{0.45\textwidth}
  \begin{align}
  p p &\rightarrow j j \, \chi_1^\pm \, \chi_2^\mp \,, \\
  p p &\rightarrow j j \, \chi_2^+ \, \chi_2^- \,, \\
  p p &\rightarrow j j \, \chi_2^\pm \, \chi_2^0 \,, \\
  p p &\rightarrow j j \, \chi_2^+ \, \chi_1^0 \,, \\
  p p &\rightarrow (j) (j) \, \widetilde{g} \, \chi_1^0 
  \end{align}
\end{minipage}
\vskip12pt
\noindent The processes of Eqs.~\eqref{ch1neu2} and \eqref{neu2neu2} give by far
the largest contributions
and are discussed in more detail. A summary of the results for the other
processes 
can be found in Section \ref{sps1asummary}.
The topologies contributing to the cross section of these processes are
similar to the ones of 
chargino pair production plus two jets (see Fig.~\ref{charprod}), only for
next-to lightest 
neutralino pair production plus two jets, Feynman diagrams with an initial
squark pair production play a more important role.
Therefore, we choose the
dynamical factorization and renormalization scales analogously to
Eqs.~\eqref{facscale} and \eqref{alphas}. 
The {\tt MadEvent} events are again fed into {\tt Herwig++}. But in contrast
to the processes 
of the irreducible background, the hadronic decay modes of the tau lepton are
needed as well: 
We have to model the contribution of soft hadronic tau lepton decays which
escape detection, and take into account tau jets which are interpreted as tagging jets.
For our analysis we construct a ``partonic tau jet'' by collecting all
visible decay 
products of the {\tt Herwig++} tau decay and combining their momenta to one
single object. 
With this prescription, we treat every  ``tau jet''  as
a normal jet  
and give an upper bound on contributions containing hadronically decaying tau
leptons. 
Further details on the event selection for the reducible background and our
definition 
of visible objects will be given in Section~\ref{reducible}.

For the generation of the signal processes we use the parton level Monte Carlo program
{\tt VBFNLO 2.6.2}~\cite{vbfnlo1, *vbfnlo2} at leading order, as this program is very fast
and efficient especially for the $h \rightarrow WW$ channel.
The tau decay for $h \rightarrow \tau\tau$ is again done by {\tt Herwig++}.
The factorization and renormalization scales for the signal process are set
to the momentum transfer of the exchanged $W/Z$ boson between the quark lines.

We checked that detector effects from finite energy resolution for jets and
leptons 
(modeled by a gaussian smearing) have only very small effects on our
distributions, 
as they do not show any sharp peaks. Therefore we neglect them.
One exception is the reconstructed tau pair
mass of the $h \rightarrow \tau\tau$ signal, which is very sensitive to fake
missing transverse energy. It can be parametrized by a Gaussian
distribution \cite{atlas_etmiss} with
\begin{equation}
 \sigma (E_x^{miss},E_y^{miss}) = 0.41 \cdot \sqrt{\sum E_T} \; .
 \label{fakeptmiss}
\end{equation}
We therefore study the effects of (\ref{fakeptmiss}) on the tau pair mass
reconstruction. In the transverse energy $E_T$ we include the energy deposit 
in calorimeters by hadronic parts of the event as well
as contributions from the underlying event, but neglect pileup effects.
From \cite{atlas_UE} we get an underlying event contribution to $\sum E_T$  of
approximately 100 GeV 
for a center of mass energy $\sqrt{s}=7\;\text{TeV}$ within the rapidity range
$-4.5 \leq \eta \leq 4.5$.  
With a value of 33 GeV for $\sqrt{s}=900\;\text{GeV}$ we extrapolate
to an underlying event contribution of approximately 130~GeV at
$\sqrt{s}=14\;\text{TeV}$.

\section{SPS1a-like Scenario: Irreducible Background}
\label{irreducible}

Several SUSY processes contribute to the background of VBF Higgs boson production.
We group them into two classes: Processes which match the signature of VBF Higgs boson production
exactly and processes which produce additional jets or leptons.
In this section we investigate the first class in detail, the other processes
will be discussed in Section \ref{reducible}. We start with the processes containing
charginos and neutralinos and continue with processes containing sleptons in 
Sect. \ref{sleptons-sps1amod}.

First, we investigate the processes $p p \rightarrow j j \chi_1^+ \chi_1^-$ 
and $p p \rightarrow j j \chi_2^0 \chi_1^0$ including the subsequent decays of 
the charginos and neutralinos. These contributions to the background mainly 
involve 
squarks and gluinos and their cross sections are
 relatively large when only minimal cuts are applied 
(see below and Fig.~\ref{tau_deltaeta}).  

In principle the two Higgs decay channels $h \rightarrow \tau^+\tau^-$ and 
$h \rightarrow W^+W^-$ lead to the same signature in the detector,
namely two hard jets with a large rapidity separation, two opposite charged 
leptons,
and missing energy (from $W$ or $\tau$ decay respectively). Therefore we have to
consider the same processes as background. But as the signal processes feature
different characteristics, we analyze them separately. Necessary cuts for
Standard Model background reduction are partially taken from \cite{ATLASTDR}
in the case of $h \rightarrow \tau^+\tau^-$. For $h \rightarrow W^+W^-$ we 
follow \cite{VBFHinWW}, where this channel was discussed for relatively light 
Higgs boson masses.

\subsection{\texorpdfstring{{\boldmath $\chi_1^\pm$} and {\boldmath$\chi_2^0$} Contributions 
to the {\boldmath$h \rightarrow \tau^+\tau^-$} Channel}
{Chargino and Next-to Lightest Neutralino Contributions to the tau+ tau- Channel}}
\label{sps1amod-tautau}

\begin{figure}[p]
  \begin{center}
    \includegraphics[height=0.30\textwidth]{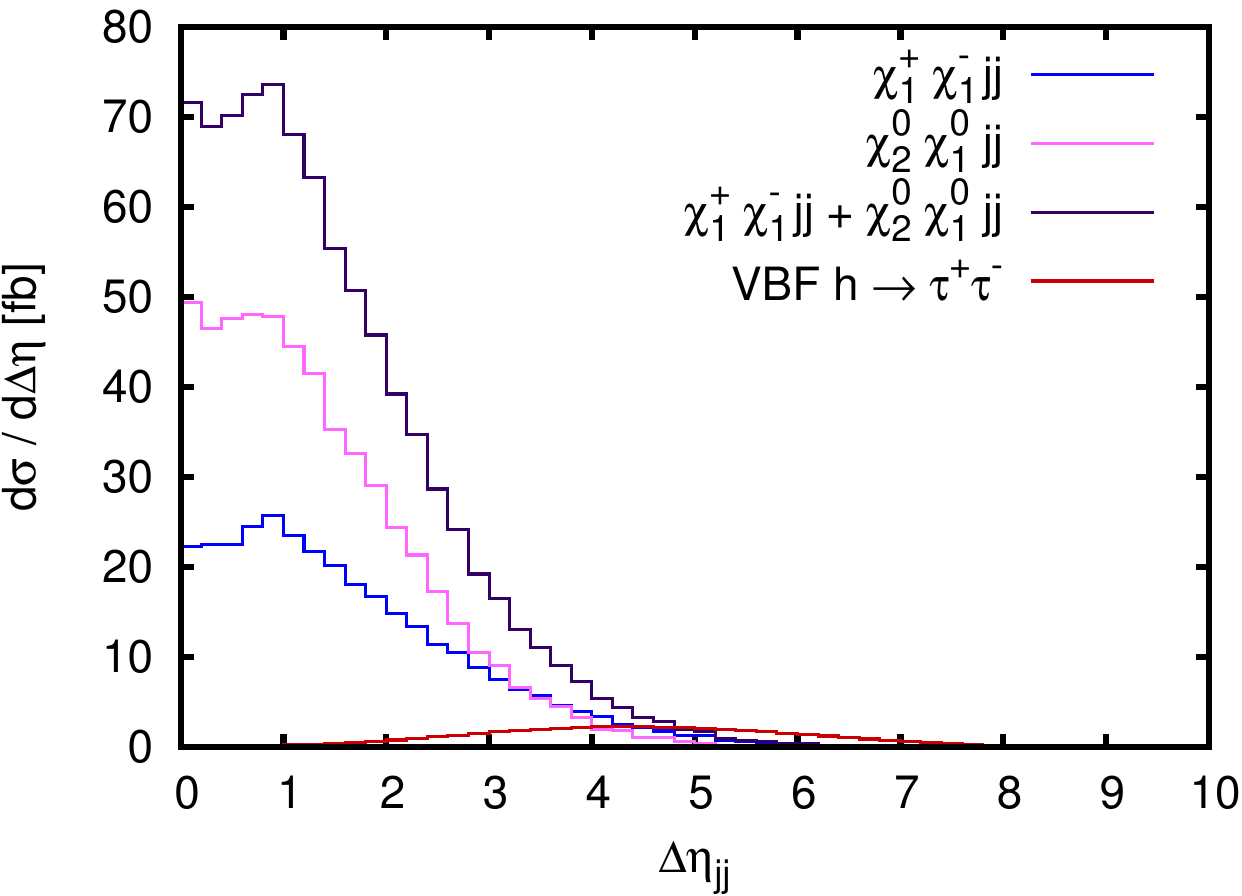}
    \hskip20pt
    \includegraphics[height=0.30\textwidth]{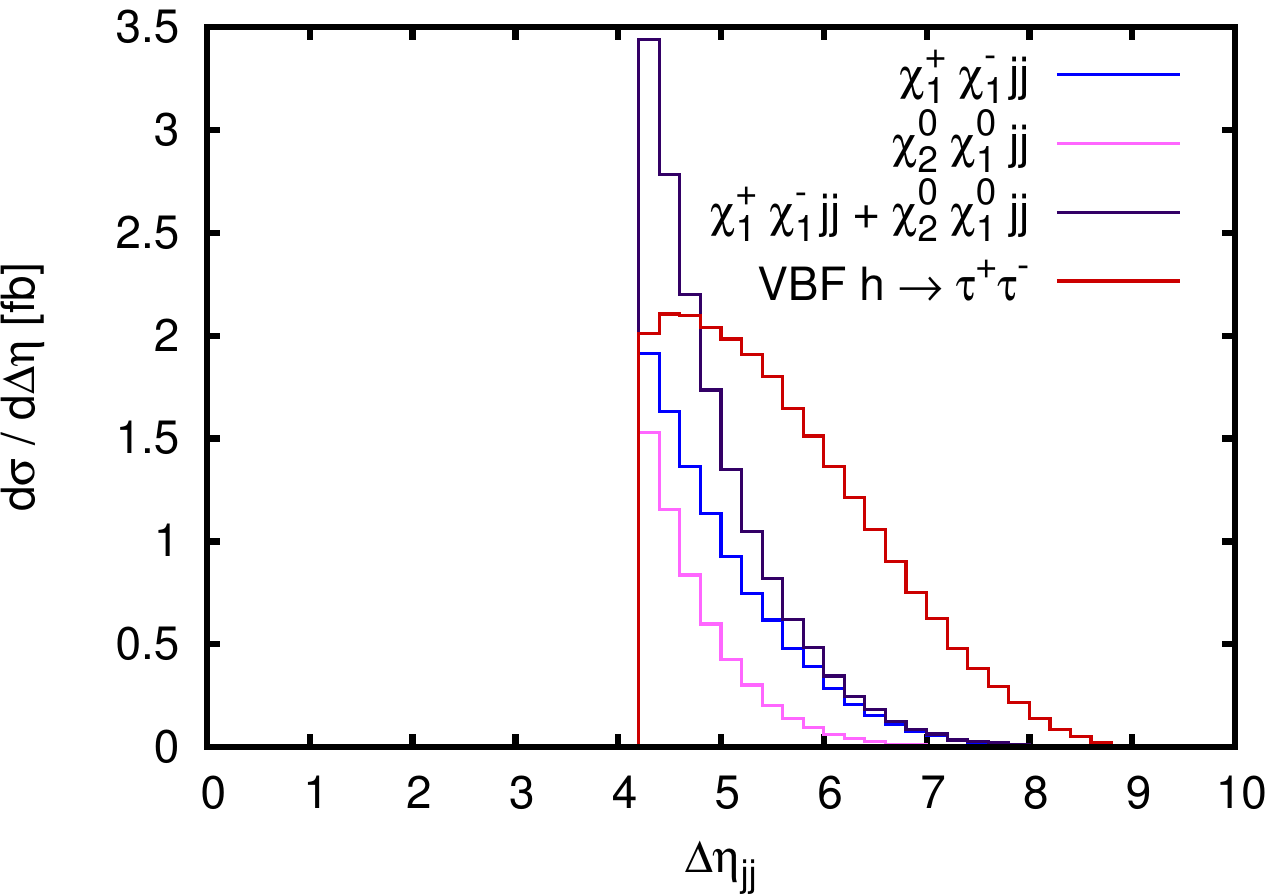}
  \end{center}
  \caption{Rapidity separation for SUSY backgrounds and $h\rightarrow \tau\tau$
           signal with basic cuts, Eq.~\eqref{cuts_tau_min}, (left) and additional
           $\Delta\eta$ plus ``leptons inside rapidity 
           gap'' cuts, Eq.~\eqref{cuts_tau_deltaeta} (right).}
  \label{tau_deltaeta}
\end{figure}

\begin{figure}[p]
  \begin{center}
    \includegraphics[height=0.30\textwidth]{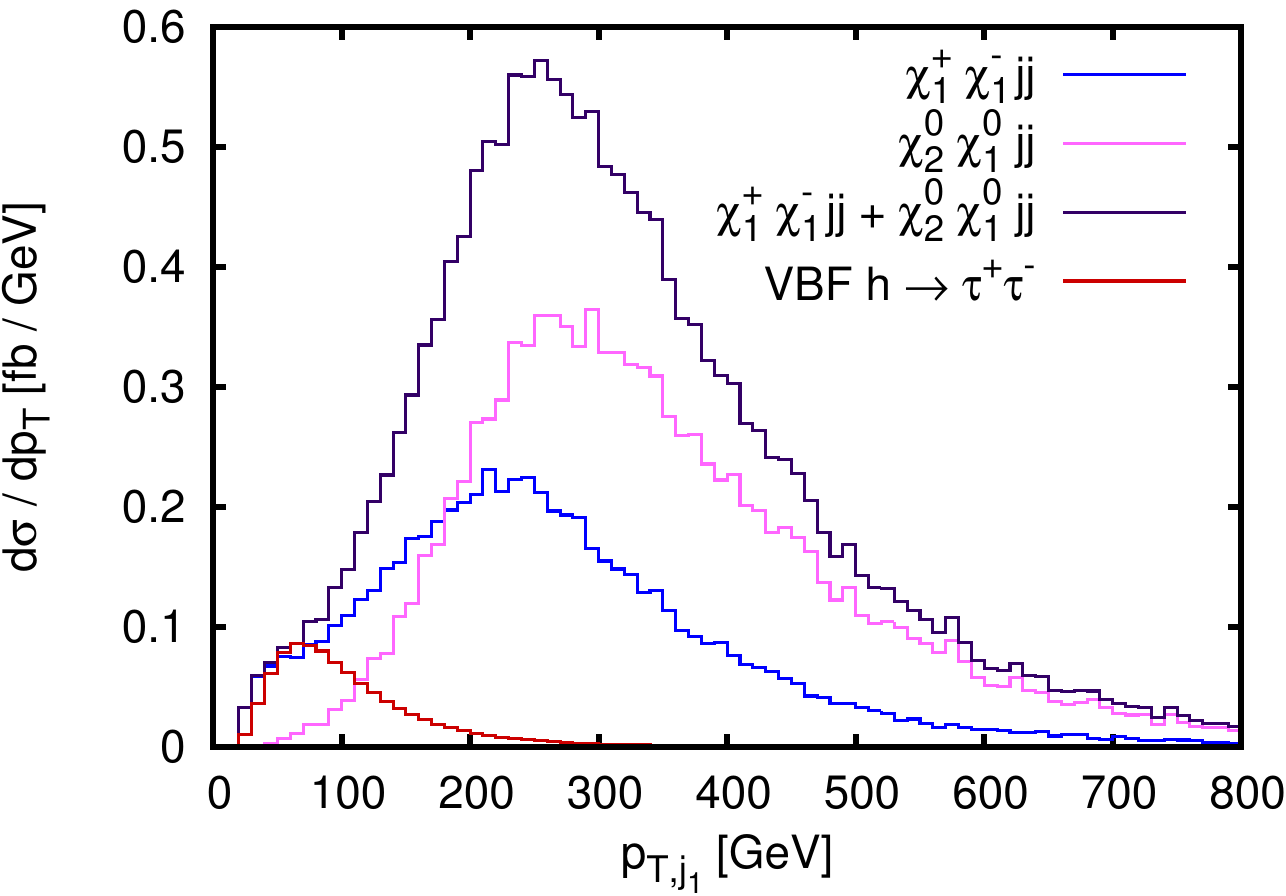}
    \hskip20pt
    \includegraphics[height=0.30\textwidth]{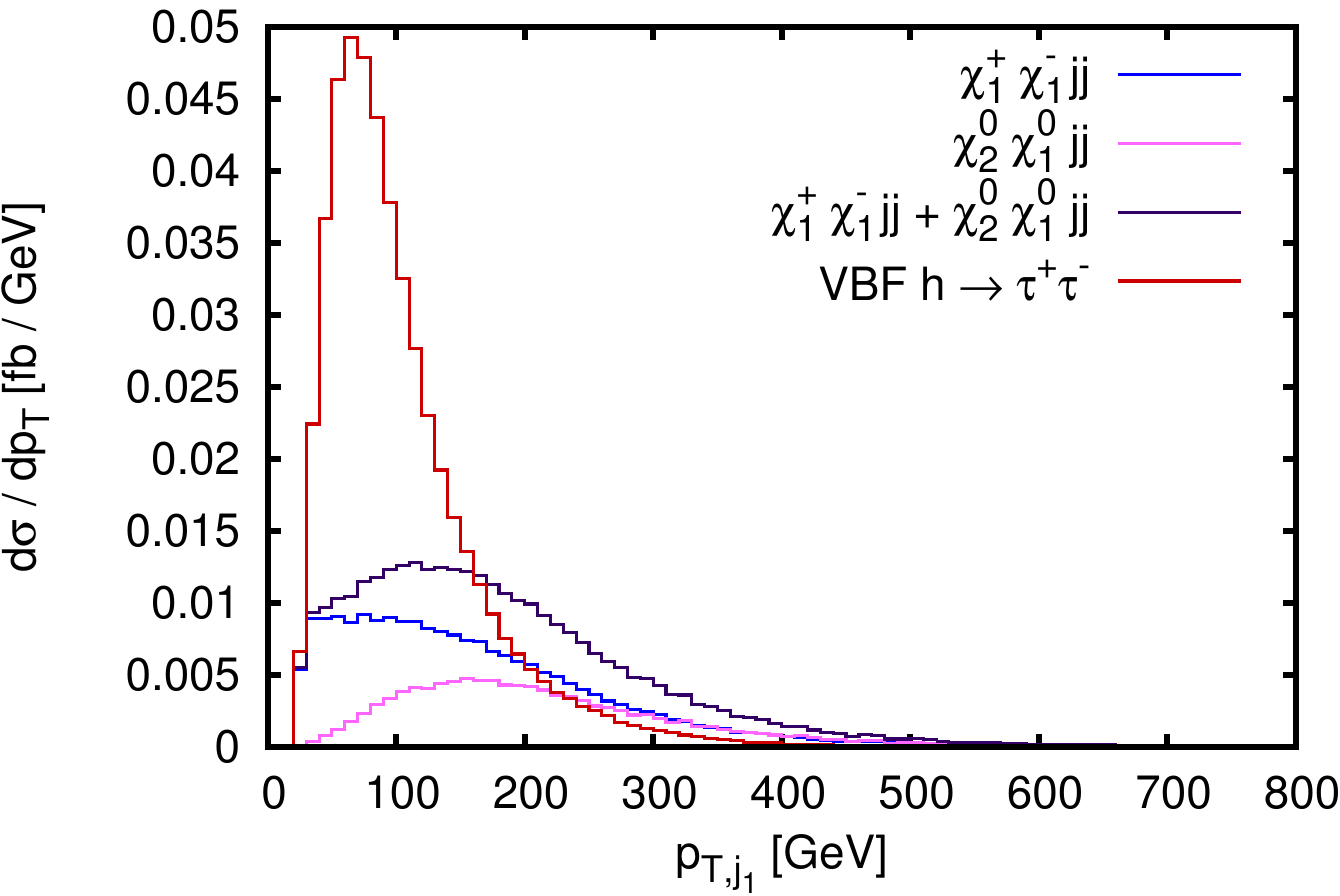}\\
    \vskip20pt
    \includegraphics[height=0.30\textwidth]{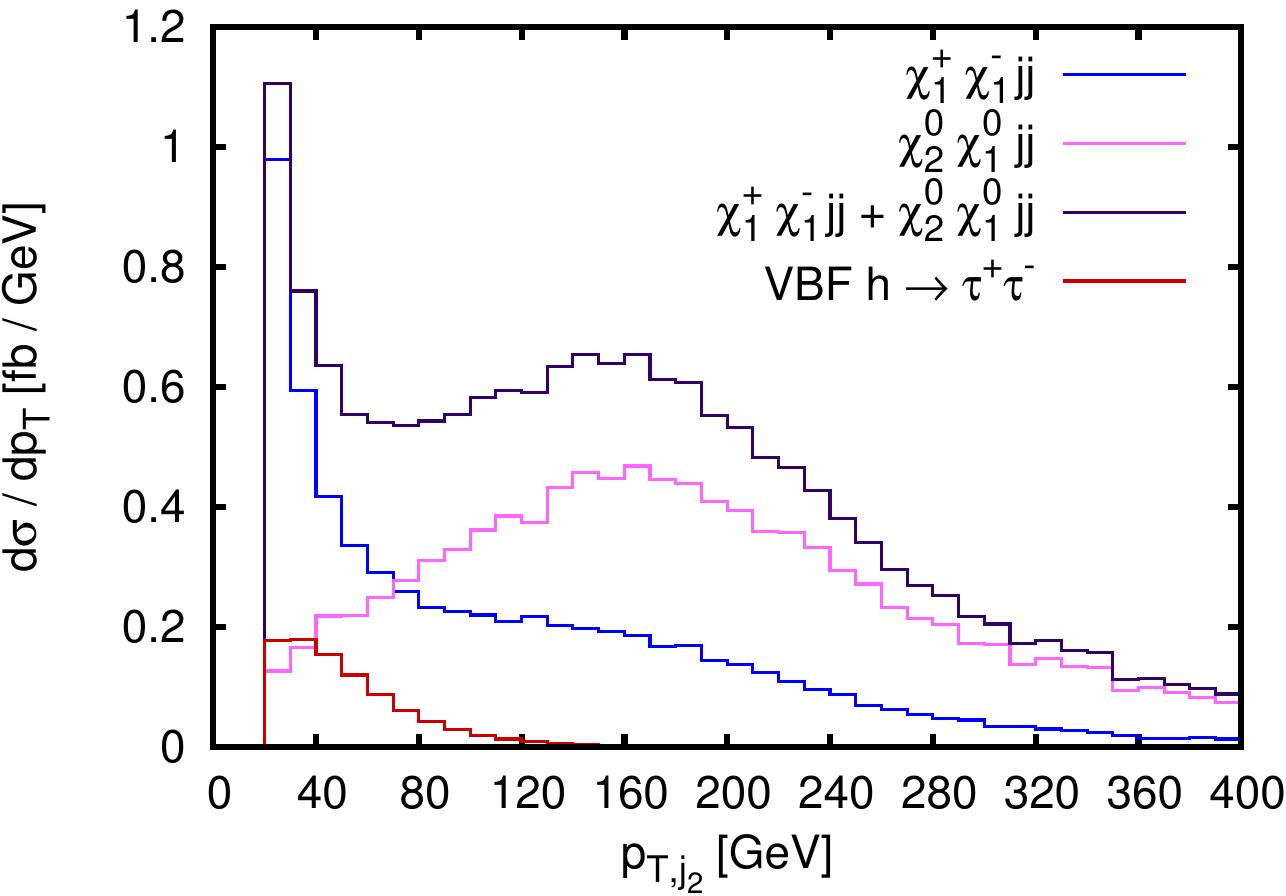}
    \hskip20pt
    \includegraphics[height=0.30\textwidth]{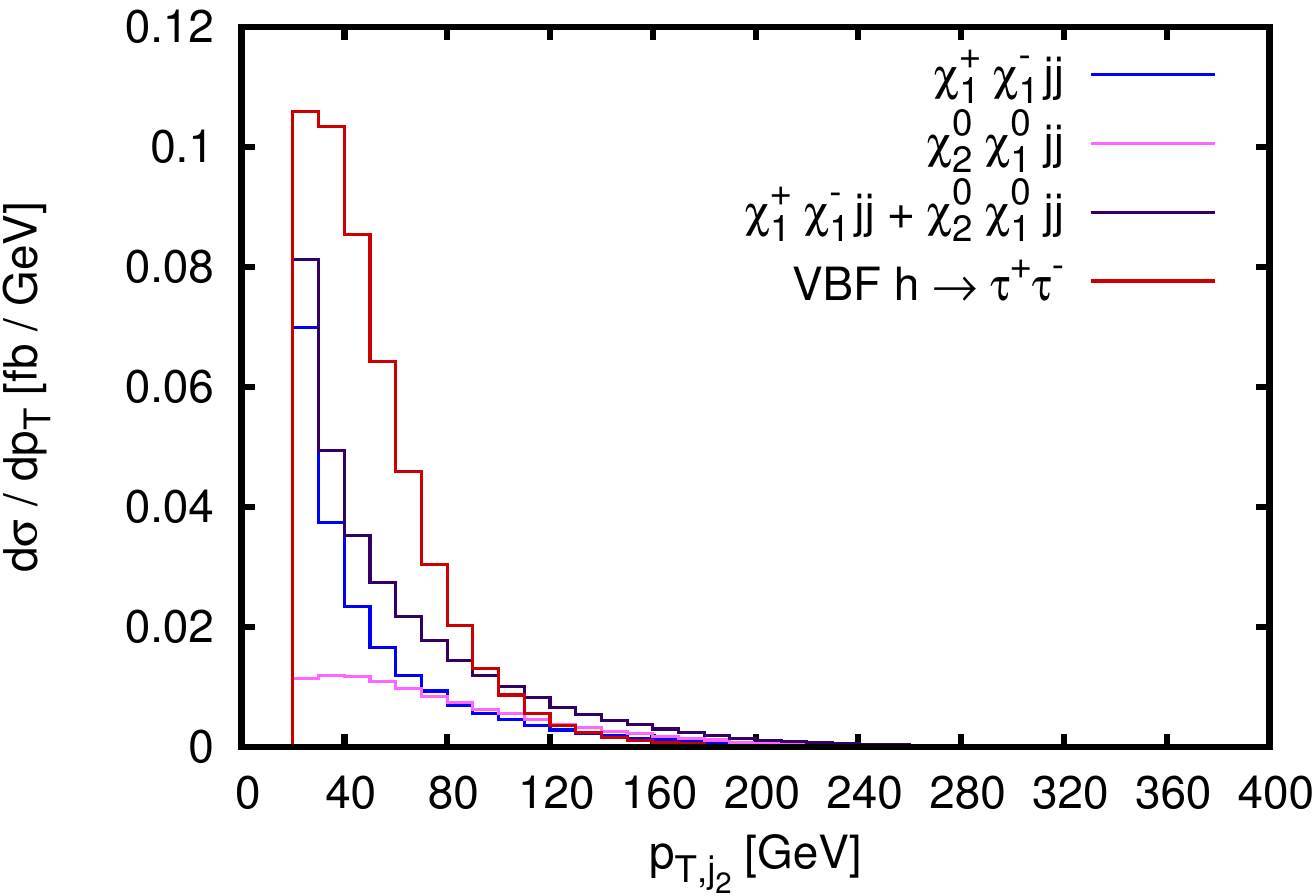}
  \end{center}
  \caption{Transverse momentum of the harder (upper plots) and softer (lower plots) tagging 
           jet for SUSY backgrounds and $h\rightarrow \tau\tau$ signal 
           with basic cuts, Eq.~\eqref{cuts_tau_min}, (left) and additional $\Delta\eta$ plus ``leptons inside rapidity 
           gap'' cuts, Eq.~\eqref{cuts_tau_deltaeta} (right).}
  \label{tau_jetpT}
\end{figure}

With the procedure discussed in Section~\ref{procedure} and some minimal cuts on the produced
jets and leptons which merely account for the detector acceptance
\begin{equation}
\begin{array}{rclcrcl}
  p_{T,j} &\geq& 20 \;\text{GeV} & \quad & p_{T,\ell} &\geq& 10 \;\text{GeV}  \\
  |\eta_j| &\leq& 4.5 & \quad & |\eta_\ell| &\leq& 2.5  \\
  R_{jj} &\geq& 0.8 & \quad & R_{\ell\ell} &\geq& 0.6  \\
  R_{j\ell} &\geq& 0.8 \;, &  & & & 
\end{array}
\label{cuts_tau_min}
\end{equation}
we get background contributions from the two SUSY processes, which are very 
large
compared to the signal. This is depicted in the left panel of 
Fig.~\ref{tau_deltaeta} and Fig.~\ref{tau_jetpT}.
($p_{T,P}$ denotes the transverse momentum of the particle $P$, $\eta_P$ the 
rapidity of the 
particle $P$ and $R_{{P_1}{P_2}}$ the R-separation of the particles $P_1$ and 
$P_2$ with 
$R_{{P_1}{P_2}} = (\Delta \eta_{{P_1}{P_2}}^2 +\phi_{{P_1}{P_2}}^2)^{1/2}$
 where $\Delta \eta_{{P_1}{P_2}}$ is the rapidity separation of the 
particles $P_1$ and $P_2$ and $\phi_{{P_1}{P_2}}$ the azimuthal angle between 
those particles.)
In the distribution of the rapidity separation between the two tagging jets,
$\Delta\eta_{jj}$, but 
 also in the distribution of the transverse momenta of the jets,
the difference between the
VBF signal and the SUSY backgrounds becomes clearly visible 
(see Fig.~\ref{tau_deltaeta} and Fig.~\ref{tau_jetpT}). The signal shows
the typical VBF shape with two tagging jets in the forward and backward regions with a large 
rapidity separation and rather small 
transverse momenta of the jets. 
The background processes tend to have jets with smaller rapidity 
separation and larger transverse momenta which is the result of the jets being 
decay products of the produced SUSY particles. While for the hardest jet of 
the background processes
 the transverse momentum of the jet is larger than in the signal case
the $\chi_1^+\, \chi_1^-\,jj$ process shows two peaks in the transverse 
momentum distribution of the second jet, one for larger and one for small 
transverse momenta. This accounts for the different production mechanisms of 
the charginos: The peak at larger transverse momenta is due to squark pair 
production and decay (see right diagram in Fig.~\ref{charprod}) while the peak at small 
transverse momenta is due to additional QCD radiation as can be seen in the 
two left diagrams in Fig.~\ref{charprod}.
Due to these differences, particularly in the $\Delta\eta_{jj}$ distribution, 
the additional cuts
\begin{equation}
\begin{array}{rclcrclcrcl}
  \Delta\eta_{jj} &\geq& 4.2 & \quad & \eta_{j1} \cdot \eta_{j2} &<& 0 & \quad 
       & \eta_{j,min} &\leq \eta_{\ell} \leq& \eta_{j,max}
\end{array}
\label{cuts_tau_deltaeta}
\end{equation}
improve the signal to background ratio tremendously, which is shown on the 
right panel of
Fig.~\ref{tau_deltaeta}  and Fig.~\ref{tau_jetpT}. Numbers for total cross 
sections at different cut levels are
shown in Table~\ref{xs_tau_table}.

Further useful cuts for Standard Model background reduction which have been included in
the analysis are a cut on the missing transverse momentum $\slashed{p}_T$ and the jet pair mass $m_{jj}$.
\begin{equation}
\begin{array}{rclcrcl}
  \slashed{p}_T &\geq& 40 \;\text{GeV} & \quad & m_{jj} &\geq& 700 \;\text{GeV} \;.
\end{array}
\label{cuts_tau_atlas}
\end{equation}
The first one reduces backgrounds without neutrinos and improves the tau pair mass reconstruction
\cite{ATLASTDR}. The second cut helps reducing the QCD backgrounds \cite{HinWW_heavy}.
With these cuts the combined cross section of $\chi_1^+\, \chi_1^-\,jj$ and $\chi_2^0\,\chi_1^0\,jj$ production
channels is roughly equal to the signal cross section.

An important technique to reduce Standard Model backgrounds in the $h\rightarrow \tau\tau$
channel is the reconstruction of the invariant tau pair mass $m_{\tau\tau}$, which corresponds to the Higgs
boson mass, from the measured lepton four-momenta and missing energy \cite{ATLASTDR, massrec}.
This geometrical reconstruction is possible, because the Higgs boson mass is much larger than the tau lepton
mass $m_\tau$, more precisely $m_h \gg 2\cdot m_\tau$. This
leads to highly boosted tau leptons. Therefore, to a good approximation, the
tau leptons and their decay products 
are collinear in the lab frame and only the momentum fraction of
the visible leptons compared to the tau leptons is needed for the reconstruction. 
As there is no other source of missing energy in the signal than from the tau decay, 
the two measured components of missing transverse energy can provide this information.
The cuts
\begin{equation}
\begin{array}{ccccc}
x_i \in [0,1] &\quad& \text{cos}\, \phi_{\ell\ell} \geq -0.9 &\quad& |m_{\tau\tau}-m_h| \leq 15 \;\textrm{GeV}
\end{array}
\label{cuts_tau_massrec}
\end{equation}
can be used for the tau pair mass reconstruction. $x_i = p_{T,\ell_i} / p_{T,\tau_i}$ 
stands for the tau momentum fraction of the detected leptons, where $0 < x_i < 1$ is the physically
allowed range. The cut on the azimuthal angle between the leptons discards events where
the two leptons are back-to-back and therefore the reconstruction is not possible.
The third cut accepts only events with a reconstructed mass around the expected Higgs
mass. This is very effective for example against tau pairs from a $Z$-boson decay.

\begin{figure}[p]
  \begin{center}
    \includegraphics[height=0.24\textwidth]{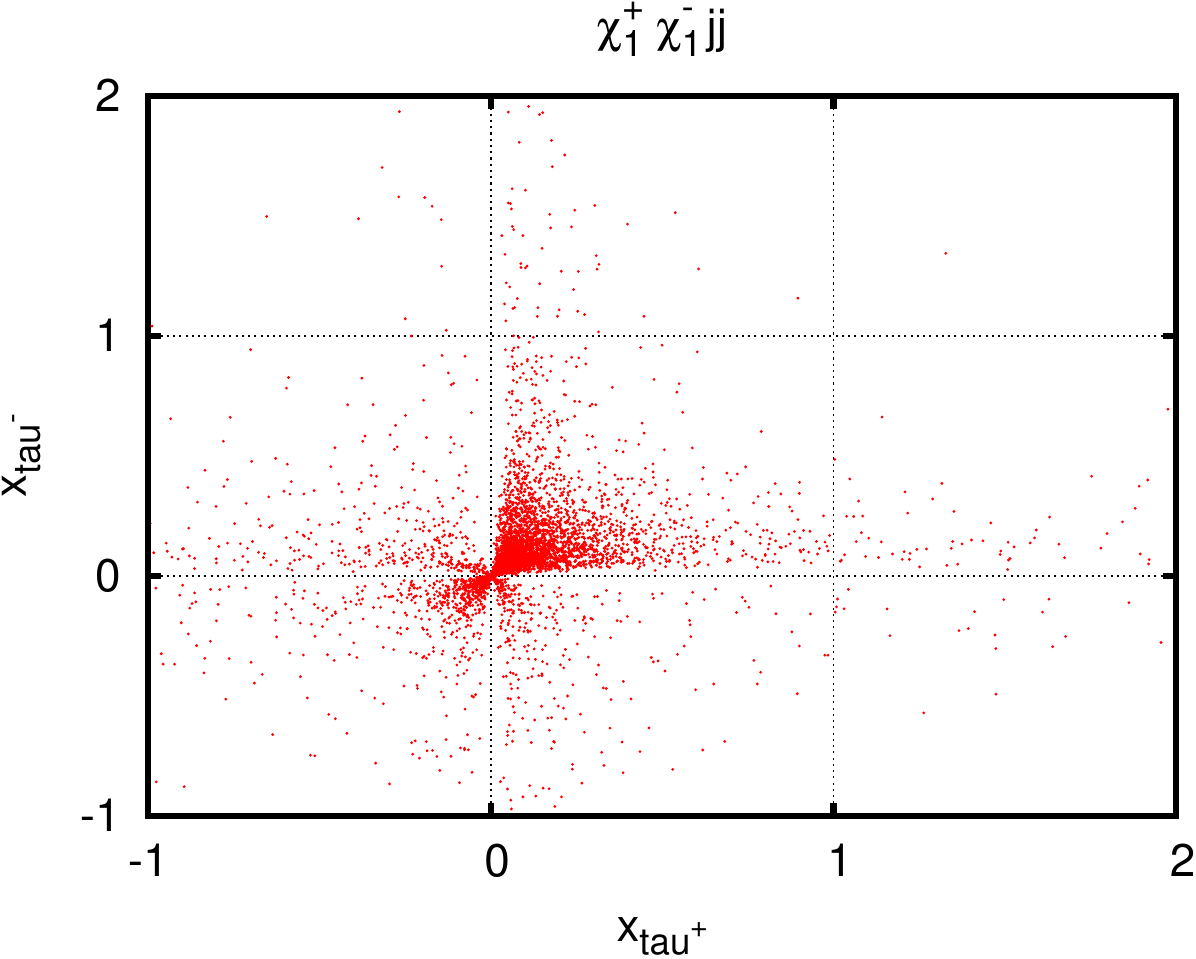}
    \hspace{2pt}
    \includegraphics[height=0.24\textwidth]{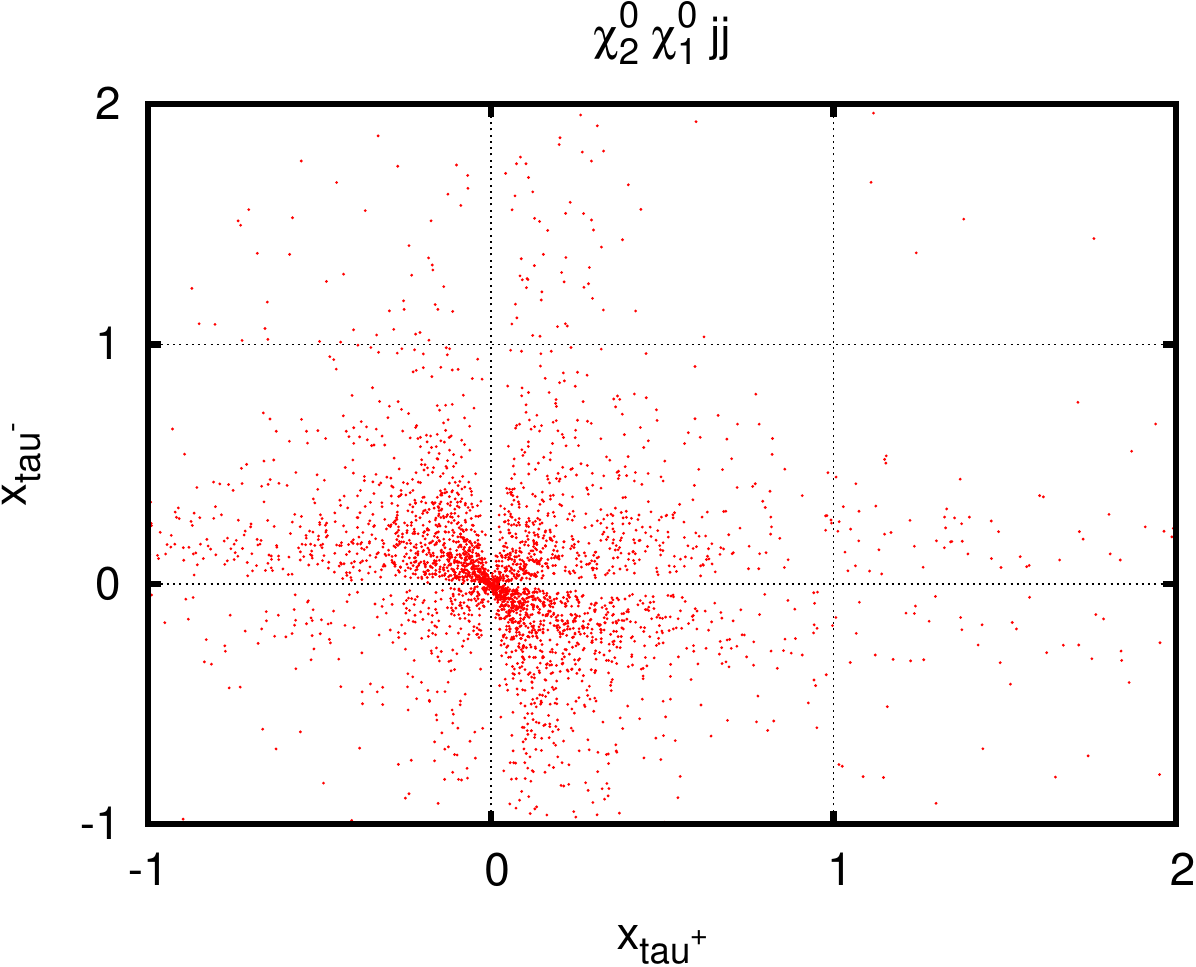}
    \hspace{2pt}
    \includegraphics[height=0.24\textwidth]{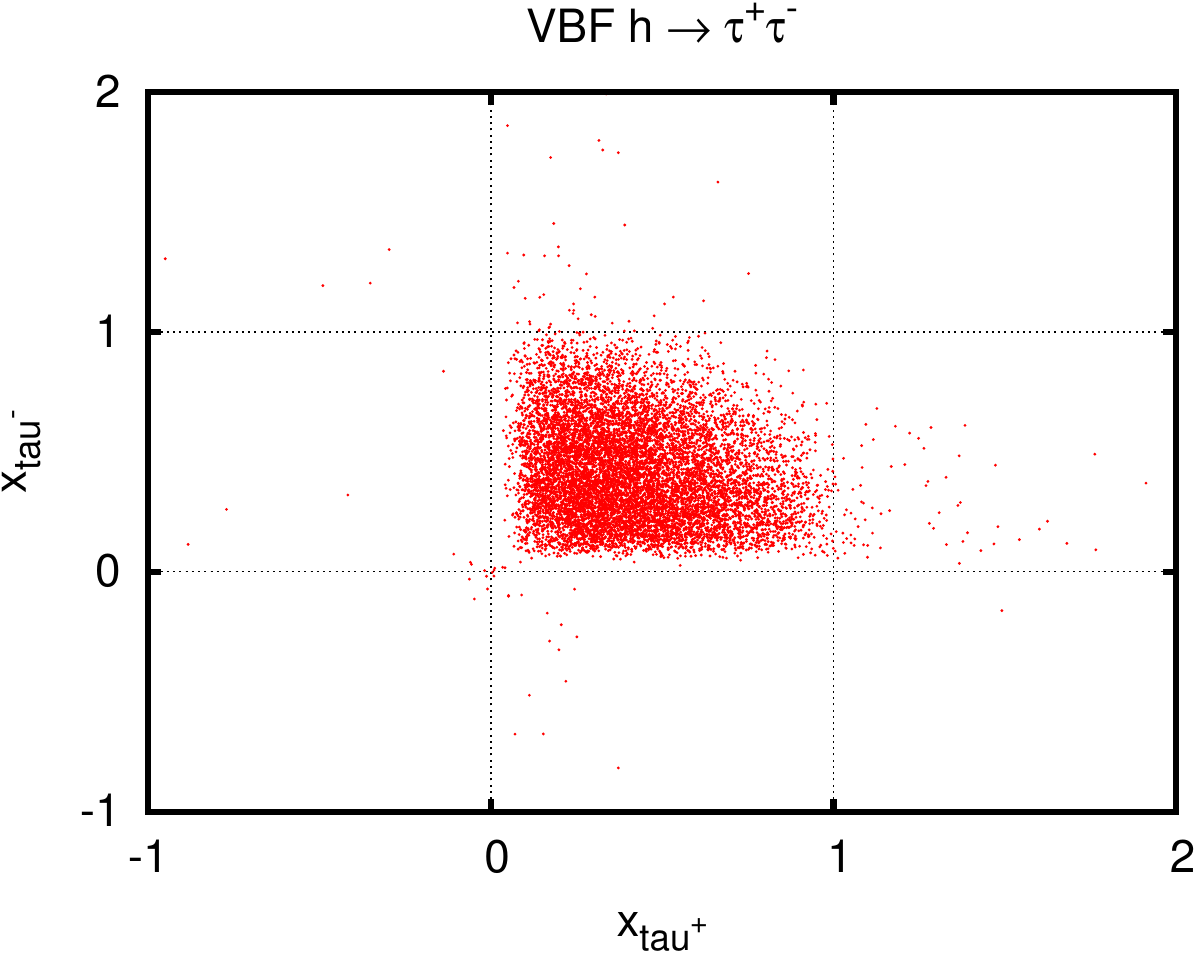}\\
  \end{center}
  \caption{Tau momentum fractions for $\chi_1^+\chi_1^-jj$, $\chi_2^0\chi_1^0jj$ and $h\rightarrow \tau\tau$ 
            with cuts from Eqs.~\eqref{cuts_tau_min}-\eqref{cuts_tau_atlas}.}
  \label{tau_masscut}
\end{figure}

\begin{figure}[p]
  \begin{center}
    \includegraphics[height=0.3\textwidth]{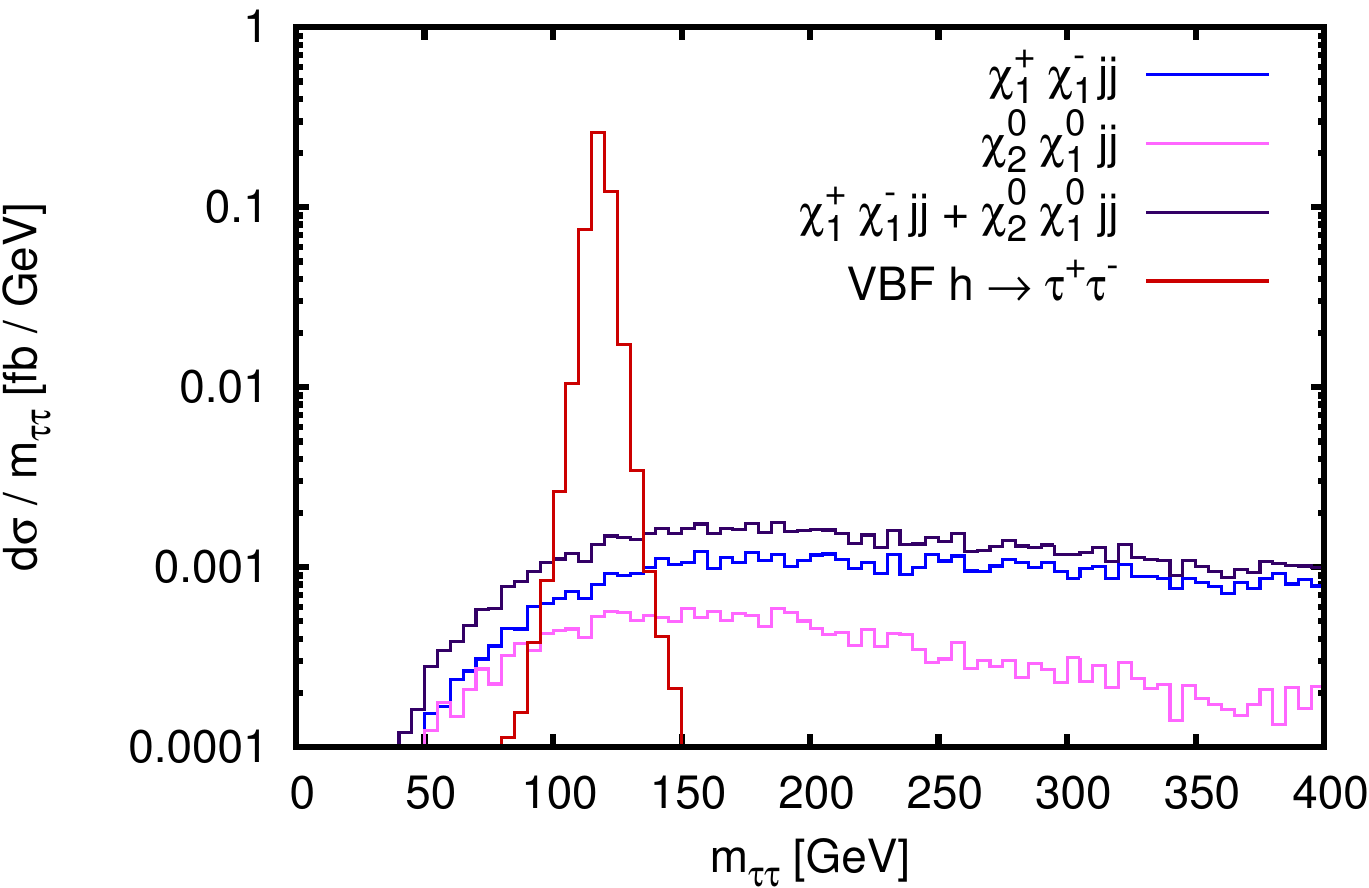}
    \hskip20pt
    \includegraphics[height=0.3\textwidth]{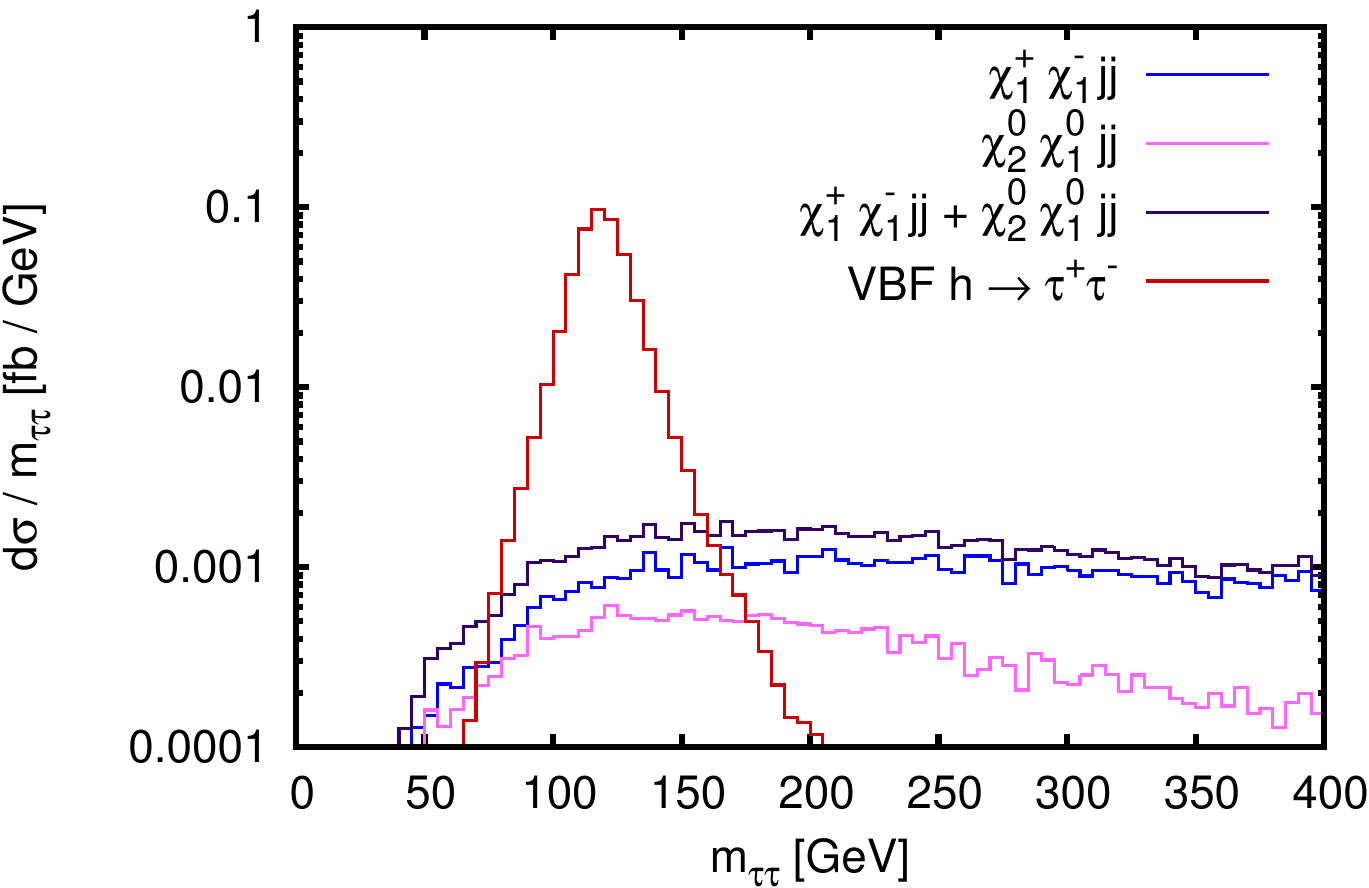}
  \end{center}
  \caption{Reconstructed tau pair mass without the "mass window" cut from Eq.~(\ref{cuts_tau_massrec}).
           The left panel is without detector effects, the plot on the right
           incorporates resolution 
           effects according to Eq.~(\ref{fakeptmiss}).}
  \label{tau_massrec}
\end{figure}

\tabcolsep3mm
\begin{table}[p]
  \begin{center}
  \begin{tabular}{|l||c|c|c|}
  \hline
  Cuts & $\chi_1^+\chi_1^-jj$ & $\chi_2^0\chi_1^0jj$ & VBF $h\rightarrow\tau\tau$ \\
  \hline
  \hline
  basics (Eq.~(\ref{cuts_tau_min})) & $64.13 \;\text{fb}$  & $109.24 \;\text{fb}$ & $9.17 \;\text{fb}$ \\
  + rapidity gap (Eq.~(\ref{cuts_tau_deltaeta})) & $2.04 \;\text{fb}$  & $1.09 \;\text{fb}$ & $4.94 \;\text{fb}$ \\
  + $\slashed{p}_{T,min}$ and $m_{jj}$ (Eq.~(\ref{cuts_tau_atlas})) & $1.35 \;\text{fb}$  & $0.96 \;\text{fb}$ & $2.67 \;\text{fb}$ \\
  + $m_{\tau\tau}$ reconstruction (Eq.~(\ref{cuts_tau_massrec})) & $0.024 \;\text{fb}$  & $0.015 \;\text{fb}$ & $2.46 \;\text{fb}$ \\
  \hline
  \hline
  + fake $\slashed{E}_T$ from Eq.~(\ref{fakeptmiss})
        & $0.025 \;\text{fb}$  & $0.015 \;\text{fb}$ & $1.93 \;\text{fb}$ \\
  \hline
  \end{tabular}
  \end{center}
  \caption{Total cross sections for $\chi_1^+\chi_1^-jj$, $\chi_2^0\chi_1^0jj$
           and VBF $h\rightarrow\tau\tau$ at different cut levels.}
  \label{xs_tau_table}
\end{table}

Processes involving SUSY particles in a R-parity conserving model always lead
to additional missing transverse energy in the detector. Therefore these processes behave
quite differently from the signal concerning the tau pair mass reconstruction as one
can see in  Fig.~\ref{tau_masscut}. The right plot of the Higgs signal 
shows the momentum fractions for the two leptons
with the cuts of Eqs.~(\ref{cuts_tau_min}) - (\ref{cuts_tau_atlas}), which is
nicely located in the allowed 
range. The situation for the chargino background, and also for the one
containing the next-to-lightest neutralino, 
is very different: For a large fraction of the events at least one of the $x_i$
has a value below zero. The physically allowed events tend to have very small
$x_i$. 
As the reconstructed tau pair mass is given by
\begin{equation}
  m_{\tau\tau} = \frac{m_{\ell^+\ell^-}}{\sqrt{x_+x_-}} \;,
\end{equation}
these events yield a very high $m_{\tau\tau}$ and get sorted out by the ``mass window'' cut.
We therefore improve the background suppression by more than a factor of 50 while the signal
cross section reduction is only 8\%.
Missing $E_T$ resolution from Eq.~(\ref{fakeptmiss}) slightly reduces the efficiency of the mass reconstruction
for SUSY background suppression as they make the sharp signal resonance broader 
(see Fig.~\ref{tau_massrec} and Table~\ref{xs_tau_table}).

To summarize, we see that even when we include fake missing transverse momentum effects, the
tau pair mass reconstruction is very efficient against SUSY processes: the additional
sizable sources of missing transverse energy effectively eliminate the SUSY background.

\subsection{\texorpdfstring{{\boldmath $\chi_1^\pm$} and {\boldmath$\chi_2^0$} 
Contributions to the {\boldmath$h \rightarrow W^+W^-$} Channel}
{Chargino and Next-to Lightest Neutralino Contributions to the W+ W- Channel}}
\label{sps1amod-WW}

\subsubsection{Scenario SPS1amod}
\label{hwwsps1amod}

A full reconstruction of the invariant $WW$ mass analogous to the $\tau$ decay 
mode of the previous section is not possible, because the
$W$ bosons are much heavier than the $\tau$ leptons and the collinear
approximation does not work in this channel. Therefore we analyze this channel
separately.
We start with the basic cuts
\begin{equation}
\begin{array}{rclcrcl}
  p_{T,j} &\geq& 20 \;\text{GeV} & \quad & p_{T,\ell} &\geq& 10 \;\text{GeV}  \\
  |\eta_j| &\leq& 4.5 & \quad & |\eta_\ell| &\leq& 2.5  \\
  R_{jj} &\geq& 0.8 & \quad & R_{j\ell} &\geq& 1.7  \\
  m_{\ell\ell} &\geq& 10 \;\text{GeV} \;, &  & & & 
\end{array}
\label{cuts_W_min}
\end{equation}
where $m_{\ell\ell}$ is the lepton pair mass, and find a similar situation 
compared to the $h\rightarrow\tau\tau$ case with
basic cuts, as we show in Table \ref{xs_W_table}.
Again, the cuts on the rapidity separation of the tagging jets are very
efficient, 
here with an additional rapidity separation between jets and leptons which was
already used for the SM background rejection in \cite{VBFHinWW}:
\begin{equation}
\begin{array}{rclcrclcrcl}
  \Delta\eta_{jj} &\geq& 4.2 & \quad & \eta_{j1} \cdot \eta_{j2} &<& 0 & \quad 
       & \eta_{j,min} + 0.6 &\leq \eta_{\ell} \leq& \eta_{j,max} - 0.6 \;.
\end{array}
\label{cuts_W_deltaeta}
\end{equation}

\begin{figure}[p]
  \begin{center}
    \includegraphics[height=0.3\textwidth]{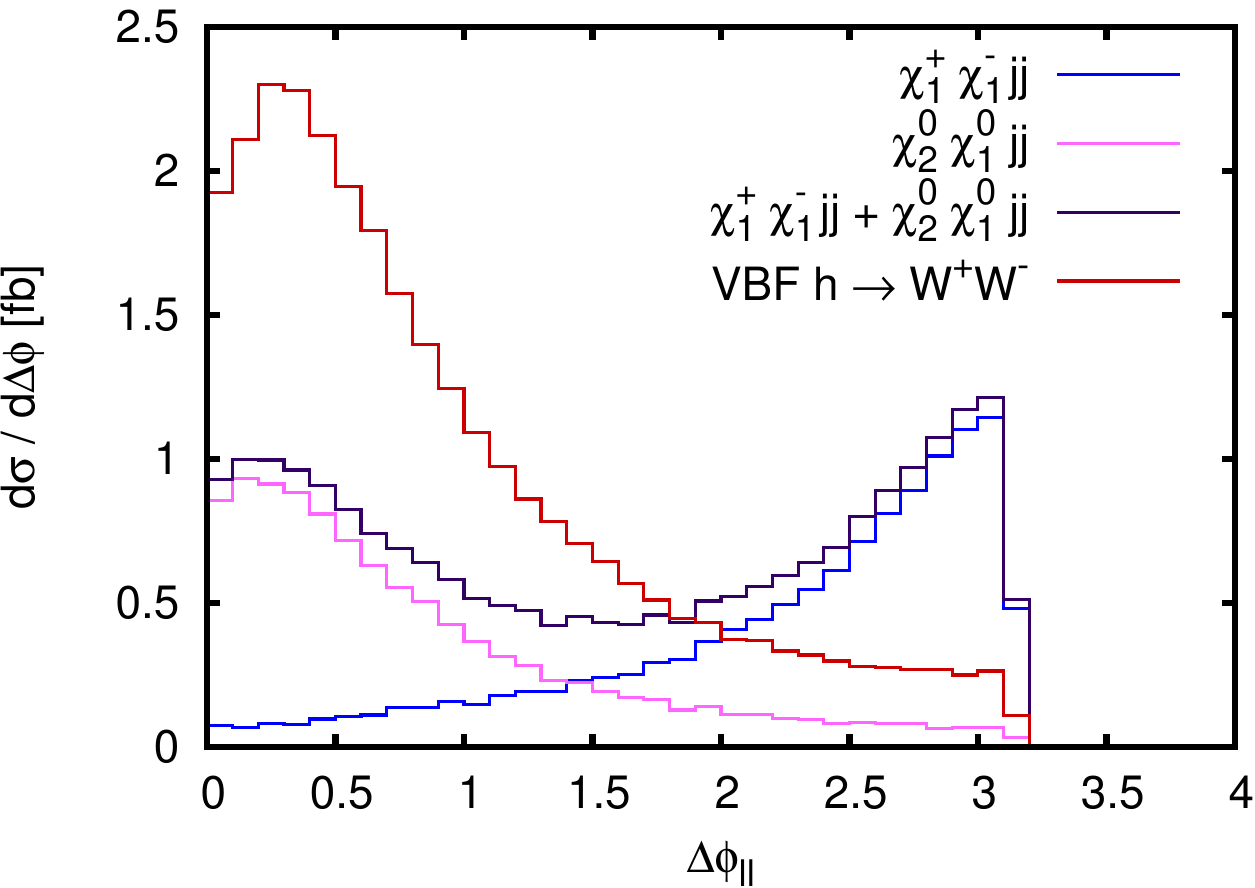}
    \hskip20pt
    \includegraphics[height=0.3\textwidth]{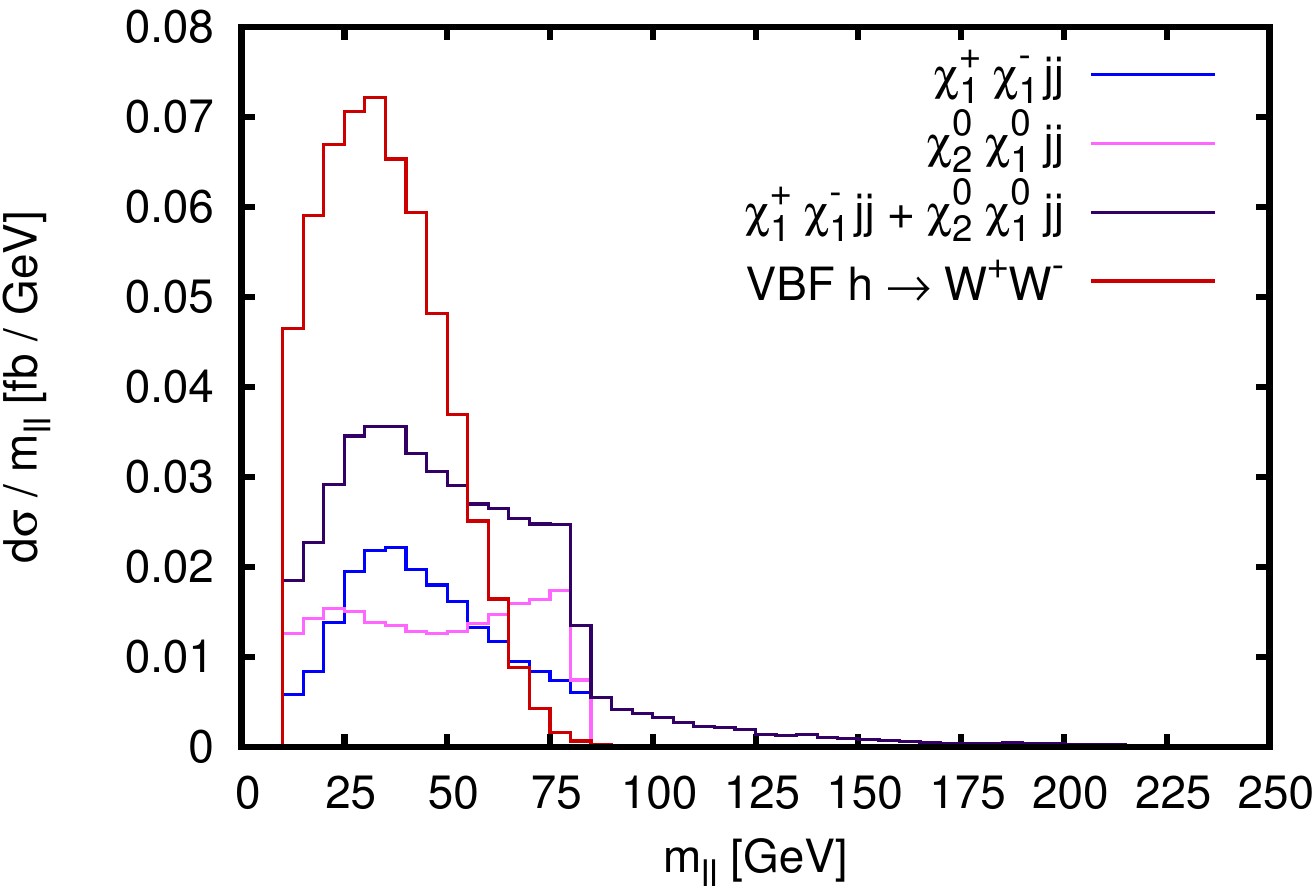}
    \vskip10pt
    \includegraphics[height=0.3\textwidth]{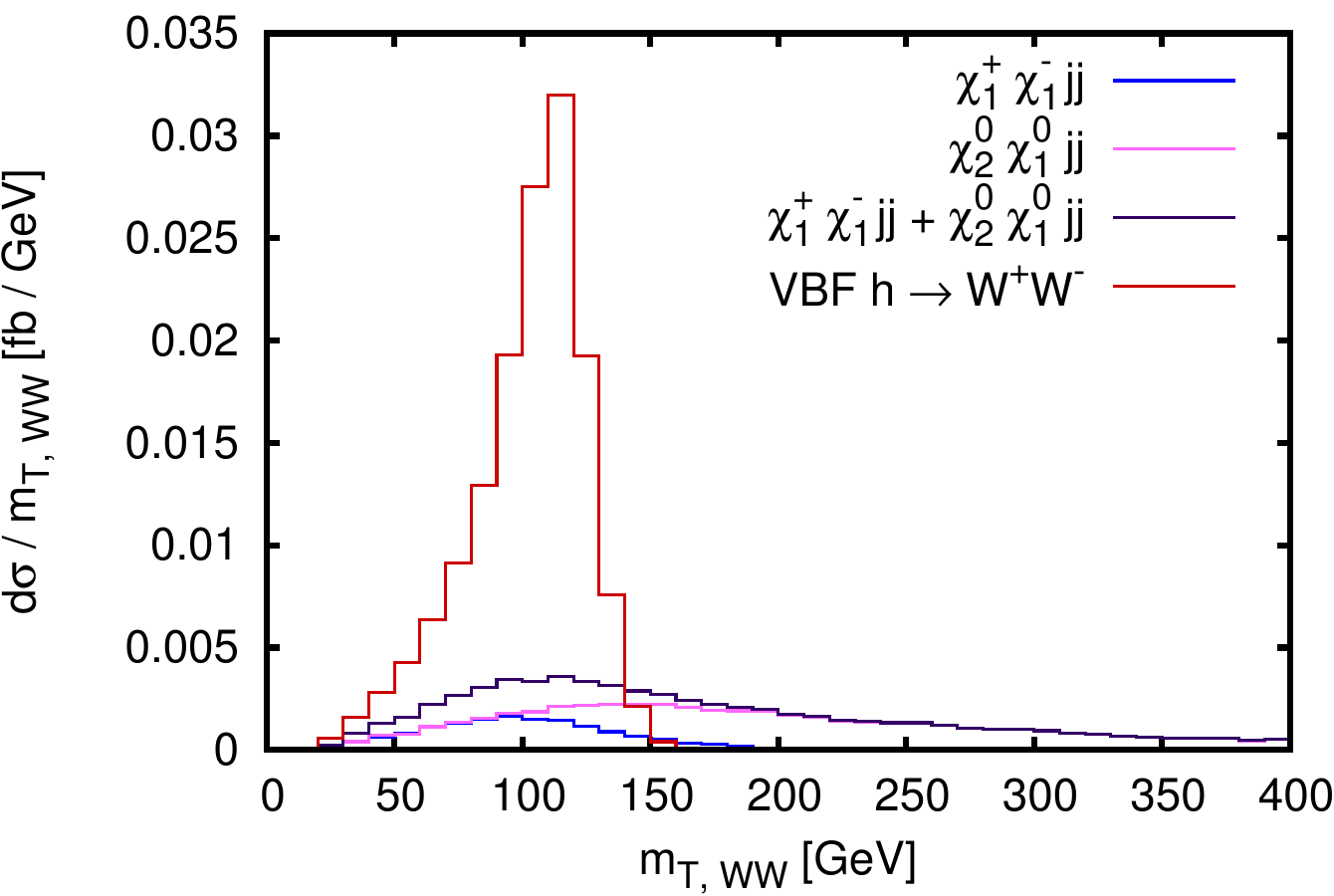}
    \hskip20pt
    \includegraphics[height=0.3\textwidth]{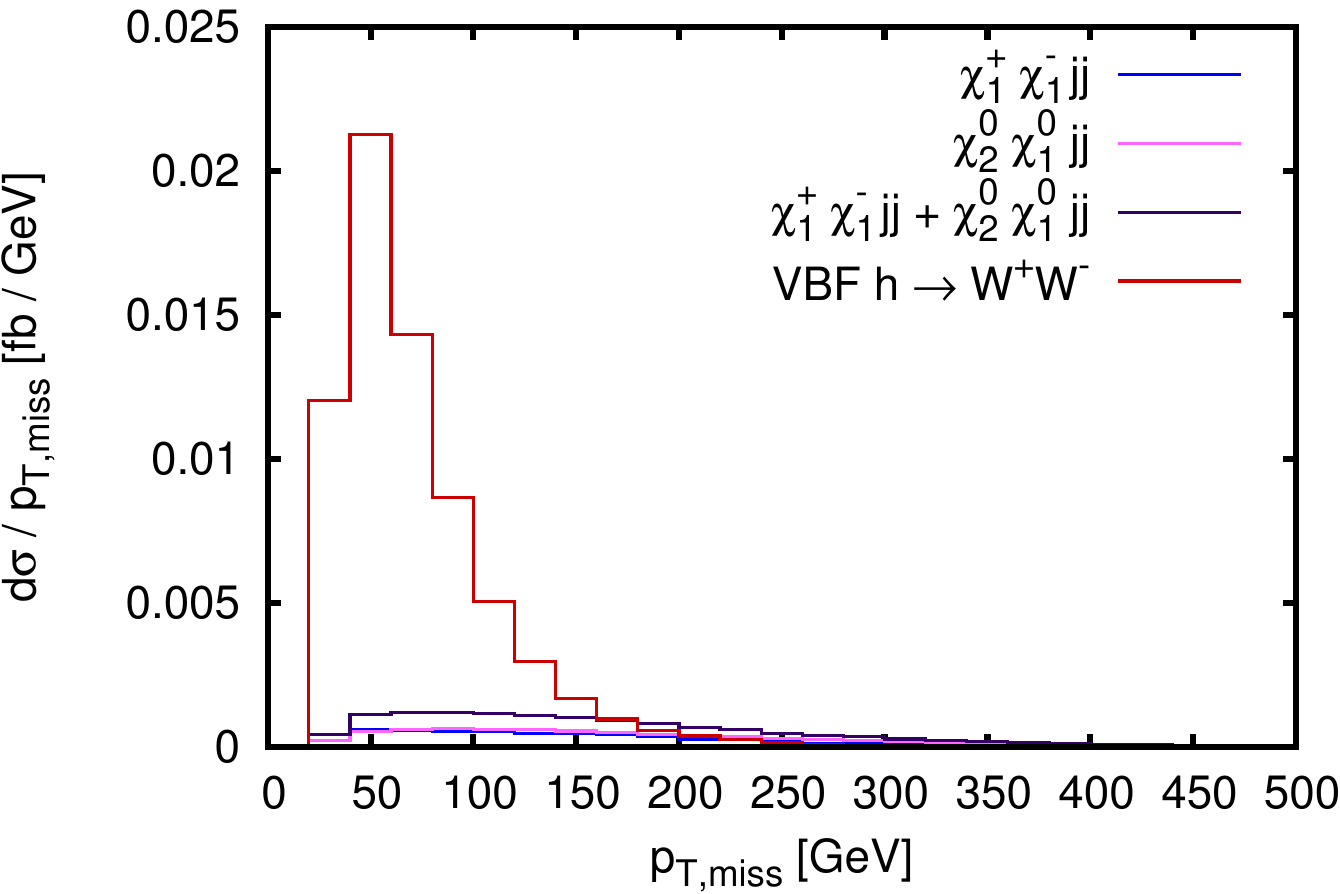}
  \end{center}
  \caption{Upper row: Azimuthal angle between leptons (left) and invariant mass of both leptons (right) 
           with cuts (\ref{cuts_W_min}) - (\ref{cuts_W_deltaeta}).
           Lower left panel: Transverse $WW$ mass distribution with cuts (\ref{cuts_W_min}) - (\ref{cuts_W_nomass}).
           Lower right panel: $\slashed{p}_T$ distribution with cuts (\ref{cuts_W_min}) - (\ref{cuts_W_nomass}), 
           (\ref{cuts_W_paper}).}
  \label{Wjj_plots}
\end{figure}

Several other cuts are needed for Standard Model background reduction
\cite{VBFHinWW} which also suppress the SUSY background relative to the signal:
\begin{equation}
\begin{array}{rclcrcl}
  \slashed{p}_T &\geq& 30 \;\text{GeV} & \quad & m_{jj} &\geq& 600 \;\text{GeV} \\
  m_{\ell\ell} &\leq& 60 \;\text{GeV} & \quad & \phi_{\ell\ell} &\leq& 140\,^\circ \\
  m_{\tau\tau, rec} &\leq& M_Z - 25 \;\text{GeV} \;.
\end{array}
\label{cuts_W_nomass}
\end{equation}
The $m_{\tau\tau, rec}$ cut rejects all events where the tau pair mass reconstruction is possible
($x_{+}>0$ and $x_{-}>0$) and the resulting reconstructed mass $m_{\tau\tau, rec}$ is greater than 
$M_Z - 25 \;\text{GeV}$, meaning they are consistent with $Z\rightarrow\tau\tau$.
Especially the
cuts on the azimuthal angle between the leptons $\phi_{\ell\ell}$ and the invariant mass 
$m_{\ell\ell}$ (see Fig.~\ref{Wjj_plots}) improve the signal to SUSY background ratio.
At this stage, the cross sections of $\chi_1^+\chi_1^-jj$ plus $\chi_2^0\chi_1^0jj$
are already suppressed by a factor of two compared to the vector boson fusion signal.

\tabcolsep3mm
\begin{table}[p]
  \begin{center}
  \begin{tabular}{|l||c|c|c|}
  \hline
  Cuts & $\chi_1^+\chi_1^-jj$ & $\chi_2^0\chi_1^0jj$ & VBF $h\rightarrow WW$ \\
  \hline
  \hline
  basics (Eq.~(\ref{cuts_W_min}))        & $25.97 \;\text{fb}$  & $66.79 \;\text{fb}$ & $5.09 \;\text{fb}$ \\
  + rapidity gap (Eq.~(\ref{cuts_W_deltaeta})) & $1.21 \;\text{fb}$  & $1.04 \;\text{fb}$ & $2.91 \;\text{fb}$ \\
  + $m_{jj}$, $m_{\ell\ell,max}$, ...
   (Eq.~(\ref{cuts_W_nomass}))   & $0.148 \;\text{fb}$  & $0.537 \;\text{fb}$ & $1.46 \;\text{fb}$ \\
  + $M_T(WW)$ (Eq.~(\ref{cuts_W_paper}))    & $0.113 \;\text{fb}$  & $0.146 \;\text{fb}$ & $1.37 \;\text{fb}$ \\
  + $\slashed{p}_{T,max}$ (Eq.~(\ref{cuts_W_paper_ptmiss}))
                                & $0.073 \;\text{fb}$  & $0.081 \;\text{fb}$ & $1.33 \;\text{fb}$ \\
  \hline
  \end{tabular}
  \end{center}
  \caption{Total cross sections for $\chi_1^+\chi_1^-jj$, $\chi_2^0\chi_1^0jj$
           and VBF $h\rightarrow WW$ at different cut levels for the scenario SPS1amod.}
  \label{xs_W_table}
\end{table}

\tabcolsep3mm
\begin{table}[p]
  \begin{center}
  \begin{tabular}{|l||c|c|c|}
  \hline
  Cuts & $\chi_1^+\chi_1^-jj$ & $\chi_2^0\chi_1^0jj$ & VBF $h\rightarrow WW$ \\
  \hline
  \hline
  basics + rapidity gap (Eqs.~(\ref{cuts_W_min})+(\ref{cuts_W_deltaeta}))
                                & $1.20 \;\text{fb}$  & $0.85 \;\text{fb}$  & $4.96 \;\text{fb}$ \\
  + $m_{jj}$, $m_{\ell\ell,max}$, ...
   (Eq.~(\ref{cuts_W_nomass}))   & $0.149 \;\text{fb}$ & $0.444 \;\text{fb}$ & $2.48 \;\text{fb}$ \\
  + $M_T(WW)$
   (Eq.~(\ref{cuts_W_paper}))    & $0.110 \;\text{fb}$ & $0.130 \;\text{fb}$ & $2.28 \;\text{fb}$ \\
  + $\slashed{p}_{T,max}$
   (Eq.~(\ref{cuts_W_paper_ptmiss}))
                                & $0.070 \;\text{fb}$ & $0.072 \;\text{fb}$ & $2.21 \;\text{fb}$ \\
  \hline
  \end{tabular}
  \end{center}
  \caption{Total cross sections for $\chi_1^+\chi_1^-jj$, $\chi_2^0\chi_1^0jj$
           and VBF $h\rightarrow WW$ at different cut levels for the scenario SPS1amod2.}
  \label{xs_W_table_mod2}
\end{table}

As mentioned before, the full $WW$ mass reconstruction is not possible, but we
can apply a cut on 
the transverse $WW$ mass, which is effective against Standard Model backgrounds and
the SUSY backgrounds.
The transverse $WW$ mass can be defined as \cite{VBFHinWW}
\begin{equation}
 M_T(WW)=\sqrt{(\slashed{E}_T+E_{T,\ell\ell})^2-(\slashed{\mathbf{p}}_T+{\mathbf{p}}_{T,\ell\ell})^2}
 \label{transverse_WW_mass}
\end{equation}
with $E_{T,\ell\ell}=\sqrt{{\mathbf{p}}_{T,\ell\ell}^2 + m_{\ell\ell}^2}$ and 
$\slashed{E}_T=\sqrt{\slashed{\mathbf{p}}_T^2 + m_{\ell\ell}^2}$,
where $\mathbf{p}_{T,\ell\ell}$ is the sum of the transverse momenta of the charged leptons.
The signal peaks around the Higgs mass as depicted in Fig.~\ref{Wjj_plots}. The cut
\begin{equation}
 50 \;\textrm{GeV} < M_T(WW) < m_h + 20 \;\textrm{GeV} \;,
 \label{cuts_W_paper}
\end{equation}
which was proposed in \cite{VBFHinWW}, reduces the chargino background by 25\% and the
next-to lightest neutralino background by 75\%, while the effect on the signal
process is much smaller.

Finally a cut on the missing transverse momentum improves the signal to background ratio
significantly. As can be seen in Fig.~\ref{Wjj_plots}, a cut of
\begin{equation}
 \slashed{p}_T \leq 170 \;\text{GeV}
 \label{cuts_W_paper_ptmiss}
\end{equation}
leaves the signal almost unaffected, but reduces the background further by 40\%.
So after all cuts, the production of $\chi_1^+\chi_1^-jj$ and $\chi_2^0\chi_1^0jj$
generate a background that accounts for 12\% of the VBF $h\rightarrow W^+W^-$ cross section.

\subsubsection{Scenario SPS1amod2 with a higher Higgs boson mass}

We also checked the background with these cuts in the scenario SPS1amod2 introduced in Section~\ref{scenarios},
which aims for a higher Higgs boson mass. With $m_h=124 \;\text{GeV}$ compared to $m_h=118 \;\text{GeV}$,
the $h\rightarrow WW$ branching ratio and therefore the signal cross section is much higher. 
As can be seen in Table \ref{xs_W_table_mod2} the background due to the two SUSY processes 
amounts to only 6.4\% of the signal cross section.

\subsubsection{Central Jet Veto on Additional Jets from QCD Radiation}
\label{qcdjetveto}

\begin{table}[b]
  \begin{center} $
    \begin{array}{|c||c|c|c|}
    \hline
    \sigma  & jj & jjj & P_{\textrm{veto}}  \\
    \hline \hline
    \chi_1^+\chi_1^-   &  0.073 \;\textrm{fb}  &  0.044 \;\textrm{fb} &  0.45  \\
    \chi_2^0\chi_1^0   &  0.081 \;\textrm{fb} &  0.109 \;\textrm{fb} &  0.74  \\
    h \rightarrow WW   &  1.38  \;\textrm{fb} &  0.139 \;\textrm{fb} &  0.10  \\
    \hline
    \end{array}$
  \end{center}
  \caption{Total cross sections and central jet veto probabilities
           for $\chi_1^+ \chi_1^- \, jj(j)$, $\chi_2^0 \chi_1^0 \, jj(j)$
           and $h\rightarrow WW$ with final cuts. For the $jjj$ case, the
           cross sections are within the veto region from Eq.~(\ref{vetocuts}).}
  \label{tab_jetveto}
\end{table}

We now want to give a rough estimate for the efficiency of a central jet veto
on the additional hadronic activity due to QCD radiation in SUSY backgrounds
from Eqs.~(\ref{eq_chargino}) and (\ref{eq_neutralino}). 
Therefore we also generated events for the
corresponding three-jet processes $\chi_1^+ \chi_1^- \, jjj$ and 
$\chi_2^0\chi_1^0 \, jjj$  (in the latter case again in the approximation of
taking only $\alpha_s^3 \, \alpha^2$-diagrams into account). As the third jet in these
processes
typically arises from gluon radiation\footnote{An exception is when gluino production 
is involved.}, the average multiplicity of additional
jets can be estimated by $\bar n=\sigma(jjj)/\sigma(jj)$, which leads to
the exponentiation model of Ref.~\cite{jetveto1, *jetveto2}.
In this model, with the final cuts from Eqs.~(\ref{cuts_W_min}) - (\ref{cuts_W_nomass}),
(\ref{cuts_W_paper}) - (\ref{cuts_W_paper_ptmiss}), we get the central jet veto
probabilities $P_{veto}=1-\exp(-\bar n)$ shown in Table \ref{tab_jetveto}. 
The three jet cross sections within the veto region are calculated with
\begin{equation}
\begin{array}{rclcrcl}
  p_{T,j_{veto}} &\geq& 20 \;\text{GeV} & \quad & \eta_{j_{tag},min} \leq &\eta_{j_{veto}}& \leq \eta_{j_{tag},max}  \\
  R_{j_{tag,i},j_{veto}} &\geq& 0.8 & \quad & R_{j_{veto},\ell} &\geq& 0.3 \;, 
\end{array}
 \label{vetocuts}
\end{equation}
where we require the two hardest jets to be the tagging jets $j_{tag,i}$ and consider the third jet 
$j_{veto}$ as a veto candidate. 
In addition the final cuts of section \ref{hwwsps1amod} for the tagging jets and leptons are applied
for the event generation. The small difference in the
$h\rightarrow WW$ cross section compared to Table \ref{xs_W_table} is due to a different
scale choice: We choose $\mu_F=\mu_R=\text{min}(p_T(j_i))$ which is reasonable for the
$jj$ and $jjj$ case.
A detailed analysis of the central jet veto is beyond the scope of this paper, 
but we can see in the approximation of the exponentiation model, that a central 
jet veto would lead to a further reduction of the SUSY backgrounds.

Comparison of the cross sections with two and three additional jets in
Table~\ref{tab_jetveto} also shows that the cross sections which we consider
are sufficiently stable perturbatively. This
situation is different from the one encountered in the $t\bar t$ background to
the VBF Higgs search, where the $t\bar tj$ cross section strongly dominates
over the $t\bar t$ cross section after final cuts~\cite{SMHiggsVBFtotau}.

\subsubsection{Contributions from b-Quarks}

As last task in this section, we want to investigate the contributions which arise
when we allow for b-quarks to appear in the initial and final state.
We separate this subprocess class into two subclasses: The first one requires at
least one $b/\bar{b}$-quark in the initial state with no $b/\bar{b}$-quarks in the
final state. Contributions from these subprocesses are checked to be very small
and will be neglected. The more interesting subprocesses contain at least one
$b/\bar{b}$-quark in the final state with no restrictions on the initial state.
We call contributions from these subprocesses ``contributions from b-quarks''
in the following paragraphs.

\begin{figure}[p]
  \begin{center}
    \includegraphics[height=0.2\textwidth]{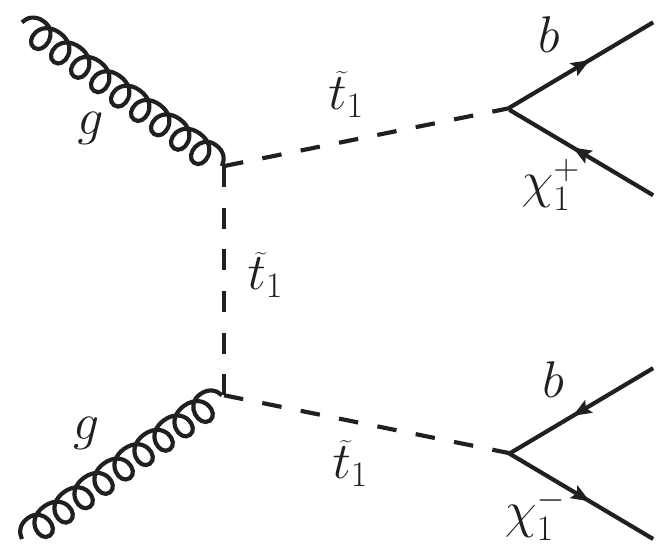}
    \hskip40pt
    \includegraphics[height=0.2\textwidth]{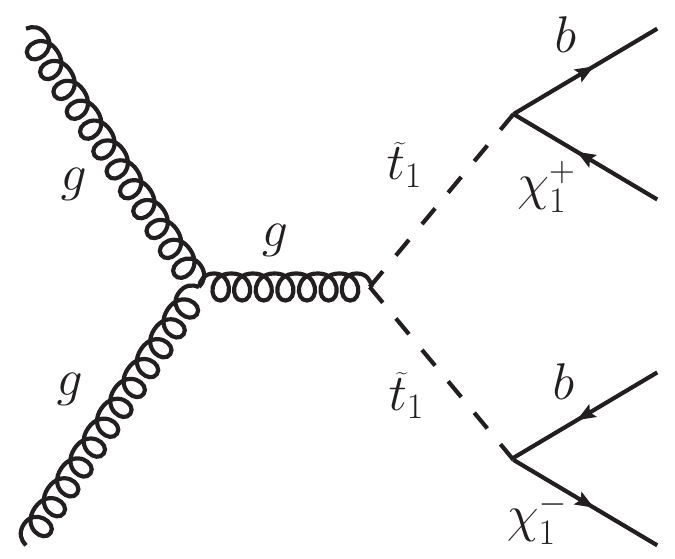}
  \end{center}
  \caption{Dominant Feynman graphs for $\chi_1^+\chi_1^-jj$ production with
           at least one $b/\bar{b}$ quark in the final state.}
  \label{graphs_b_charginos}
\end{figure}

\begin{figure}[p]
  \begin{center}
    \includegraphics[height=0.3\textwidth]{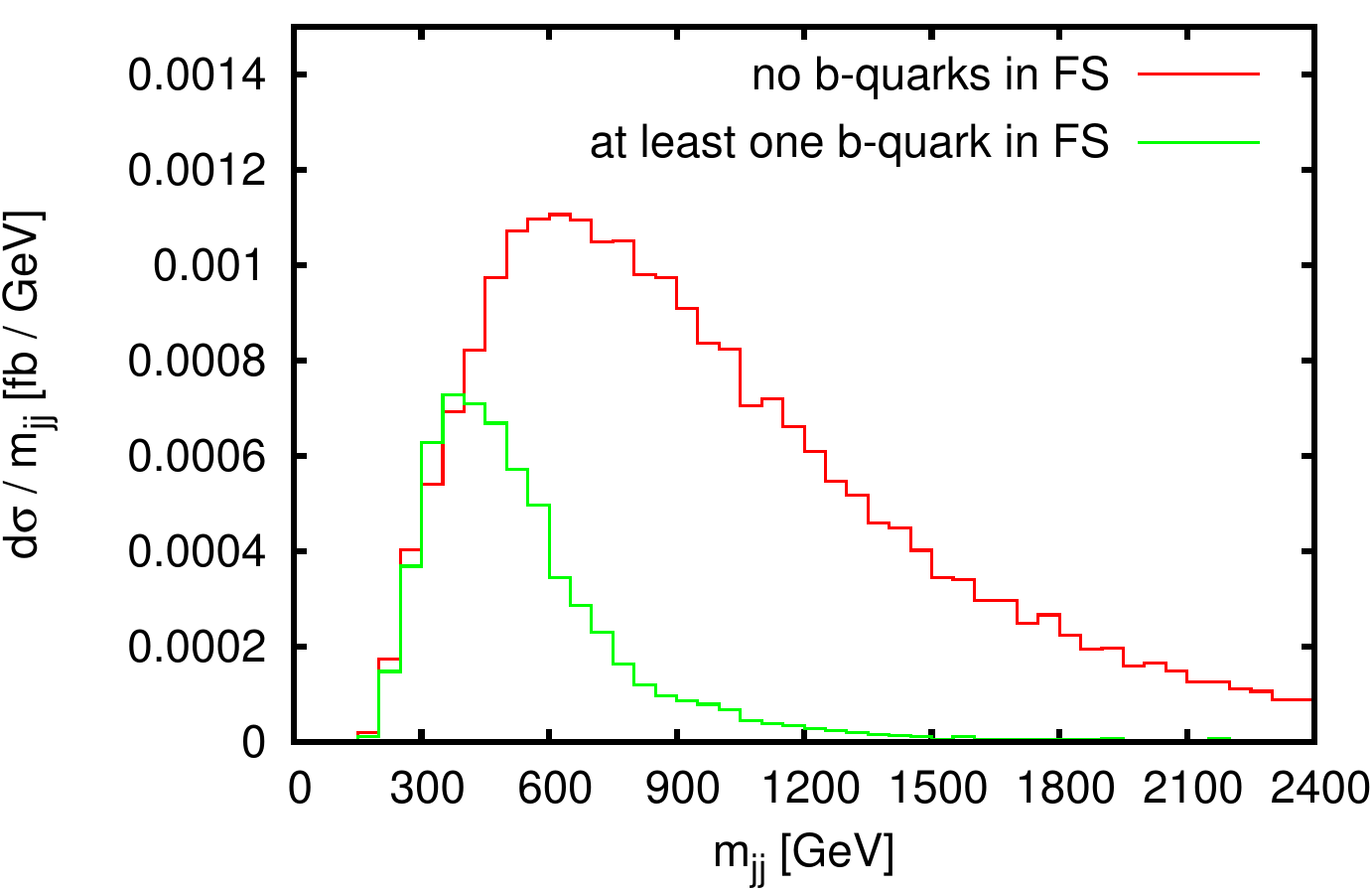}
    \hskip20pt
    \includegraphics[height=0.3\textwidth]{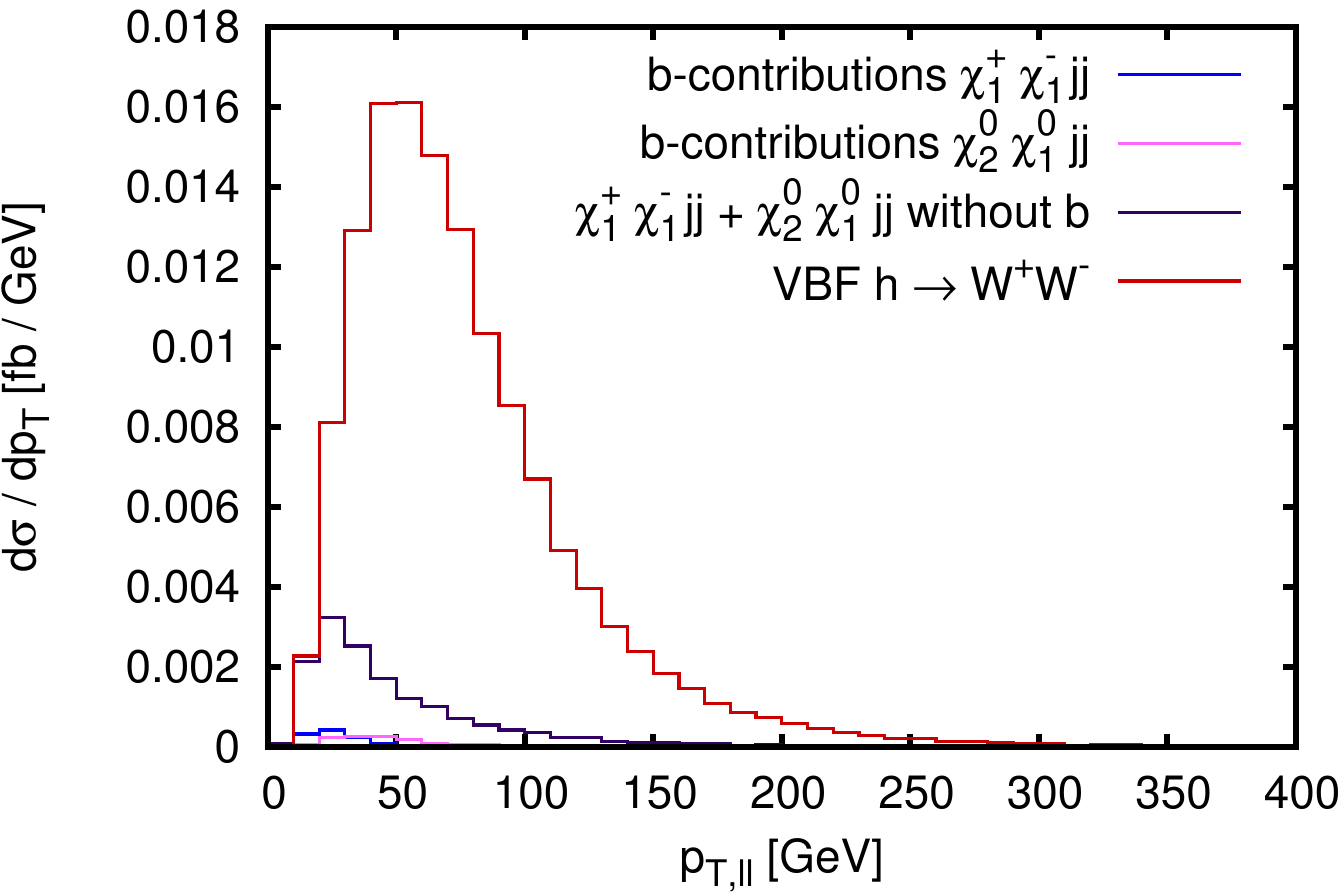}
  \end{center}
  \caption{Left panel: Dijet mass distribution of $\chi_1^+\chi_1^-jj$ with cuts (\ref{cuts_W_min}) + (\ref{cuts_W_deltaeta}).
           Right panel: Dilepton transverse momentum distribution with cuts (\ref{cuts_W_min}) - (\ref{cuts_W_nomass}), 
           (\ref{cuts_W_paper}) - (\ref{cuts_W_paper_ptmiss}) and b-quark veto from Eq.~(\ref{btag}).}
  \label{bquarks_plots}
\end{figure}

For the chargino pair production, b-quark contributions with only basic cuts
from Eq.~(\ref{cuts_W_min}) are of the same order as the previously discussed ones:
Without b-quarks we have a total cross section of 26.0 fb (see
Table~\ref{xs_W_table}), and b-quarks add 14.5 fb 
on top of this. The dominant b-quark contributions stem from the production of a
$\tilde t_1$ pair with two gluons in the initial state where $\tilde t_1$ is
the lighter top squark (Fig.~\ref{graphs_b_charginos}) and 
account for more than 75\% of the b-quark cross section. Because of the light
$\tilde t_1$-mass,
compared to the other squark masses, the dijet mass of these events is much
smaller than the one from non-b-contributions (see Fig.~\ref{bquarks_plots}) and the rapidity
separation of the two jets is also smaller. Therefore the cuts from 
Eqs.~(\ref{cuts_W_deltaeta}) and (\ref{cuts_W_nomass}) are much more efficient.
After applying all cuts (Eqs.~(\ref{cuts_W_min}) - (\ref{cuts_W_nomass}),
(\ref{cuts_W_paper}) -  
(\ref{cuts_W_paper_ptmiss})) the b-quarks add only 25\% to the chargino background.
This can still be reduced by means of a b-jet veto. As b-quarks identified as tagging 
jets are widely separated in rapidity, they are unlikely to lie both in the central 
detector where the b-quark tagging is possible. Nevertheless as most of the b-jet contributions
contain a $b\bar{b}$ pair, at least one of these quarks may lie in the central
region of the detector.
If we assume a b-tagging efficiency of
\begin{equation}
 P_\text{b-tag} = 0.6 \quad \text{for} \quad |\eta_b| \leq 2.5 \;,
 \label{btag}
\end{equation}
we face a mistag rate of at most 1\%\cite{ATLASTDR,b-tag-cms1, *b-tag-cms2}.
We include this 1\% mistag rate in our calculations.
Vetoing events with at least one tagged b-quark,
the b-quark contributions from charginos reduce to 13\% of the chargino
cross section without b-quarks for final cuts. Details are given in
Table \ref{tab_bquarks} and Fig.~\ref{bquarks_plots}.

\begin{figure}[p]
  \begin{center}
    \includegraphics[height=0.2\textwidth]{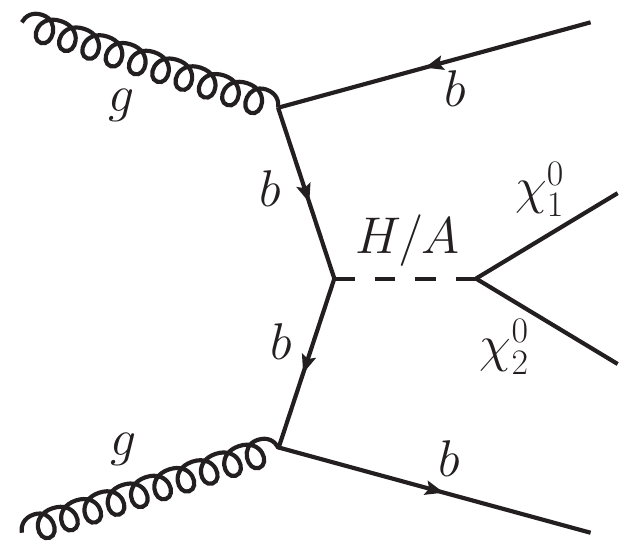}
    \hskip40pt
    \includegraphics[height=0.2\textwidth]{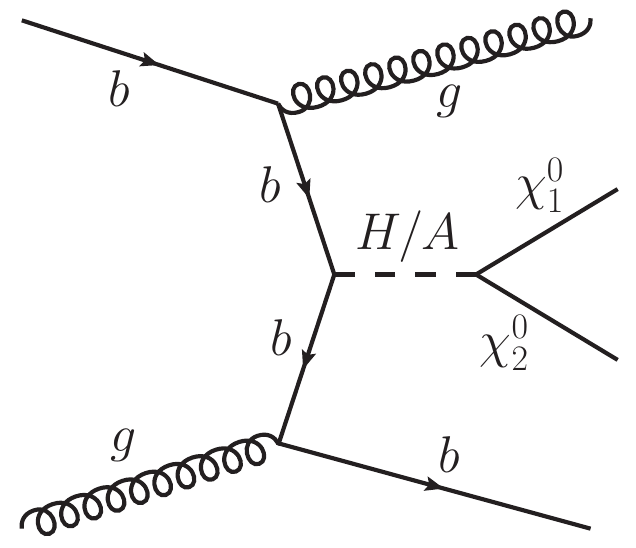}
  \end{center}
  \caption{Additional Feynman graphs for $\chi_2^0\chi_1^0jj$ production with
           at least one $b/\bar{b}$ quark in the final state.}
  \label{graphs_b_neutralino}
\end{figure}

\tabcolsep2.5mm
\begin{table}[p]
  \begin{center}
  \begin{tabular}{|l||c|c|c|c|}
  \hline
  & \multicolumn{2}{|c|}{b-quark contr. in} & no b-quarks & VBF \\
  Cuts & $\chi_1^+\chi_1^-jj$ & $\chi_2^0\chi_1^0jj$ & $\chi_1^+\chi_1^-jj + \chi_2^0\chi_1^0jj$ & $h\rightarrow WW$ \\
  \hline
  \hline
  Eq.~(\ref{cuts_W_min})        & $14.50 \;\text{fb}$  & $7.65 \;\text{fb}$ & $92.76 \;\text{fb}$ & $5.09 \;\text{fb}$ \\
  + Eq.~(\ref{cuts_W_deltaeta}) & $0.31 \;\text{fb}$   & $0.29 \;\text{fb}$ & $2.25 \;\text{fb}$  & $2.91 \;\text{fb}$ \\
  + Eqs.~(\ref{cuts_W_nomass}),(\ref{cuts_W_paper}),(\ref{cuts_W_paper_ptmiss})
                                & $0.022 \;\text{fb}$  & $0.022 \;\text{fb}$ & $0.154 \;\text{fb}$ & $1.33 \;\text{fb}$ \\
  + b-tagging (Eq.~\ref{btag})
                                & $0.012 \;\text{fb}$  & $0.014 \;\text{fb}$ & $0.153 \;\text{fb}$ & $1.32 \;\text{fb}$ \\
  \hline
  \end{tabular}
  \end{center}
  \caption{b-quark contributions for $\chi_1^+\chi_1^-jj$ \& $\chi_2^0\chi_1^0jj$ in the scenario SPS1amod with different
           cuts and a b-quark veto.}
  \label{tab_bquarks}
\end{table}

In contrast to the chargino production, squark production plays a minor role for
the b-quark contributions to the $\chi_2^0\chi_1^0 jj$ production.
There are several reasons for this: First, the higher
bottom squark mass compared to the small $\tilde t_1$-mass. Second, the much  
smaller branching ratio of the decay channel bottom squark $\tilde b_i$, $i =
1,2$ to bottom quarks and neutralinos. 
Also, the analogous Feynman graph to the dominant graphs of the chargino case
is not allowed: 
Since a $g$-$\widetilde{b}_1$-$\widetilde{b}_2$ coupling does not exist, only
$\widetilde{b}_1\,\widetilde{b}_1$ and $\widetilde{b}_2\,\widetilde{b}_2$
production needs to be considered. According to Table~\ref{brtable} their decay
favors the production of either zero or two $\chi_2^0$ states, whose decays do
not typically result in exactly two charged leptons.
Therefore the cross section is no longer dominated by squark pair  
production contributions, especially after the $\Delta \eta$ cut, and major
contributions come from graphs like the ones shown in Fig.~\ref{graphs_b_neutralino}. 
As a consequence, we choose the 
factorization and renormalization scales as for the chargino production,
described in Section~\ref{procedure}.
Detailed numbers are given in Table \ref{tab_bquarks}.
In total, the b-quark contributions increase the SUSY background by 17\% of
the background due to
chargino and next-to lightest neutralino production without b-quark
contributions.

\subsection{Backgrounds with Sleptons}
\label{sleptons-sps1amod}

  \begin{figure}[p]
  \begin{center}
      \includegraphics[height=0.30\textwidth]{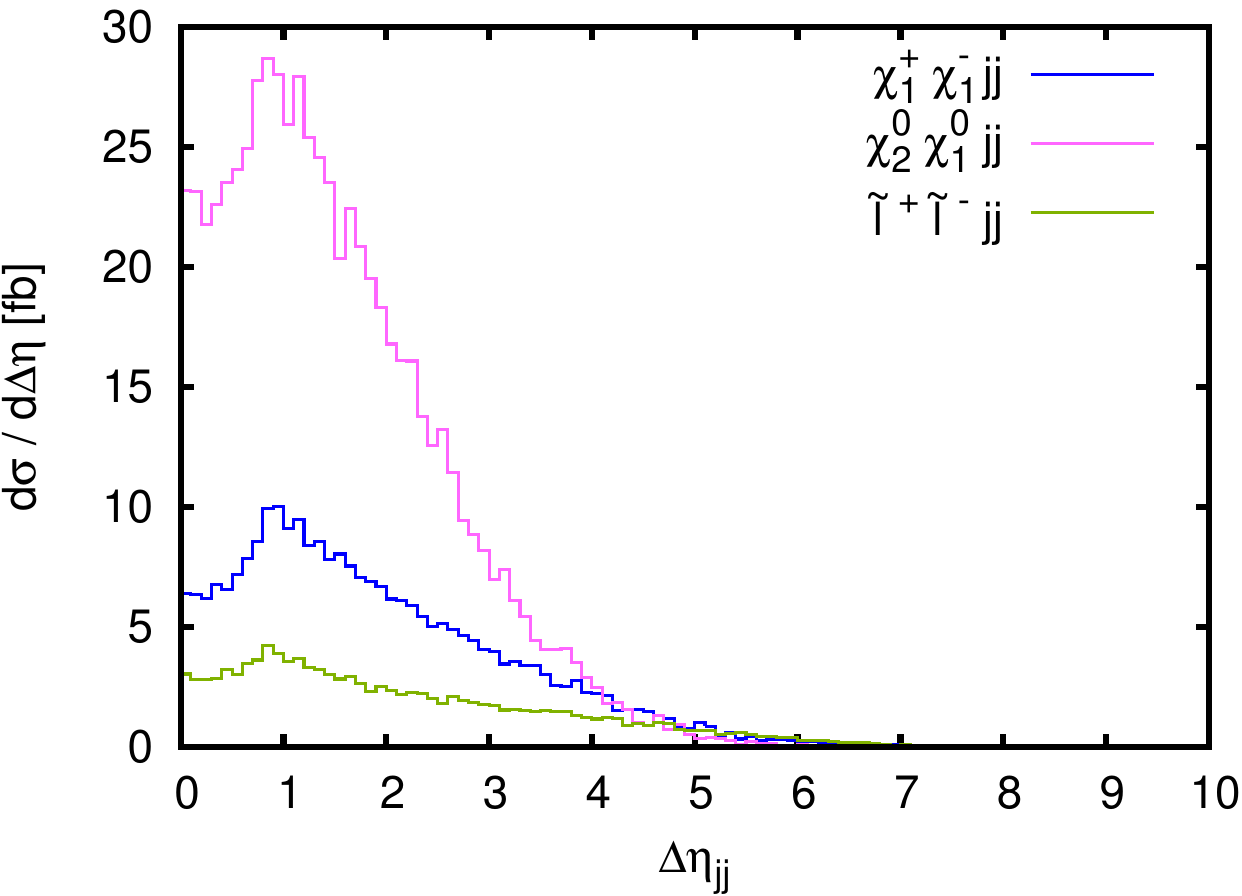}
      \hskip5pt
      \includegraphics[height=0.30\textwidth]{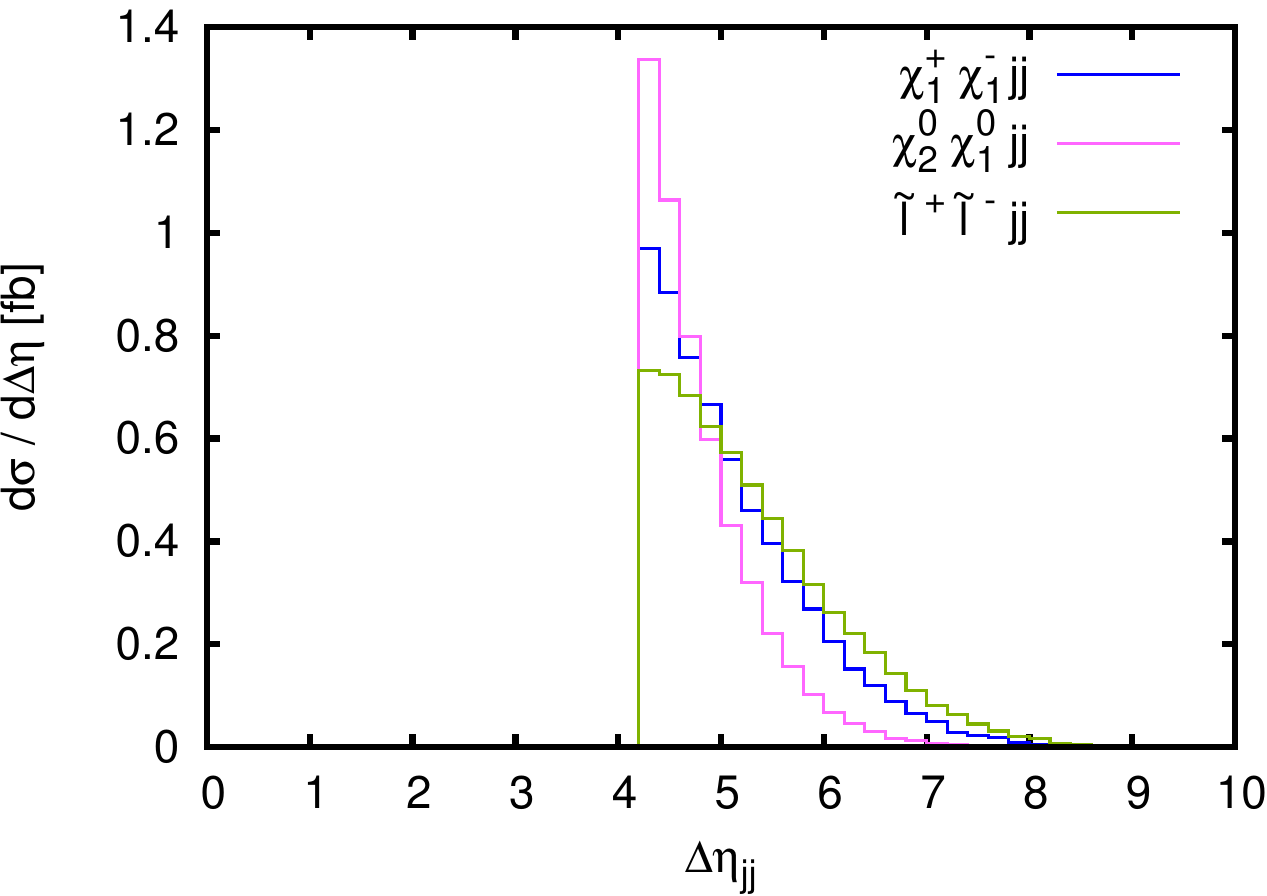}\\
  \end{center}
  \caption{$\Delta \eta_{jj}$ distribution of the channels
   $\widetilde{\ell}^+\,\widetilde{\ell}^-\,jj$, $\chi_1^+\chi_1^-jj$ and
   $\chi_2^0\chi_1^0jj$ without b-quark contributions, applying the cuts of 
   Eq.~\eqref{cuts_W_min} (left) and of Eq.~\eqref{cuts_W_min} and
   Eq.~\eqref{cuts_W_deltaeta} (right).}
  \label{slepton_plots_deta}
  \end{figure}

  \begin{figure}[p]
  \begin{center}
      \includegraphics[height=0.30\textwidth]{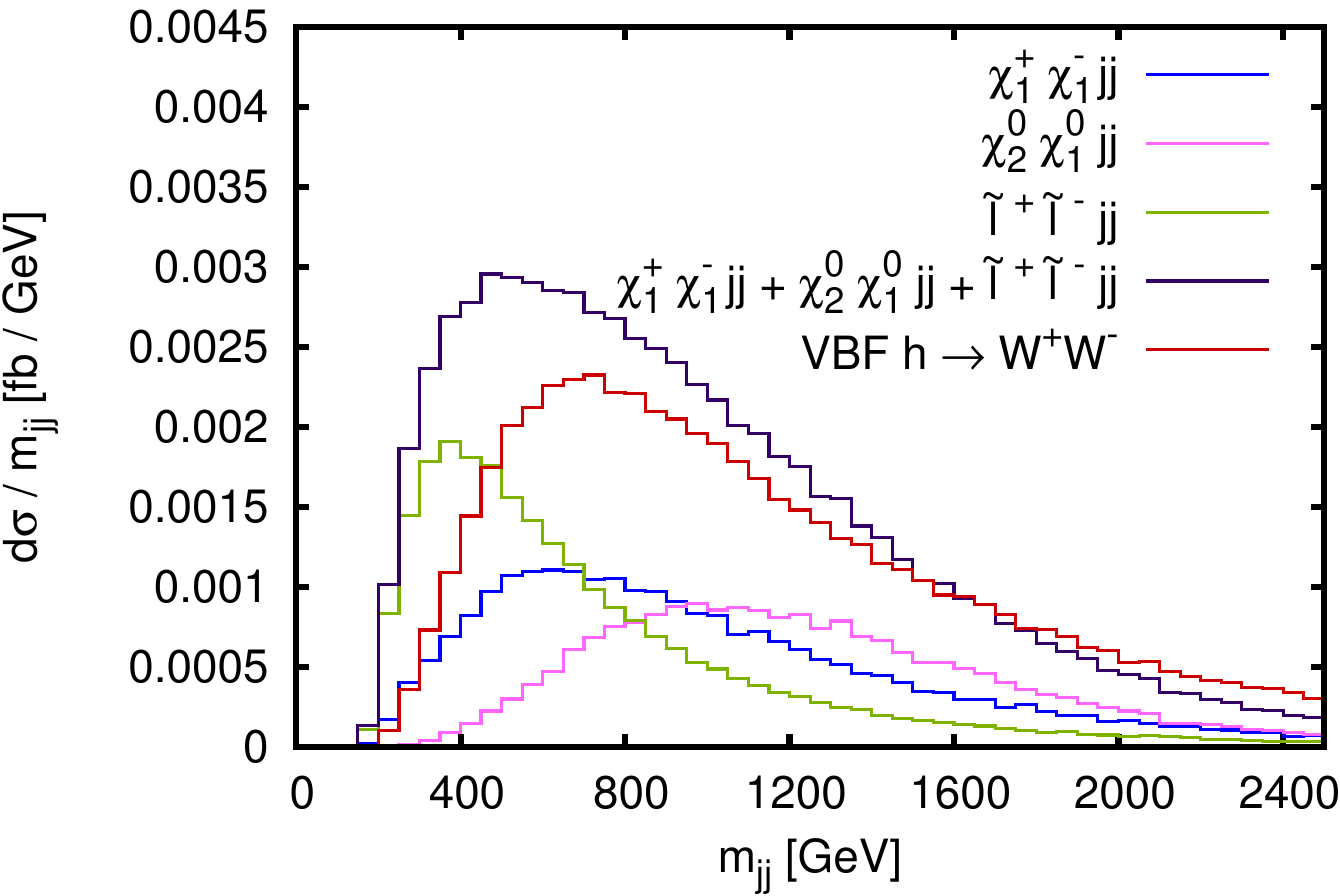}
      \hskip5pt
      \includegraphics[height=0.30\textwidth]{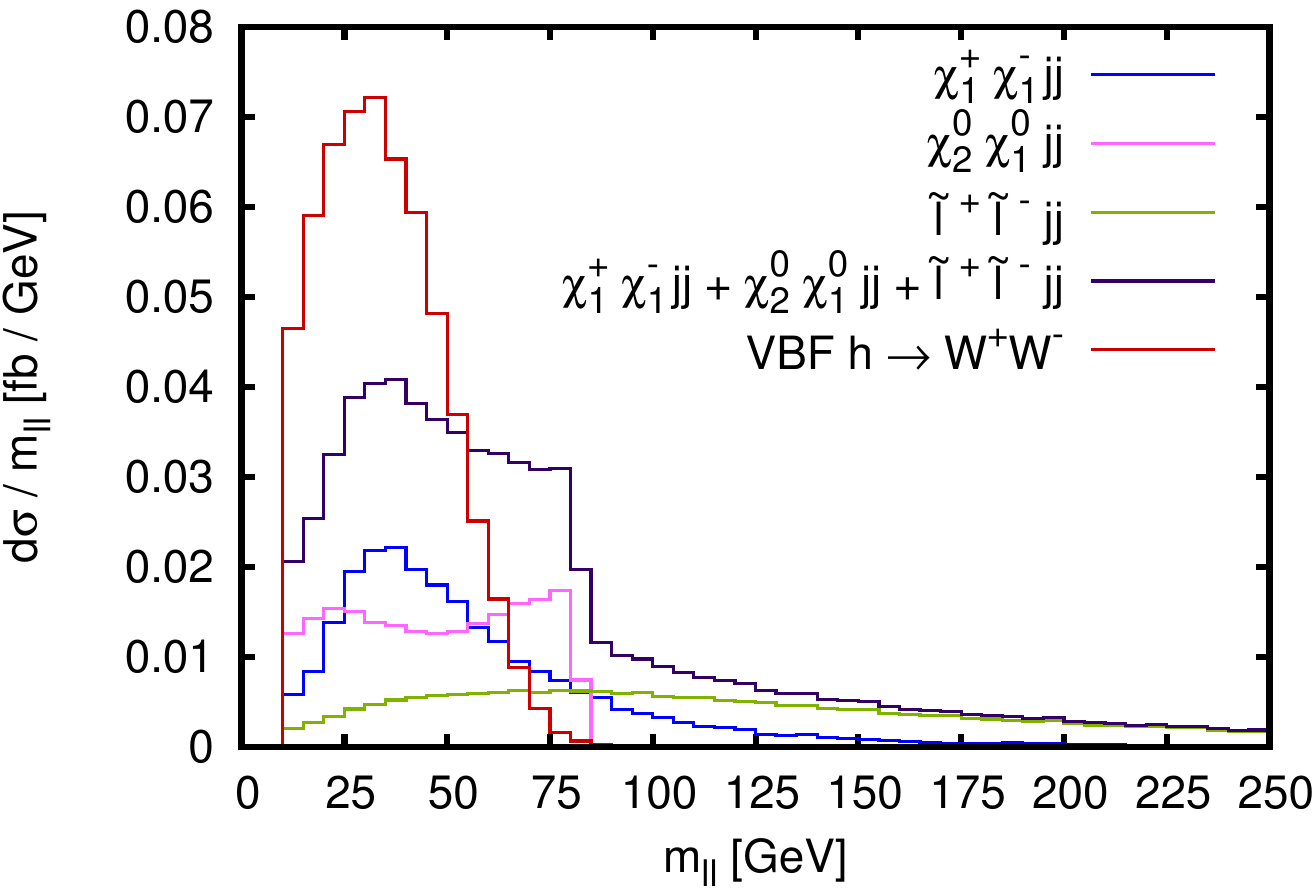}
  \end{center}
  \caption{Invariant jet pair
   mass distribution (left) and  invariant lepton pair mass
   distribution (right) of the
   $\widetilde{\ell}^+\,\widetilde{\ell}^-\,jj$, $\chi_1^+\chi_1^-jj$ and 
   $\chi_2^0\chi_1^0jj$ background channel, of the sum of the considered
   background channels and the signal process. Cuts of
   Eq.~\eqref{cuts_W_min} and \eqref{cuts_W_deltaeta} are applied, b-quark contributions
   are not included.}
  \label{slepton_plots}
  \end{figure}

\tabcolsep3mm
\begin{table}[p]
  \begin{center}
  \begin{tabular}{|l||c|c|c||c||c|}
  \hline
  Cuts & $\chi_1^+\chi_1^-jj$ & $\chi_2^0\chi_1^0jj$ & $\widetilde{\ell}^+\, \widetilde{\ell}^-\, jj$ & $\sum_{irred}^{SUSY}$ & VBF $h\rightarrow WW$ \\
  \hline
  \hline
  Eq.~(\ref{cuts_W_min})        & $40.47 \;\text{fb} $ & $74.44 \;\text{fb}$ & $11.55 \;\text{fb}$ & $ 126.46 \;\text{fb}$ & $5.09 \;\text{fb}$ \\
  + Eq.~(\ref{cuts_W_deltaeta}) & $1.52 \;\text{fb} $  & $1.33 \;\text{fb}$  &  $1.23 \;\text{fb}$ & $   4.08 \;\text{fb}$ & $2.91 \;\text{fb}$ \\
  + Eq.~(\ref{cuts_W_nomass})   & $0.177 \;\text{fb}$  & $0.610 \;\text{fb}$ & $0.080 \;\text{fb}$ & $  0.867 \;\text{fb}$ & $1.46 \;\text{fb}$ \\
  + Eq.~(\ref{cuts_W_paper})    & $0.137 \;\text{fb}$  & $0.172 \;\text{fb}$ & $0.031 \;\text{fb}$ & $  0.340 \;\text{fb}$ & $1.37 \;\text{fb}$ \\
  + Eq.~(\ref{cuts_W_paper_ptmiss})
                                & $0.095 \;\text{fb}$  & $0.103 \;\text{fb}$ & $0.028 \;\text{fb}$ & $  0.226 \;\text{fb}$ & $1.33 \;\text{fb}$ \\
  + Eq.~(\ref{btag})
                                & $0.085 \;\text{fb}$  & $0.094 \;\text{fb}$ & $0.028 \;\text{fb}$ & $  0.207 \;\text{fb}$ & $1.32 \;\text{fb}$ \\
  \hline
  \end{tabular}
  \end{center}
  \caption{Total cross sections for $\chi_1^+\chi_1^-jj$, $\chi_2^0\chi_1^0jj$, $\widetilde{\ell}^+\, \widetilde{\ell}^-\, jj$
           and VBF $h\rightarrow WW$ at various cut levels for the scenario SPS1amod, including 
           b-quark contributions for $\chi_1^+\chi_1^-jj$ and $\chi_2^0\chi_1^0jj$.}
  \label{tab_sleptons}
\end{table}

Additionally to the background processes discussed in Sect.~\ref{sps1amod-tautau} and Sect.~\ref{sps1amod-WW}
there are further contributions to the background (already mentioned in 
Sect.~\ref{scenarios} and Sect.~\ref{procedure}) which involve the production
 of a slepton pair accompanied by two jets 
$pp \rightarrow \widetilde{\ell}^+\,\widetilde{\ell}^-\,jj$, see 
Eq.~\eqref{sleptonproc}. The cross section of this
 process compared to the production processes of charginos and neutralinos
 plus two jets is small, 85~fb for all slepton combinations versus 2.6~pb for
 the chargino production and 1.4~pb for the production of a lightest and a
 next-to-lightest neutralino (only the minimal jet cuts of
 Eq.~\eqref{cuts_W_min} are applied). However, when taking into account the
 decays of the produced SUSY particles, the fraction of the slepton induced
 SUSY background becomes larger as the decay of  $\widetilde{e}_R$ and
 $\widetilde{\mu}_R$ produces directly detectable leptons. In contrast, the
 chargino and the next-to-lightest neutralino channels involve a cascade
 decay. Including the decays and again applying only the jet cuts of 
 Eq.~\eqref{cuts_W_min} the slepton channel cross
 section is 46~fb versus 312~fb and 321~fb for the cross section of the
 chargino and the neutralino channel, respectively. 

 Incorporating also the lepton cuts for the $h \rightarrow WW$ analysis, Eq.~\eqref{cuts_W_min}, the relative
 contribution of the sleptons rises again, see Table~\ref{tab_sleptons}:
 The channels including the production and decay of $\widetilde{e}_R$ and $\widetilde{\mu}_R$ 
 do not contain a tau-lepton decay which leads to charged final state 
 leptons with larger transverse momentum 
 than in the case where the detected leptons originate from a tau-lepton decay.
 Therefore, with respect to the chargino and next-to-lightest neutralino case,
 the acceptance of the events is increased when
 the cut on the transverse momentum of the lepton is applied.

 The rapidity separation of the jets, $\Delta\eta_{jj}$, is larger for the slepton
 channel than for chargino and next-to-lightest neutralino ones, which 
 leads to a slepton channel cross section which is comparable in size to the
 ones of the chargino and the neutralino channel at the $\Delta\eta$-cut level (see
 Fig.~\ref{slepton_plots_deta}). 

On the other hand the cut on the invariant jet pair
mass is more efficient for the sleptons, as $m_{jj}$
is rather small compared to the other channels which can be seen in
Fig.~\ref{slepton_plots}. 
Also, the cut on the  
invariant lepton pair mass decreases the contribution of the slepton channel as
the invariant lepton pair mass distribution is rather flat, yielding sizable
contributions at larger masses (see Fig.~\ref{slepton_plots}).

 Finally, 
the reconstructed transverse $WW$ mass tends to be larger than for the signal 
which means that the corresponding cut is efficient in reducing this 
background contribution. Overall, taking all cuts into account, the cross 
section of the slepton channel is 0.028~fb whereas the cross section of the 
chargino and next-to-lightest neutralino channel including b-quark contributions
is 0.179~fb.
B-quark contributions to the slepton channel account for only 3\% of the corresponding 
cross section without b-quarks (including decays) in the final state and are therefore neglected.
Summing the irreducible background contributions from the chargino, neutralino and slepton channels yields 
$\sigma^{SUSY}_{irred}=0.207 \;\text{fb}$ while the $h\rightarrow WW$ signal process
has a cross section of $\sigma^{h\rightarrow WW} = 1.32 \;\text{fb}$.

\vskip6pt

For the $h \rightarrow \tau\tau$ analysis the situation is still very good:
With all cuts of Eqs.~\eqref{cuts_tau_min}-\eqref{cuts_tau_massrec}, including
detector effects of Eq.~\eqref{fakeptmiss}, the sleptons contribute with
a cross section of 0.0034~fb, compared to 0.0274~fb and 0.0174~fb of the
chargino and the next-to-lightest neutralino channel cross section (with b-quark 
contributions), respectively. Therefore the sum of these three processes 
($\sigma^{SUSY}_{irred}=0.048 \;\text{fb}$) is still
much smaller than the $h \rightarrow \tau\tau$ cross section 
$\sigma^{h\rightarrow \tau\tau}=1.93 \;\text{fb}$.

\section{SPS1a-like Scenario: Reducible Background}
\label{reducible}

In this section, we want to focus on processes that lead to additional jets or
leptons 
compared to the signature of the considered VBF Higgs boson
signal channels. In principle, these events can be
eliminated by vetoing additional particles. However, these additional particles
can remain undetected so that vetoing is not possible and the events
contribute significantly to the SUSY background.
At the beginning of this section, we will discuss the treatment of the
additional 
particles. Then, the focus will be on the processes which dominantly
contribute to 
the reducible background. In the last part of the section we will summarize
the processes giving rise to the reducible and irreducible SUSY background for
the considered VBF Higgs boson signal channels. The signal to
SUSY background ratios will be given
and the investigated processes that give only very small background
contributions will also be listed.

\subsection{Particle and Event Selection}
\label{particleandeventselection}

Up to now, the particle selection of our parton level analysis was fairly
simple: 
As, on parton level, the previously discussed processes produce exactly the
same number of 
visible particles as appear in the final state of the signal processes, it was
sufficient to apply some separation and $p_T$ cuts  
for the matrix element calculation
and to identify each particle with a detector signal.
The analysis routine contained the other cuts which were used for the background
suppression.
This approach is no longer valid for the reducible background: For example, an
additional tau 
lepton could decay hadronically and point into the same direction as a hard
quark 
from the production process. Those two objects have to be considered as one jet.

For the new processes, we start again with some basic cuts on the jets applied
already at {\tt MadEvent} level
($p_{T,j} > 20 \; \textrm{GeV}$, $|\eta_,j| < 4.5$, $R_{jj} > 0.8$) which ensure
a finite cross section. After performing the decay part with the help of {\tt
  Herwig++}, the events  
contain up to two light partons generated already at {\tt MadEvent} level plus
leptons, additional jets (for 
example from tau lepton or W boson decay) and invisible particles from the decay
of the SUSY particles.

The missing transverse momentum, $\slashed{p}_T$, is calculated via a vectorial
sum of all particles 
that deposit their energy in the calorimeter: This includes light partons,
tau jets and electrons up to $|\eta| \leq 4.5$ and muons up to $|\eta| \leq
2.5$. For all of these particles, a threshold $p_T$ is applied, $p_T \geq
3$~GeV. 

For the jet definition, we use the anti-kt jet clustering algorithm
\cite{antikt}. 
Leptons which are in the vicinity of a jet with $R_{jl} \leq 0.3$ are
included into the jet.  
The resulting jets are defined as visible when they fulfill
\vskip-10pt
\begin{equation}\label{visiblejet}
 p_{T,j} > 20 \; \textrm{GeV}\,, \quad \quad |\eta_j| < 4.5 \,.
\end{equation}
For the leptons, we require
\vskip-10pt
\begin{equation}
 p_{T,l} > 10 \; \textrm{GeV}\,, \quad \quad |\eta_l| < 2.5 \,.
\end{equation}
Events are kept when they finally have at least two visible jets and
exactly one positively and one negatively charged visible lepton.

To get a well defined cross section, we have
to discard an event, if it meets the following criteria:
\begin{itemize}
 \item One of the jets generated at {\tt MadEvent} level is not  a
   tagging jet, but
       instead, a jet resulting from the following SUSY particle decay chain
       serves as 
       tagging jet. 
 \item A jet at {\tt MadEvent} level has to be recombined with 
   a jet originating from the further decays.
\end{itemize}
To explain the problems with events of this type it should be noted that  the
jets 
generated at {\tt MadEvent} level can be characterized as two different kinds
of jets,  
jets originating from a QCD splitting and jets produced via a decay of a heavy
particle. These two contributions, however, can only be separated using a
narrow width approximation but not in a calculation taking into account the
full matrix elements as it is done in our approach:
\begin{itemize}
  \item If the jet at {\tt MadEvent} level originates from a QCD-splitting
         this event is part of the real emission contribution of the
         corresponding {\tt MadEvent} level process with the number of
              jets reduced by one. This contribution is not IR safe but
              diverges for small $p_T$. Taking into account the virtual NLO
              corrections to the {\tt MadEvent} level process with the number
              of jets reduced by one would yield an IR finite result. This
              means that the real emission contribution at parton-level is
              part of the NLO corrections. Another problem, going beyond the
              pure parton level, is that part of the 
              real emission would be generated also by a parton shower and
              would lead to double counting.  
 
              In our approach, the IR divergences are regulated
              by the $p_T$ cut of 20~GeV which is somewhat arbitrary
              but justified as long as the jets generated at {\tt MadEvent}
              level are tagging jets and the parton-level process with exactly
              this number of jets is considered. 
  \item The jet at {\tt MadEvent} level can also be produced by the
          decay of a heavy particle. Recombined jets
     with the jet at {\tt MadEvent} level being of this type are not
     fully included due to the $p_T$ cut at {\tt MadEvent} level. While the
     jet originating from the further cascade decays can have arbitrarily low
     values of $p_T$ the jets at {\tt MadEvent} level with low values of $p_T$,
     are not included.
\end{itemize}
Discarding events where the pure {\tt MadEvent} level jets are no tagging jets
leads to an error that can be approximated requiring $p_{T,j} > 20 \;
              \textrm{GeV}$ for all jets and investigating how large the
              contribution of the discarded events is in that sample: 
              In the first case, where a jet generated at {\tt MadEvent} level
              is not a tagging jet, this contribution accounts for 1-3\% of
              the total cross section after applying the tagging jet rapidity
              separation cut used in the analysis. In the second case, where a
              jet originating from the {\tt MadEvent} level process  is
              recombined with a jet produced via the further decay chain, 
               the size of the discarded contributions can be estimated 
               to be below 1\%. Considering the precision of our analysis, the
               overall size of the error due to the event selection described
               above is small.

After this procedure, the analysis cuts presented in the previous section are
applied.

\subsection{\texorpdfstring{Contributions to the {\boldmath$h \rightarrow W^+W^-$} Channel}
{Contributions to the W+ W- Channel}}

In this subsection, the dominant reducible background channels to the VBF Higgs production signal channel 
with a subsequent decay to W bosons are investigated. The considered channels comprise the production of a 
lightest chargino and a next-to lightest neutralino plus 2 jets as well as of
a pair of next-to lightest neutralinos
plus 2 jets including the subsequent decays of the SUSY particles,
Eq.~\eqref{eq_CH1pN2}~-~\eqref{eq_N2N2}. 
Background channels with, at {\tt MadEvent} level, the same SUSY particles  but fewer jets in the final state 
play a much less important role and are listed, together with other less dominant channels, in 
Sect.~\ref{sps1asummary}.

\tabcolsep3mm
\begin{table}[b]
  \begin{center}
  \begin{tabular}{|l||c|c|c|c||c||c|}
  \hline
       &  &  & \multicolumn{2}{|c||}{$\chi_2^0\chi_2^0jj$} & & VBF \\
  Cuts & $\chi_1^+\chi_2^0jj$ & $\chi_1^-\chi_2^0jj$ & no b & b-contr. & $\sum_{red}^{SUSY}$ & $h\rightarrow WW$ \\
  \hline
  \hline
  Eq.~(\ref{cuts_W_min})
                                & $100.8 \;\text{fb}$  & $ 63.2 \;\text{fb}$  & $ 46.4 \;\text{fb}$  & $ 5.02 \;\text{fb}$ & $215.4 \;\text{fb}$   & $5.09  \;\text{fb}$ \\
  + Eq.~(\ref{cuts_W_deltaeta})
                                & $ 3.94 \;\text{fb}$  & $ 2.20 \;\text{fb}$  & $ 1.35 \;\text{fb}$  & $0.168 \;\text{fb}$ & $7.66 \;\text{fb}$   & $2.91  \;\text{fb}$ \\
  + Eq.~(\ref{cuts_W_nomass})   & $ 1.46 \;\text{fb}$  & $0.814 \;\text{fb}$  & $0.537 \;\text{fb}$  & $0.037 \;\text{fb}$ & $2.85 \;\text{fb}$   & $1.46  \;\text{fb}$ \\
  + Eq.~(\ref{cuts_W_paper})    & $0.562 \;\text{fb}$  & $0.310 \;\text{fb}$  & $0.231 \;\text{fb}$  & $0.020 \;\text{fb}$ & $1.123 \;\text{fb}$   & $1.37 \;\text{fb}$ \\
  + Eq.~(\ref{cuts_W_paper_ptmiss})
                                & $0.406 \;\text{fb}$  & $0.225 \;\text{fb}$  & $0.150 \;\text{fb}$  & $0.018 \;\text{fb}$ & $0.799 \;\text{fb}$   & $1.33 \;\text{fb}$ \\
  + Eq.~(\ref{btag})
                                & $0.403 \;\text{fb}$  & $0.222 \;\text{fb}$  & $0.149 \;\text{fb}$  & $0.010 \;\text{fb}$ & $0.784 \;\text{fb}$   & $1.32 \;\text{fb}$ \\
  + Eq.~(\ref{cuts_jetveto})
                                & $0.275 \;\text{fb}$  & $0.144 \;\text{fb}$  & $0.059 \;\text{fb}$  & $0.006 \;\text{fb}$ & $0.484 \;\text{fb}$   & $1.32 \;\text{fb}$ \\
  \hline
  \end{tabular}
  \end{center}
  \caption{Total cross sections for $\chi_1^\pm\,\chi_2^0jj$, $\chi_2^0\chi_2^0jj$
           and VBF $h\rightarrow WW$ at different cut levels for the scenario SPS1amod.}
  \label{xs_W_table_reducible}
\end{table}

\begin{figure}[p]
  \begin{center}
    \includegraphics[height=0.30\textwidth]{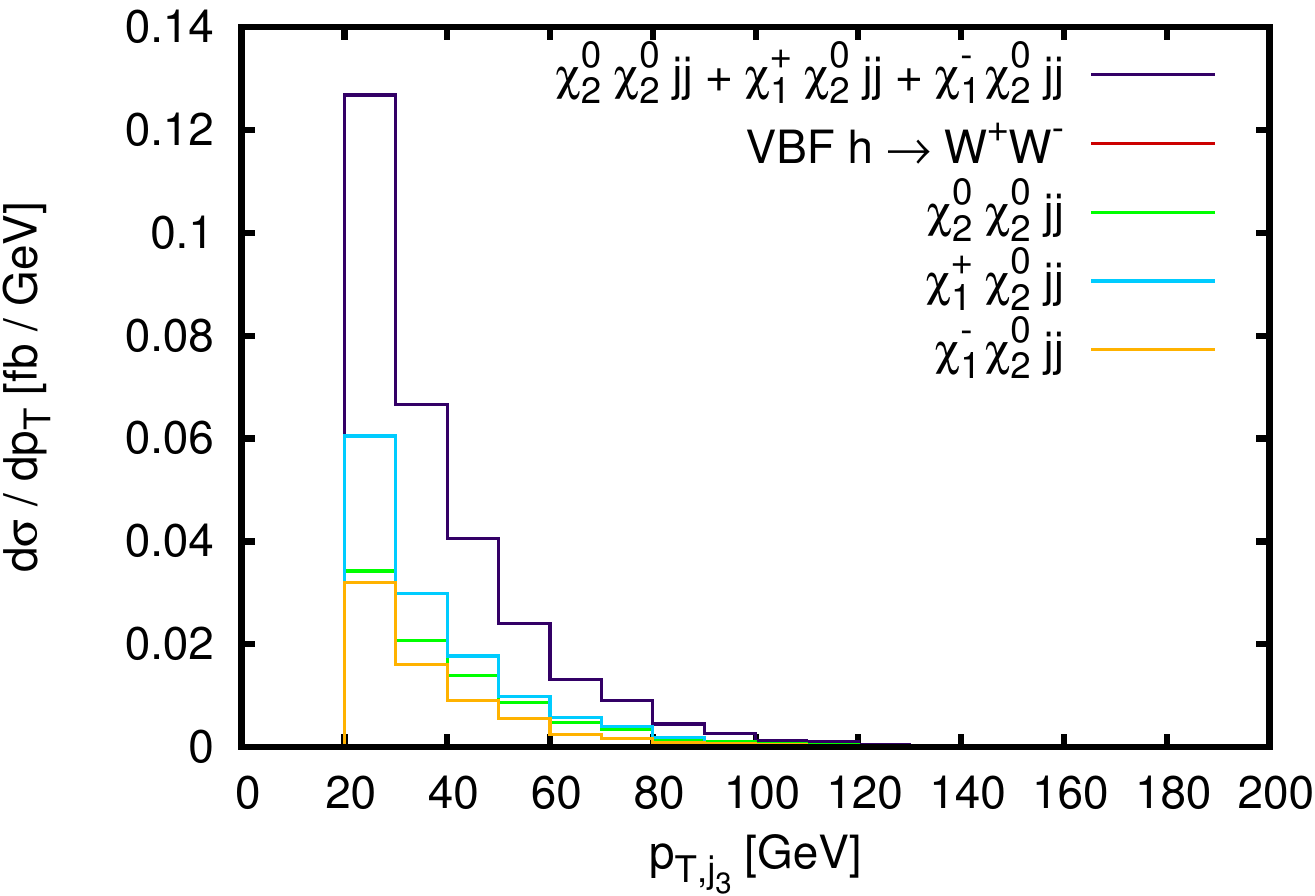}
    \hskip20pt
    \includegraphics[height=0.30\textwidth]{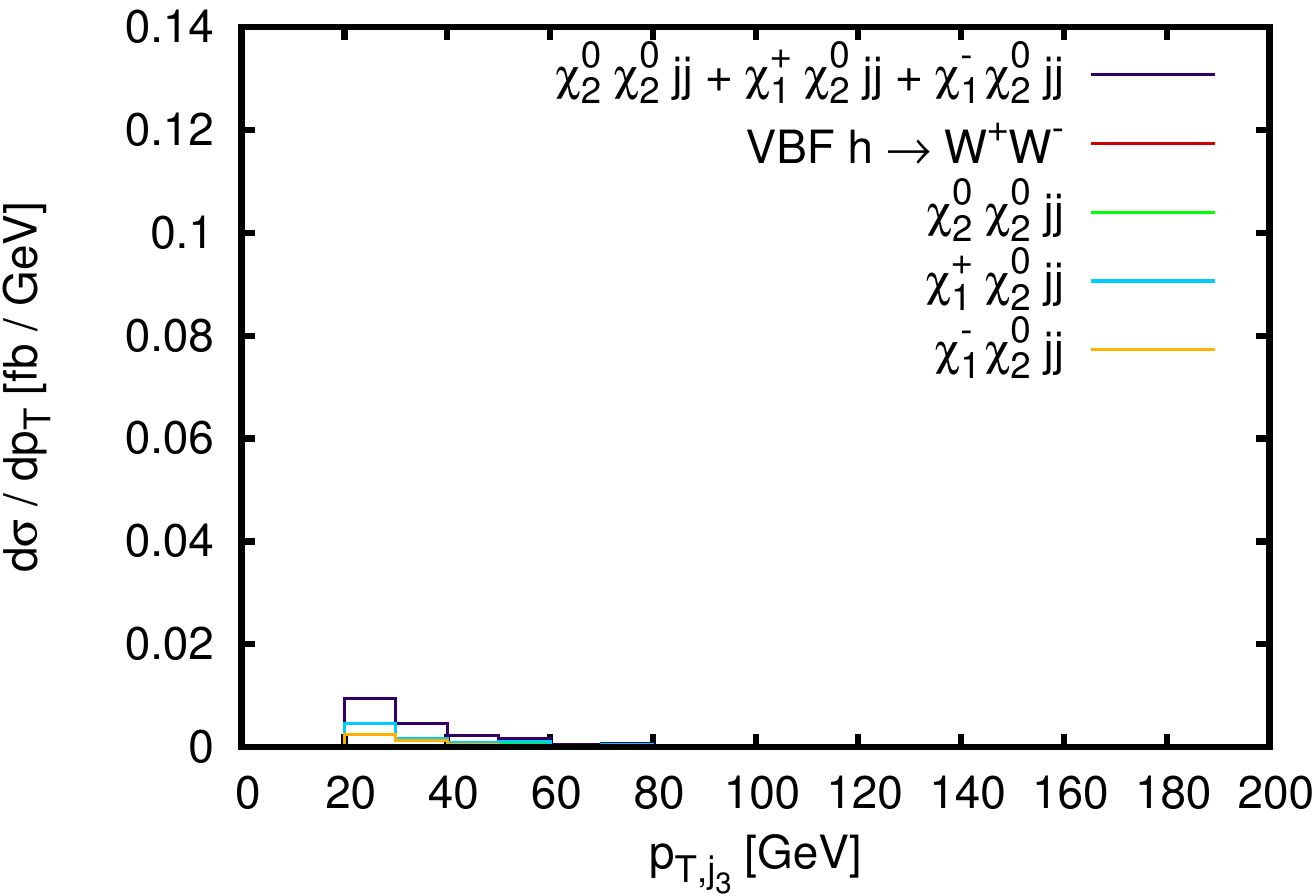}\\
    \vskip20pt
    \includegraphics[height=0.30\textwidth]{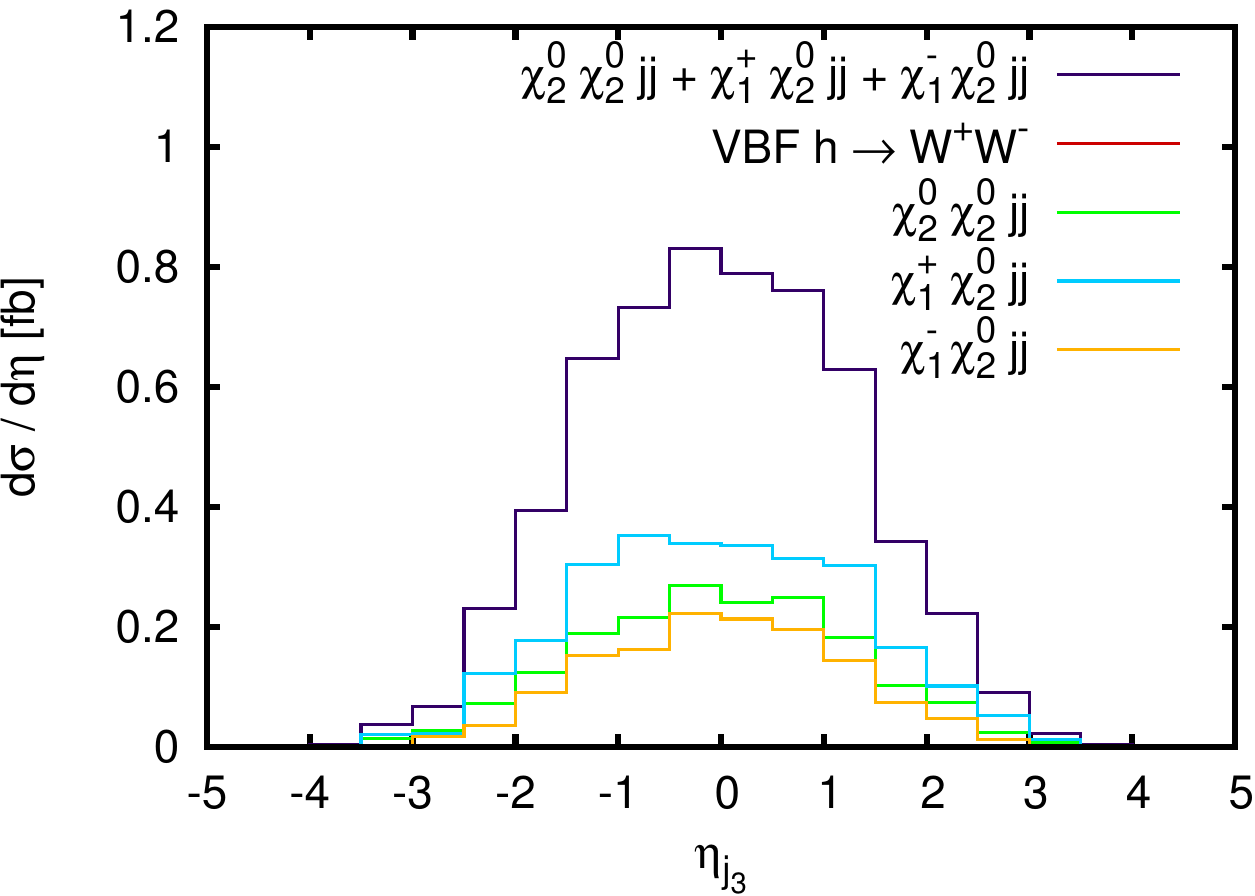}
    \hskip20pt
    \includegraphics[height=0.30\textwidth]{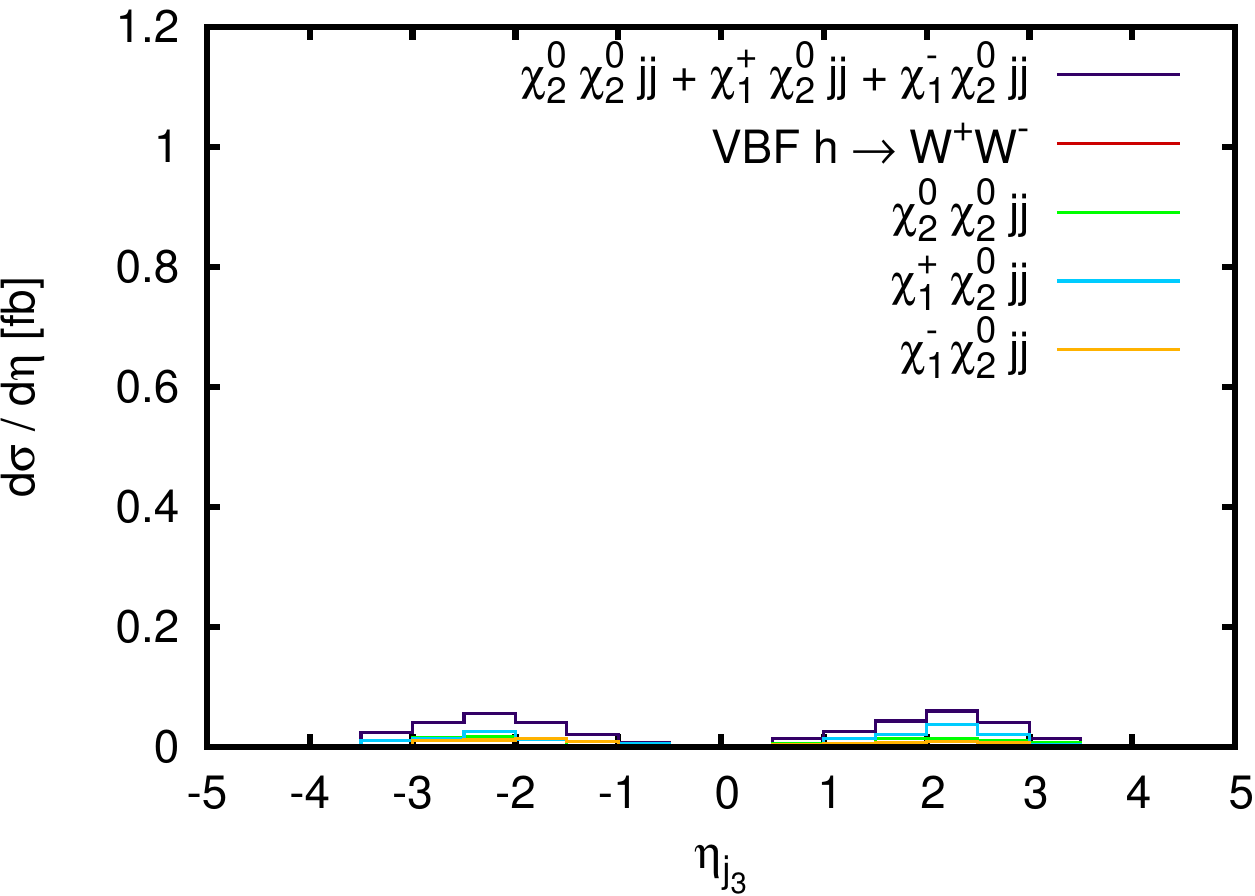}\\
    \vskip20pt
    \includegraphics[height=0.30\textwidth]{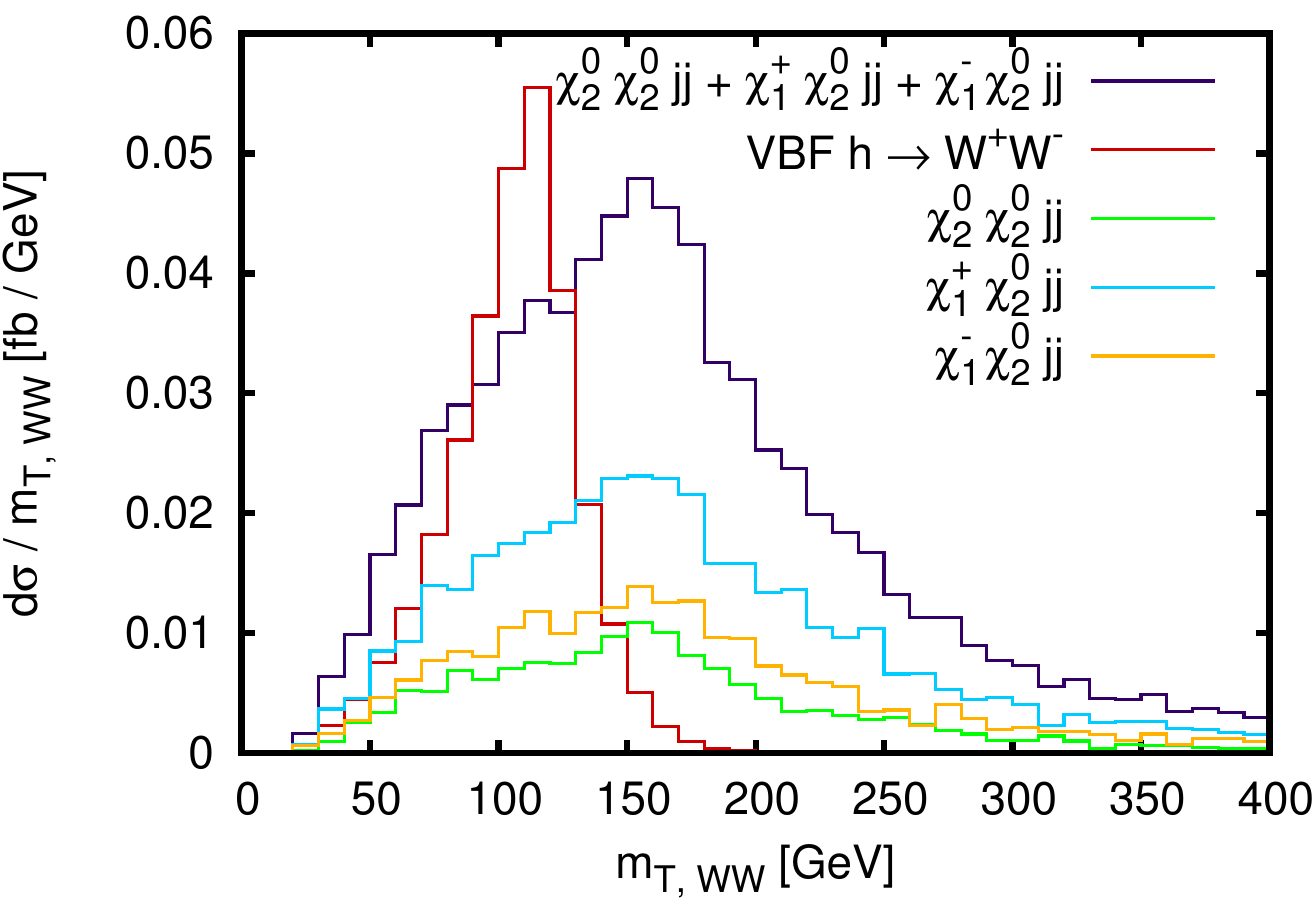}
    \hskip20pt
    \includegraphics[height=0.30\textwidth]{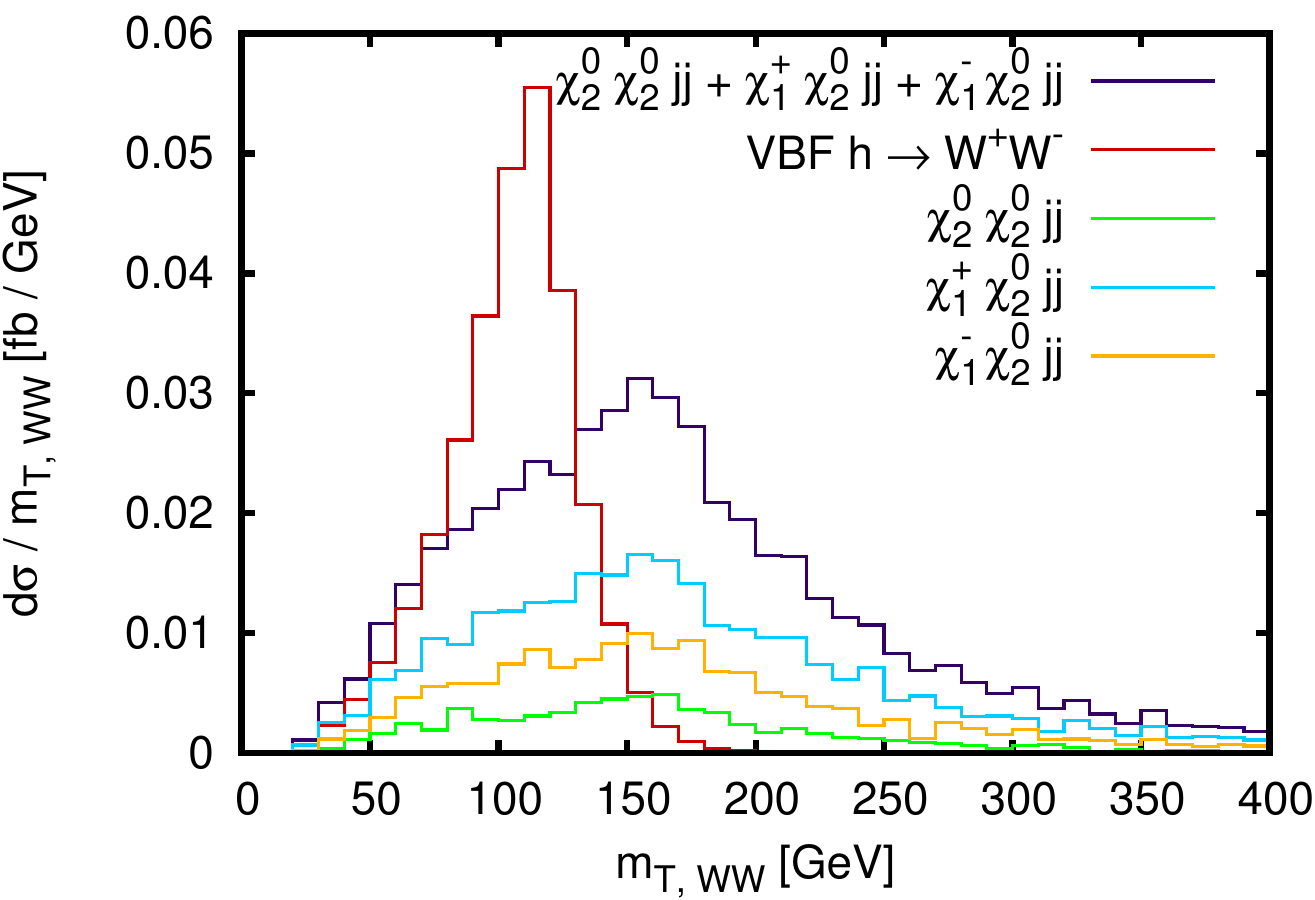}
  \end{center}
  \caption{Transverse momentum (top row) and rapidity (middle row) of the third jet occuring in
           $\chi_1^\pm\,\chi_2^0\,jj$ and $\chi_2^0\,\chi_2^0\,jj$. The bottom row shows 
           the transverse $WW$ mass distribution of these processes.
           Cuts for the left column are from Eqs.~\eqref{cuts_W_min} - \eqref{cuts_W_deltaeta}; 
           the plots in the right column also include the veto from Eq.~\eqref{cuts_jetveto}.
           Distributions are shown for the scenario SPS1amod.}
  \label{reducible_plots_W}
\end{figure}

In Table~\ref{xs_W_table_reducible} the contributions of the different channels to the SUSY background are 
shown at different cut levels. The cuts of Eqs.~\eqref{cuts_W_min} -- \eqref{btag} have been applied before 
to the irreducible background channels and, of course, have to be applied also to the reducible ones. As 
before, the typical VBF type cuts of Eq.~\eqref{cuts_W_deltaeta} including the cut on the rapidity separation 
of the jets reduces the background considerably. The b-quark contributions to the $\chi_1^\pm \chi_2^0 jj$ 
channel are tiny, less than 1 \% of the contributions without b quarks. As this is within the error 
originating from the Monte Carlo generation of the events these contributions
are neglected. 
 
Additionally to the cuts discussed before, a veto on additional visible jets as defined in 
Eq.~\eqref{visiblejet} is applied,
\begin{equation}
 \eta_{j_{tag},min} < \eta_{j_{decay}} < \eta_{j_{tag},max} \quad \text{with}
 \quad p_{T,j_{decay}} > 20~\text{GeV}\,, \quad |\eta_{j_{decay}}| < 4.5~.
 \label{cuts_jetveto}
\end{equation}
It should be noted that due to the event selection as described in Sect.~\ref{particleandeventselection} this
cut is only efficient on jets originating from the decay of a heavy
particle. The central jet veto as described 
 in Eq.~\eqref{vetocuts}, however, was used to reduce background events with
 extra jets from QCD radiation. It is clear that in an experimental analysis
 this  
discrimination between jets from QCD radiation and from decays of heavy
particles  cannot be performed. This 
means that the central jet veto, Eq.~\eqref{cuts_jetveto}, would also reduce the events with QCD radiation that 
we have neglected by our event selection from the beginning. In
Fig.~\ref{reducible_plots_W} the effect of the central jet veto,  
Eq.~\eqref{cuts_jetveto}, is shown. On the left hand side, distributions with
only basic and typical VBF cuts, 
Eq~\eqref{cuts_W_min}  and \eqref{cuts_W_deltaeta}, are depicted while the
ones presented on the right hand  
side include the central jet veto cut, Eq.~\eqref{cuts_jetveto}. The $p_T$ and
$\eta$ distribution   
of the  third hardest visible jet show that the number of events with a third
jet is efficiently reduced. In the  
last row, the distribution of the transverse WW mass is presented. The
reducible SUSY background is  
efficiently reduced by the central jet veto, Eq.~\eqref{cuts_jetveto},
especially in the case of the   
$\chi_2^0\chi_2^0 j j$ channel, see also
Table~\ref{xs_W_table_reducible}. However, this cut is not as efficient as it
might appear by just considering   
the  $p_T$ and the $\eta$ distribution of the third jet which is due to the
events with no visible additional jets. Even after the central jet veto cut,
the contribution of these channels to the SUSY background is larger than the
one of the irreducible channels.

\subsection{\texorpdfstring{Contributions to the {\boldmath$h \rightarrow \tau^+\tau^-$} Channel}
{Contributions to the tau+ tau- Channel}}

\tabcolsep3mm
\begin{table}[b]
  \begin{center}
  \begin{tabular}{|l||c|c|c|c||c||c|}
  \hline
       &  &  & \multicolumn{2}{|c||}{$\chi_2^0\chi_2^0jj$} & & VBF \\
  Cuts & $\chi_1^+\chi_2^0jj$ & $\chi_1^-\chi_2^0jj$ & no b & b-contr. & $\sum_{red}^{SUSY}$ & $h\rightarrow \tau\tau$ \\
  \hline
  \hline
  Eq.~(\ref{cuts_tau_min})
                                & $163.8 \;\text{fb}$  & $104.4 \;\text{fb}$  & $ 76.2 \;\text{fb}$  & $ 9.22 \;\text{fb}$ & $353.6 \;\text{fb}$   & $9.17 \;\text{fb}$ \\
  + Eq.~(\ref{cuts_tau_deltaeta})
                                & $ 5.16 \;\text{fb}$  & $ 2.81 \;\text{fb}$  & $ 1.76 \;\text{fb}$  & $0.228 \;\text{fb}$ & $9.96 \;\text{fb}$   & $4.94 \;\text{fb}$ \\
  + Eq.~(\ref{cuts_tau_atlas})  & $ 3.60 \;\text{fb}$  & $ 1.90 \;\text{fb}$  & $ 1.44 \;\text{fb}$  & $0.080 \;\text{fb}$ & $7.02 \;\text{fb}$   & $2.67 \;\text{fb}$ \\
  + Eq.~(\ref{cuts_tau_massrec})& $0.086 \;\text{fb}$  & $0.047 \;\text{fb}$  & $0.037 \;\text{fb}$  & $0.002 \;\text{fb}$ & $0.172 \;\text{fb}$   & $2.46 \;\text{fb}$ \\
  + Eq.~(\ref{fakeptmiss})
                                & $0.089 \;\text{fb}$  & $0.046 \;\text{fb}$  & $0.036 \;\text{fb}$  & $0.002 \;\text{fb}$ & $0.173 \;\text{fb}$   & $1.93 \;\text{fb}$ \\
  + Eq.~(\ref{cuts_jetveto})
                                & $0.060 \;\text{fb}$  & $0.033 \;\text{fb}$  & $0.016 \;\text{fb}$  & $0.001 \;\text{fb}$ & $0.110 \;\text{fb}$   & $1.93 \;\text{fb}$ \\
  \hline
  \end{tabular}
  \end{center}
  \caption{Total cross sections for $\chi_1^\pm\,\chi_2^0jj$, $\chi_2^0\chi_2^0jj$
           and VBF $h\rightarrow \tau\tau$ at different cut levels for the scenario SPS1amod.}
  \label{xs_tau_table_reducible}
\end{table}

Considering the same background channels as in the subsection before, we now
turn to the case of VBF Higgs boson production with a subsequent decay of the
Higgs boson into $\tau$~leptons. The contributions of the channels to the
SUSY background are listed in Table~\ref{xs_tau_table_reducible} for different
cut levels. The first four cut levels, Eqs.~\eqref{cuts_tau_min} --
\eqref{cuts_tau_massrec}, have also been applied in the analysis of the
irreducible background channels. As before, using the tau mass reconstruction,
Eq.~\eqref{cuts_tau_massrec}, reduces the SUSY background very efficiently,
leading to a much cleaner signal than in the case of the signal channel with a
subsequent decay into W bosons. Taking into account the missing $p_T$ resolution,
Eq.~\eqref{fakeptmiss}, mainly affects the number of signal events as it leads
to a broadening of the tau pair mass peak. Employing the central jet veto cut,
Eq.~\eqref{cuts_jetveto}, yields a further reduction of the background,
in particular of the contribution of the $\chi_2^0 \chi_2^0 jj$
channel. Again, we have neglected the effect of the central jet veto
on additional QCD radiation jets, which would lead to a
further reduction of the cross section.

Overall, even after the central jet veto cut, the contribution of these
reducible background channels is larger than the one coming from the
irreducible ones. Due to the tau pair mass cut, however, the SUSY background
contributions are still much smaller than the signal cross section.

\newpage

\subsection{Summary of Reducible and Irreducible Background}
\label{sps1asummary}

This subsection summarizes the results of Sections \ref{irreducible} and \ref{reducible}, which
discussed the dominant SUSY background processes to the leptonic Higgs Boson channels
$h\rightarrow WW$ and $h\rightarrow \tau\tau$ in vector boson production.
Further sub-dominant processes are also listed and briefly discussed.

\subsubsection{\texorpdfstring{Contributions to the {\boldmath$h \rightarrow W^+W^-$} Channel}
{Contributions to the W+ W- Channel}}

\begin{figure}[p]
  \begin{center}
    \includegraphics[height=0.3\textwidth]{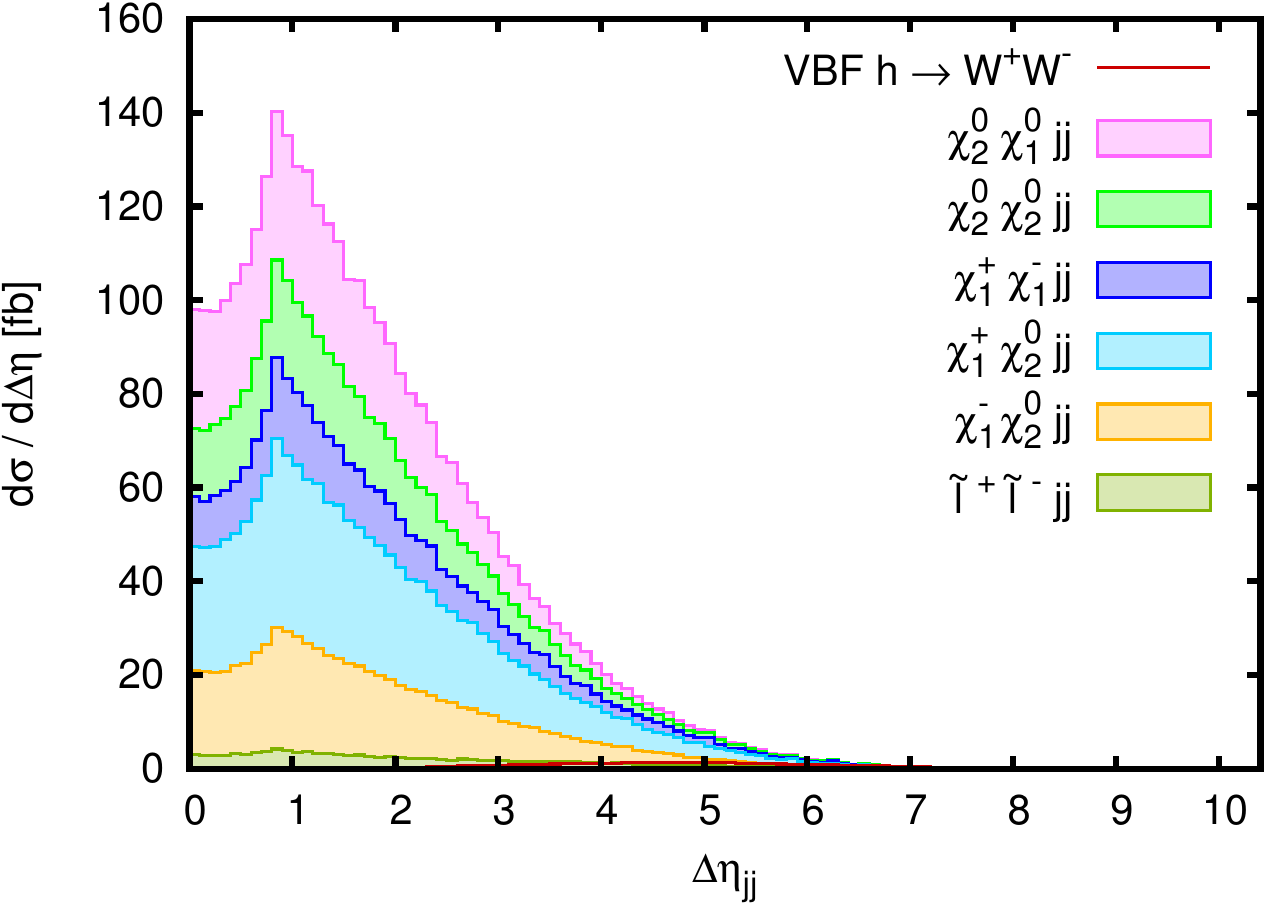}
    \hskip20pt
    \includegraphics[height=0.3\textwidth]{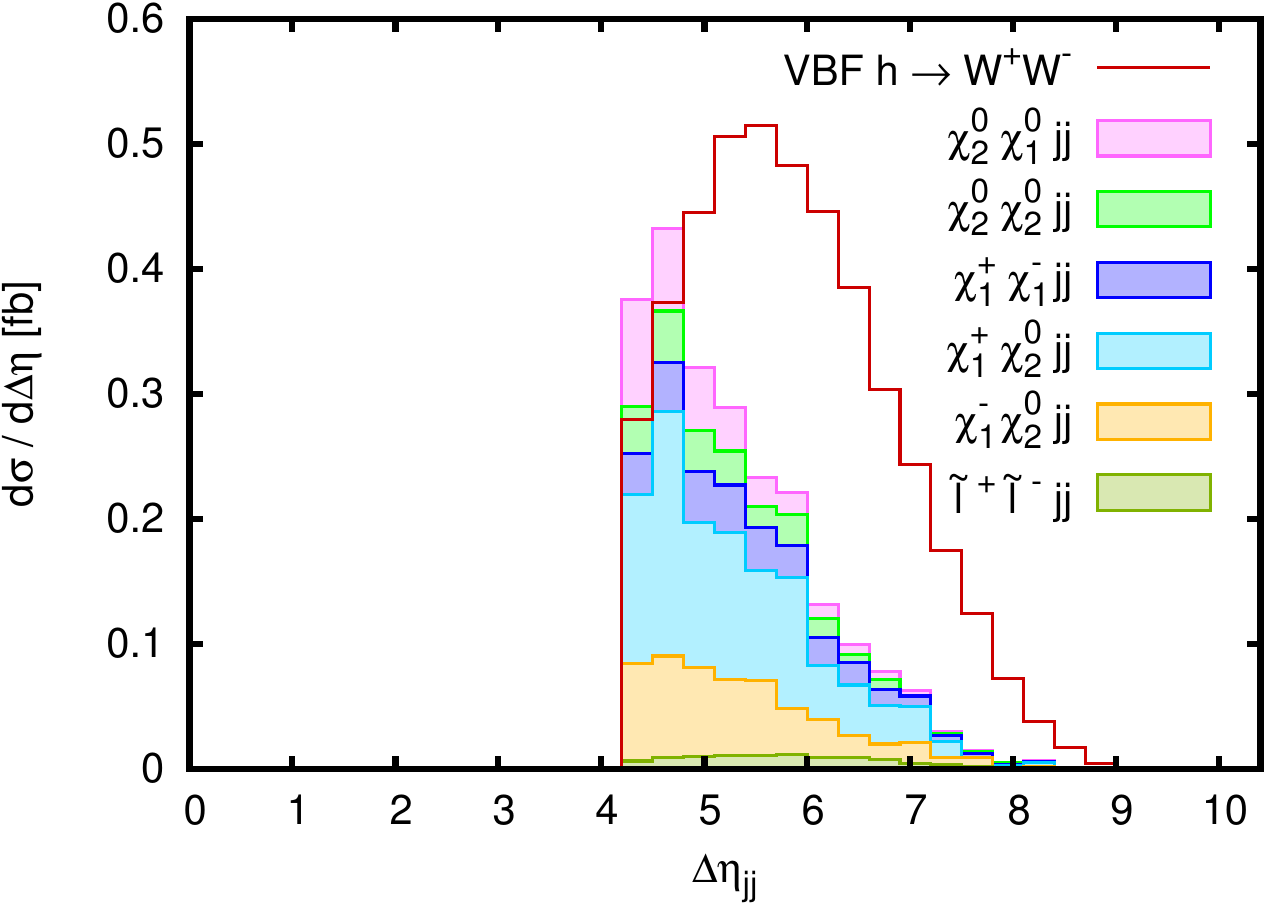}
  \end{center}
  \caption{Rapidity separation for SUSY backgrounds and $h\rightarrow WW$ signal in the scenario
           SPS1amod. The left plot is for minimal cuts from Eq,~\eqref{cuts_W_min}.
           The right plot includes all discussed cuts from Eqs.~\eqref{cuts_W_min} - \eqref{cuts_W_nomass},
           \eqref{cuts_W_paper} - \eqref{cuts_W_paper_ptmiss}, \eqref{btag}, \eqref{cuts_jetveto}. } 
  \label{plots_summary}
\end{figure}

\tabcolsep2.3mm
\begin{table}[p]
  \begin{center}
  \begin{tabular}{llcccc}
  \toprule
  & Cuts & basic & + rap. gap & + $m_{inv}$, $\slashed{p}_T$, $\phi_{\ell\ell}$, $\textrm{b-tag}$ & + CJV on $j_{decay}$ \\
  Processes && Eq.~(\ref{cuts_W_min}) & + (\ref{cuts_W_deltaeta}) & + \eqref{cuts_W_nomass}, \eqref{cuts_W_paper}, \eqref{cuts_W_paper_ptmiss}, \eqref{btag} & + \eqref{cuts_jetveto} \\
  \midrule
  \midrule
  $\chi_1^+ \, \chi_1^- \, jj$ & no b                   & $25.97 \;\text{fb}$ & $1.21 \;\text{fb}$ & $0.073 \;\text{fb}$ & $0.073 \;\text{fb}$ \\
  $\chi_1^+ \, \chi_1^- \, jj$ & b-contr.               & $14.50 \;\text{fb}$ & $0.31 \;\text{fb}$ & $0.012 \;\text{fb}$ & $0.012 \;\text{fb}$ \\
  $\chi_2^0 \, \chi_1^0 \, jj$ & no b                   & $66.79 \;\text{fb}$ & $1.04 \;\text{fb}$ & $0.080 \;\text{fb}$ & $0.080 \;\text{fb}$ \\
  $\chi_2^0 \, \chi_1^0 \, jj$ & b-contr.               & $ 7.65 \;\text{fb}$ & $0.29 \;\text{fb}$ & $0.014 \;\text{fb}$ & $0.014 \;\text{fb}$ \\
  $\widetilde{\ell}^+\, \widetilde{\ell}^-\, jj$ & no b & $11.55 \;\text{fb}$ & $1.23 \;\text{fb}$ & $0.028 \;\text{fb}$ & $0.028 \;\text{fb}$ \\
  \midrule
  $\chi_1^+ \, \chi_2^0 \, jj$ & no b                   & $100.8 \;\text{fb}$ & $3.94 \;\text{fb}$ & $0.403 \;\text{fb}$ & $0.275 \;\text{fb}$ \\
  $\chi_1^- \, \chi_2^0 \, jj$ & no b                   & $63.22 \;\text{fb}$ & $2.20 \;\text{fb}$ & $0.222 \;\text{fb}$ & $0.144 \;\text{fb}$ \\
  $\chi_2^0 \, \chi_2^0 \, jj$ & no b                   & $46.40 \;\text{fb}$ & $1.35 \;\text{fb}$ & $0.149 \;\text{fb}$ & $0.059 \;\text{fb}$ \\
  $\chi_2^0 \, \chi_2^0 \, jj$ & b-contr.               & $ 5.02 \;\text{fb}$ & $0.168 \;\text{fb}$ & $0.010 \;\text{fb}$ & $0.006 \;\text{fb}$ \\
  \midrule
  \multicolumn{2}{l}{$\sum B^{SUSY}$}                   & $341.9 \;\text{fb}$ & $11.74 \;\text{fb}$ & $0.991 \;\text{fb}$ & $0.700 \;\text{fb}$ \\
  \midrule
  \multicolumn{2}{l}{VBF $h\rightarrow WW$}             & $5.09 \;\text{fb}$ & $2.91 \;\text{fb}$ & $1.32 \;\text{fb}$ & $1.32 \;\text{fb}$ \\
  \midrule
  \multicolumn{2}{l}{$S / B^{SUSY}$}                    &      0.015      &      0.25      &      1.3     &            1.9  \\
  \bottomrule
  \end{tabular}
  \end{center}
  \caption{Total cross sections of SUSY backgrounds dominant at low squark masses 
           to the VBF $h\rightarrow WW$ channel for the scenario SPS1amod.}
  \label{xs_W_table_summary}
\end{table}

\tabcolsep2.3mm
\begin{table}[p]
  \begin{center}
  \begin{tabular}{llcccc}
  \toprule
  & Cuts & basic & + rap. gap & + $m_{inv},\, \slashed{p}_T,\, \phi_{\ell\ell},\, \textrm{b-tag}$ & + CJV on $j_{decay}$ \\
  Processes && Eq.~(\ref{cuts_W_min}) & + (\ref{cuts_W_deltaeta}) & + \eqref{cuts_W_nomass}, \eqref{cuts_W_paper}, \eqref{cuts_W_paper_ptmiss}, \eqref{btag} & + \eqref{cuts_jetveto} \\
  \midrule
  \midrule
  $\chi_1^+ \, \chi_2^0 \, j$       & no b     & $15.96 \;\text{fb}$ & $0.230 \;\text{fb}$ & $0.008 \;\text{fb}$ & $0.007 \;\text{fb}$ \\
  $\chi_1^- \, \chi_2^0 \, j$       & no b     & $8.93 \;\text{fb}$ & $0.115 \;\text{fb}$ & $0.004 \;\text{fb}$ & $0.003 \;\text{fb}$ \\
  $\chi_2^0 \, \chi_2^0 \, j$       & no b     & $7.76 \;\text{fb}$ & $0.052 \;\text{fb}$ & $0.003 \;\text{fb}$ & $0.002 \;\text{fb}$ \\
  $\chi_2^0 \, \chi_2^0 \, j$       & b-contr. & $0.95 \;\text{fb}$ & $0.010 \;\text{fb}$ & $\ll 0.001 \;\text{fb}$ & $\ll 0.001 \;\text{fb}$ \\
  $\chi_2^0 \, \chi_2^0$            & no b     & $0.25 \;\text{fb}$ & $\ll 0.001 \;\text{fb}$ & $\ll 0.001 \;\text{fb}$ & $\ll 0.001 \;\text{fb}$ \\
  $\chi_2^0 \, \chi_2^0$            & b-contr. & $0.15 \;\text{fb}$ & $\ll 0.001 \;\text{fb}$ & $\ll 0.001 \;\text{fb}$ & $\ll 0.001 \;\text{fb}$ \\
  \midrule
  $\widetilde{g} \, \chi_1^0 \, jj$ & no b     & $54.33 \;\text{fb}$ & $3.15 \;\text{fb}$ & $0.117 \;\text{fb}$ & $< 0.01 \;\text{fb}$ \\
  $\widetilde{g} \, \chi_1^0 \, j$  & no b     & $64.98 \;\text{fb}$ & $0.54 \;\text{fb}$ & $0.046 \;\text{fb}$ & $0.007 \;\text{fb}$ \\
  $\widetilde{g} \, \chi_1^0     $  & no b     & $0.32 \;\text{fb}$ & $\ll 0.001 \;\text{fb}$ & $\ll 0.001 \;\text{fb}$ & $\ll 0.001 \;\text{fb}$ \\
  \midrule
  $\chi_1^+ \, \chi_2^- \, jj$      & no b     & $2.03 \;\text{fb}$ & $0.074 \;\text{fb}$ & $0.003 \;\text{fb}$ & $0.002 \;\text{fb}$ \\
  $\chi_1^+ \, \chi_2^- \, jj$      & b-contr. & $0.67 \;\text{fb}$ & $0.028 \;\text{fb}$ & $< 0.001 \;\text{fb}$ & $< 0.001 \;\text{fb}$ \\
  $\chi_1^- \, \chi_2^+ \, jj$      & no b     & $2.06 \;\text{fb}$ & $0.095 \;\text{fb}$ & $0.005 \;\text{fb}$ & $0.003 \;\text{fb}$ \\
  $\chi_1^- \, \chi_2^+ \, jj$      & b-contr. & $0.69 \;\text{fb}$ & $0.029 \;\text{fb}$ & $< 0.001 \;\text{fb}$ & $< 0.001 \;\text{fb}$ \\
  $\chi_2^+ \, \chi_2^- \, jj$      & no b     & $0.27 \;\text{fb}$ & $0.021 \;\text{fb}$ & $< 0.001 \;\text{fb}$ & $< 0.001 \;\text{fb}$ \\
  $\chi_2^+ \, \chi_2^- \, jj$      & b-contr. & $0.11 \;\text{fb}$ & $0.002 \;\text{fb}$ & $\ll 0.001 \;\text{fb}$ & $\ll 0.001 \;\text{fb}$ \\
  $\chi_3^0 \, \chi_1^0 \, jj$      & no b     & $0.33 \;\text{fb}$ & $0.009 \;\text{fb}$ & $< 0.001 \;\text{fb}$ & $< 0.001 \;\text{fb}$ \\
  $\chi_3^0 \, \chi_1^0 \, jj$      & b-contr. & $0.14 \;\text{fb}$ & $0.002 \;\text{fb}$ & $\ll 0.001 \;\text{fb}$ & $\ll 0.001 \;\text{fb}$ \\
  $\chi_4^0 \, \chi_1^0 \, jj$      & no b     & $1.76 \;\text{fb}$ & $0.041 \;\text{fb}$ & $< 0.001 \;\text{fb}$ & $< 0.001 \;\text{fb}$ \\
  $\chi_4^0 \, \chi_1^0 \, jj$      & b-contr. & $0.24 \;\text{fb}$ & $0.004 \;\text{fb}$ & $\ll 0.001 \;\text{fb}$ & $\ll 0.001 \;\text{fb}$ \\
  $\chi_4^0 \, \chi_2^0 \, jj$      & no b     & $2.40 \;\text{fb}$ & $0.066 \;\text{fb}$ & $0.005 \;\text{fb}$ & $0.003 \;\text{fb}$ \\
  $\chi_4^0 \, \chi_2^0 \, jj$      & b-contr. & $0.34 \;\text{fb}$ & $0.003 \;\text{fb}$ & $\ll 0.001 \;\text{fb}$ & $\ll 0.001 \;\text{fb}$ \\
  $\chi_2^+ \, \chi_2^0 \, jj$      & no b     & $1.97 \;\text{fb}$ & $0.061 \;\text{fb}$ & $0.005 \;\text{fb}$ & $0.002 \;\text{fb}$ \\
  $\chi_2^+ \, \chi_2^0 \, jj$      & b-contr. & $0.06 \;\text{fb}$ & $< 0.001 \;\text{fb}$ & $\ll 0.001 \;\text{fb}$ & $\ll 0.001 \;\text{fb}$ \\
  $\chi_2^- \, \chi_2^0 \, jj$      & no b     & $1.83 \;\text{fb}$ & $0.041 \;\text{fb}$ & $0.004 \;\text{fb}$ & $0.002 \;\text{fb}$ \\
  $\chi_2^- \, \chi_2^0 \, jj$      & b-contr. & $0.04 \;\text{fb}$ & $< 0.001 \;\text{fb}$ & $\ll 0.001 \;\text{fb}$ & $\ll 0.001 \;\text{fb}$ \\
  $\chi_2^+ \, \chi_1^0 \, jj$      & no b     & $0.68 \;\text{fb}$ & $0.013 \;\text{fb}$ & $< 0.001 \;\text{fb}$ & $< 0.001 \;\text{fb}$ \\
  $\chi_2^+ \, \chi_1^0 \, jj$      & b-contr. & $0.01 \;\text{fb}$ & $\ll 0.001 \;\text{fb}$ & $\ll 0.001 \;\text{fb}$ & $\ll 0.001 \;\text{fb}$ \\
  \midrule
  \multicolumn{2}{l}{$\sum$ further processes} & $169.3 \;\text{fb}$ & $4.59 \;\text{fb}$ & $0.207 \;\text{fb}$ & $< 0.048 \;\text{fb}$ \\
  \bottomrule
  \end{tabular}
  \end{center}
  \caption{Total cross sections of further SUSY backgrounds to
           the VBF $h\rightarrow WW$ channel for the scenario SPS1amod.}
  \label{xs_W_table_other}
\end{table}

We investigated several SUSY particle production processes with subsequent decay of the SUSY particles
in the scenario SPS1amod which can give rise to the same signature as the one of the $h\rightarrow WW$ channel
in vector boson fusion. The complete list of processes is shown in Tables~\ref{xs_W_table_summary}
and~\ref{xs_W_table_other}.

The processes from Table~\ref{xs_W_table_summary} give the largest contributions to the background and we 
discussed them in detail in the last sections. Figure~\ref{plots_summary} once again shows 
the distribution of the rapidity separation  between
the two hardest jets for the SUSY processes and the signal, once for only minimal cuts and once including
the full set of cuts used in our analysis.
With basic cuts, the SUSY background $B^{SUSY}$ is larger by a factor of almost 70 compared to the signal 
process $S$.
With all cuts except for the central jet veto on additional jets from SUSY particle decays,
this factor is reduced and yields a signal of background ratio of
\begin{equation}
 S / B^{SUSY} = 1.3 \,,
\end{equation}
where $B^{SUSY}$ includes all processes listed in Table~\ref{xs_W_table_summary} from now on.
Vetoing the additional jets from SUSY particle decays this ratio increases to
\begin{equation}
 S / B^{SUSY} = 1.9 \,.
\end{equation}
In the experiment, a jet veto would also affect additional jets due to QCD radiation.
We estimated this effect for the processes $\chi_1^+ \, \chi_1^- \, jj$ and $\chi_2^0 \, \chi_1^0 \, jj$
within the exponentiation model and found that the reduction for the SUSY processes
is much larger than for the vector boson fusion process.
Therefore we expect the signal to background ratio to improve further.

The processes with small contributions of Table~\ref{xs_W_table_other} can be split into
three classes:
\begin{itemize}
 \item Processes of Table~\ref{xs_W_table_summary}, but with fewer light jets at matrix
       element level
 \item Processes involving gluinos
 \item Processes with two jets at matrix element level and other chargino and neutralino combinations
\end{itemize}

In the first process class one (or two for $\chi_2^0\chi_2^0$) of the tagging jets has to be
created by a hadronic tau lepton decay, where the tau lepton comes from the decay of one
of the charginos or neutralinos. These "one-jet" contributions are much smaller than the "two jet" contributions,
because no resonant squark pair production is possible which produces fairly hard jets,
while the tau jets are rather soft.
Furthermore the rapidity separation between the hardest jets after the decay of the SUSY particles
is smaller, increasing the efficiency of the rapidity gap cut of Eq.~\eqref{cuts_W_deltaeta}.

The production cross sections with minimal cuts are fairly large for the second process class,
but the analysis cuts eliminate their contribution efficiently:
For $\widetilde{g} \, \chi_1^0$ the two jets of the gluino decay are very close in rapidity
and no event survives the rapidity gap cut.
For $\widetilde{g} \, \chi_1^0 \,j$ and $\widetilde{g} \, \chi_1^0 \,jj$ the tail of the
tagging jet rapidity separation is still smaller than for the electroweak gaugino pair plus two jets processes.
Another characteristic is the occurrence of additional hard jets from the gluino decay which
can be well detected and vetoed.

The third class of processes involves mostly higher mass charginos and neutralinos with smaller production
cross sections. Furthermore, due to their branching ratios a
signature as the one of $h\rightarrow WW$ in vector boson fusion is not favoured.

\subsubsection{\texorpdfstring{Contributions to the {\boldmath$h \rightarrow \tau^+\tau^-$} Channel}
{Contributions to the tau+ tau- Channel}}

\tabcolsep2.1mm
\begin{table}[b]
  \begin{center}
  \begin{tabular}{llcccc}
  \toprule
  & Cuts & basic & + rap. gap & + $m_{jj}$, $\slashed{p}_T$, $m_{\tau\tau}$, $\slashed{E}_{T,fake}$ & + CJV on $j_{decay}$ \\
  Processes && Eq.~(\ref{cuts_tau_min}) & + (\ref{cuts_tau_deltaeta}) & + \eqref{cuts_tau_atlas}, \eqref{cuts_tau_massrec}, \eqref{fakeptmiss} & + \eqref{cuts_jetveto} \\
  \midrule
  \midrule
  $\chi_1^+ \, \chi_1^- \, jj$ & no b                   & $64.13 \;\text{fb}$ & $2.04 \;\text{fb}$ & $0.025 \;\text{fb}$ & $0.025 \;\text{fb}$ \\
  $\chi_1^+ \, \chi_1^- \, jj$ & b-contr.               & $52.53 \;\text{fb}$ & $0.47 \;\text{fb}$ & $0.003 \;\text{fb}$ & $0.003 \;\text{fb}$ \\
  $\chi_2^0 \, \chi_1^0 \, jj$ & no b                   & $109.2 \;\text{fb}$ & $1.09 \;\text{fb}$ & $0.015 \;\text{fb}$ & $0.015 \;\text{fb}$ \\
  $\chi_2^0 \, \chi_1^0 \, jj$ & b-contr.               & $12.88 \;\text{fb}$ & $0.37 \;\text{fb}$ & $0.003 \;\text{fb}$ & $0.003 \;\text{fb}$ \\
  $\widetilde{\ell}^+\, \widetilde{\ell}^-\, jj$ & no b & $26.35 \;\text{fb}$ & $1.89 \;\text{fb}$ & $0.003 \;\text{fb}$ & $0.003 \;\text{fb}$ \\
  \midrule
  $\chi_1^+ \, \chi_2^0 \, jj$ & no b                   & $163.8 \;\text{fb}$ & $5.16 \;\text{fb}$ & $0.089 \;\text{fb}$ & $0.060 \;\text{fb}$ \\
  $\chi_1^- \, \chi_2^0 \, jj$ & no b                   & $104.4 \;\text{fb}$ & $2.81 \;\text{fb}$ & $0.046 \;\text{fb}$ & $0.033 \;\text{fb}$ \\
  $\chi_2^0 \, \chi_2^0 \, jj$ & no b                   & $76.15 \;\text{fb}$ & $1.76 \;\text{fb}$ & $0.036 \;\text{fb}$ & $0.016 \;\text{fb}$ \\
  $\chi_2^0 \, \chi_2^0 \, jj$ & b-contr.               & $ 9.22 \;\text{fb}$ & $0.23 \;\text{fb}$ & $0.002 \;\text{fb}$ & $0.001 \;\text{fb}$ \\
  \midrule
  \multicolumn{2}{l}{$\sum B^{SUSY}$}                   & $618.7 \;\text{fb}$ & $15.82 \;\text{fb}$ & $0.222 \;\text{fb}$ & $0.159 \;\text{fb}$ \\
  \midrule
  \multicolumn{2}{l}{VBF $h\rightarrow \tau\tau$}       & $9.17 \;\text{fb}$ & $4.94 \;\text{fb}$ & $1.93 \;\text{fb}$ & $1.93 \;\text{fb}$ \\
  \midrule
  \multicolumn{2}{l}{$S / B^{SUSY}$}                    &         0.015  &      0.31      &     8.7      &            12 \\
  \midrule
  \midrule
  \multicolumn{2}{l}{$\sum$ further processes}          &            &             &                &               \\
  \multicolumn{2}{l}{(see Table \ref{xs_W_table_other})}& $374.9 \;\text{fb}$ & $6.41 \;\text{fb}$ & $0.115 \;\text{fb}$ & $< 0.016 \;\text{fb}$ \\
  \bottomrule
  \end{tabular}
  \end{center}
  \caption{Total cross sections of SUSY backgrounds dominant at low squark masses 
           to the VBF $h\rightarrow \tau\tau$ channel for the scenario SPS1amod.}
  \label{xs_tau_table_summary}
\end{table}

For the $h\rightarrow\tau\tau$ analysis the same processes as before give the largest contributions to the SUSY 
background, 
see Table~\ref{xs_tau_table_summary}. The contributions from the processes
listed in Table~\ref{xs_W_table_other} are again much smaller, the sum of their contribution
is listed in Table~\ref{xs_tau_table_summary} as well.

SUSY backgrounds are much less troublesome in this search channel, due to the possibility
to reconstruct the Higgs boson mass from the decay products. Therefore the overall
signal to background ratio with all cuts is much higher than for $h\rightarrow WW$:
It is
\begin{equation}
 S / B^{SUSY} = 8.7 \,
\end{equation}
without a veto on additional jets from the SUSY particle decay, including the veto leads to
\begin{equation}
 S / B^{SUSY} = 12 \,.
\end{equation}

\section{Sparticle Mass Dependence}
\label{massdep}

In this section we want to discuss the influence of larger sparticle masses
on the SUSY background processes without changing the characteristic of our
scenario which means that the branching ratios should only change slightly. 
In order to keep these characteristics, the gluino has to remain heavier than
all squarks~\footnote{If the gluino is lighter than the squarks with
  $m_{\tilde{g}} < 
  m_{\tilde{q}} - m_q$, the decay channel $\tilde q \rightarrow q \tilde g$ will
  open up 
  which, finally, does not lead to the considered final state.}.
As the gluino mass is only slightly larger than the squark masses, we
have to increase it by the same factor as the squark masses.

The b-quark contributions involve mostly third generation squarks, while
the other contributions have only first and second generation squark contributions.
Therefore we check the b-quark contributions separately.

We focus on the dominant processes identified at the end of the previous section.

\subsection{Squark + Gluino Mass Dependence}

\begin{figure}[b]
  \begin{center}
    \includegraphics[height=0.33\textwidth]{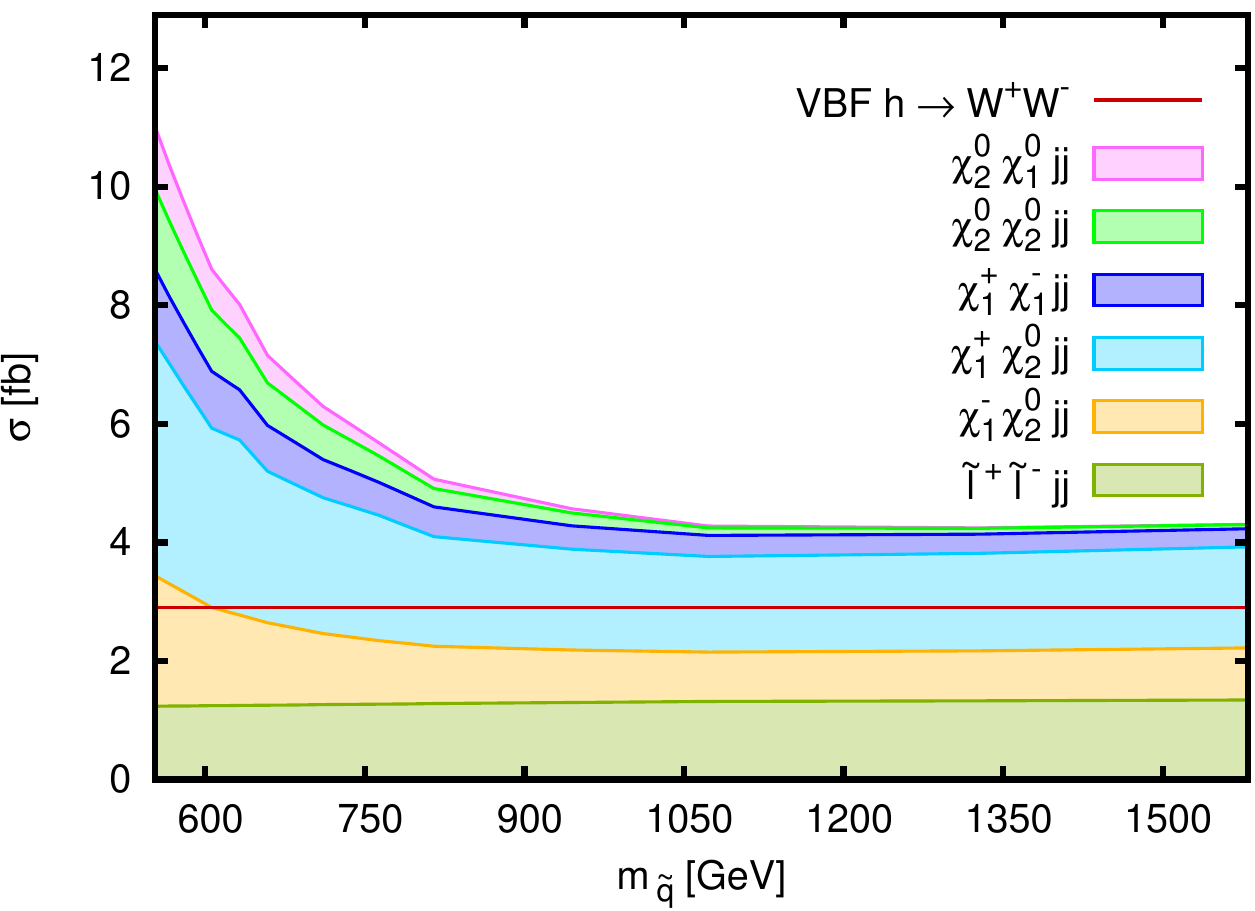}
    \hskip15pt
    \includegraphics[height=0.33\textwidth]{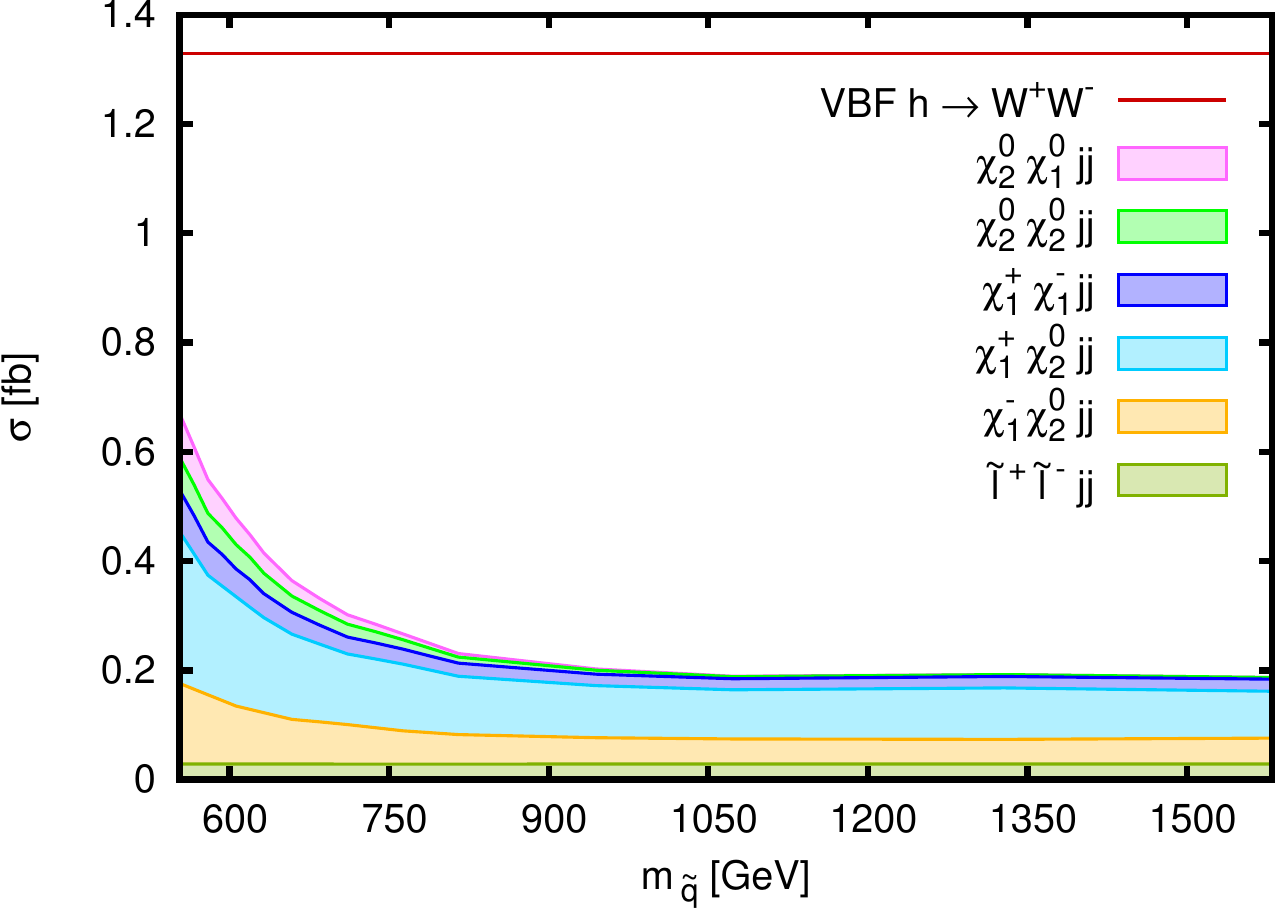}
  \end{center}
  \caption{Squark+gluino mass dependence for the SUSY-backgrounds as a function of the average squark
           mass without b-quark contributions. Left panel: with cuts from Eqs.~(\ref{cuts_W_min}) + (\ref{cuts_W_deltaeta}).
           Right panel: with additional cuts from Eqs.~(\ref{cuts_W_nomass}), (\ref{cuts_W_paper}), 
           (\ref{cuts_W_paper_ptmiss}) and (\ref{cuts_jetveto}).}
  \label{massdep_plot_squark}
\end{figure}

\tabcolsep2.5mm
\begin{table}[p]
  \begin{center}
  \begin{tabular}{llccc}
  \toprule
  & Cuts & basic + rap. gap & + $m_{inv} + \slashed{p}_T + \phi_{\ell\ell} + \textrm{b-tag}$ & + CJV on $j_{decay}$ \\
  Processes && Eqs.~(\ref{cuts_W_min}) + (\ref{cuts_W_deltaeta}) & + \eqref{cuts_W_nomass}, \eqref{cuts_W_paper}, \eqref{cuts_W_paper_ptmiss}, \eqref{btag} & + \eqref{cuts_jetveto} \\
  \midrule
  \midrule
  $\chi_1^+ \, \chi_1^- \, jj$ & no b                   & $0.353 \;\text{fb}$ & $0.020 \;\text{fb}$ & $0.020 \;\text{fb}$ \\
  $\chi_1^+ \, \chi_1^- \, jj$ & b-contr.               & $0.227 \;\text{fb}$ & $0.010 \;\text{fb}$ & $0.010 \;\text{fb}$ \\
  $\chi_2^0 \, \chi_1^0 \, jj$ & no b                   & $0.032 \;\text{fb}$ & $0.001 \;\text{fb}$ & $0.001 \;\text{fb}$ \\
  $\chi_2^0 \, \chi_1^0 \, jj$ & b-contr.               & $0.228 \;\text{fb}$ & $0.010 \;\text{fb}$ & $0.010 \;\text{fb}$ \\
  $\widetilde{\ell}^+\, \widetilde{\ell}^-\, jj$ & no b & $1.311 \;\text{fb}$ & $0.028 \;\text{fb}$ & $0.028 \;\text{fb}$ \\
  \midrule
  $\chi_1^+ \, \chi_2^0 \, jj$ & no b                   & $1.611 \;\text{fb}$ & $0.122 \;\text{fb}$ & $0.090 \;\text{fb}$ \\
  $\chi_1^- \, \chi_2^0 \, jj$ & no b                   & $0.836 \;\text{fb}$ & $0.061 \;\text{fb}$ & $0.046 \;\text{fb}$ \\
  $\chi_2^0 \, \chi_2^0 \, jj$ & no b                   & $0.128 \;\text{fb}$ & $0.011 \;\text{fb}$ & $0.004 \;\text{fb}$ \\
  $\chi_2^0 \, \chi_2^0 \, jj$ & b-contr.               & $0.137 \;\text{fb}$ & $0.008 \;\text{fb}$ & $0.004 \;\text{fb}$ \\
  \midrule
  \multicolumn{2}{l}{$\sum B^{SUSY}$}                   & $4.86 \;\text{fb}$ & $0.271 \;\text{fb}$ & $0.213 \;\text{fb}$ \\
  \midrule
  \multicolumn{2}{l}{VBF $h\rightarrow WW$}             & $2.91 \;\text{fb}$ & $1.32 \;\text{fb}$ & $1.32 \;\text{fb}$ \\
  \midrule
  \multicolumn{2}{l}{$S / B^{SUSY}$}                    &     0.60       &     4.9       &            6.2  \\
  \bottomrule
  \end{tabular}
  \end{center}
  \caption{Total cross sections of SUSY backgrounds dominant at low squark masses to
           the VBF $h\rightarrow WW$ channel for average squark masses of 1.1 TeV.}
  \label{tab_msq1tev}
\end{table}

\tabcolsep2.5mm
\begin{table}[p]
  \begin{center}
  \begin{tabular}{llccc}
  \toprule
  & Cuts & basic + rap. gap & + $m_{inv} + \slashed{p}_T + \phi_{\ell\ell} + \textrm{b-tag}$ & + CJV on $j_{decay}$ \\
  Processes && Eqs.~(\ref{cuts_W_min}) + (\ref{cuts_W_deltaeta}) & + \eqref{cuts_W_nomass}, \eqref{cuts_W_paper}, \eqref{cuts_W_paper_ptmiss}, \eqref{btag} & + \eqref{cuts_jetveto} \\
  \midrule
  \midrule
  $\chi_1^+ \, \chi_2^0 \, j$       & no b & $0.350 \;\text{fb}$ & $0.013 \;\text{fb}$ & $0.012 \;\text{fb}$ \\
  $\chi_1^- \, \chi_2^0 \, j$       & no b & $0.168 \;\text{fb}$ & $0.004 \;\text{fb}$ & $0.004 \;\text{fb}$ \\
  $\chi_2^0 \, \chi_2^0 \, j$       & no b & $0.012 \;\text{fb}$ & $0.001 \;\text{fb}$ & $< 0.001 \;\text{fb}$ \\
  $\widetilde{g} \, \chi_1^0 \, jj$ & no b & $0.038 \;\text{fb}$ & $0.001 \;\text{fb}$ & $< 0.001 \;\text{fb}$ \\
  $\widetilde{g} \, \chi_1^0 \, j$  & no b & $0.008 \;\text{fb}$ & $\ll 0.001 \;\text{fb}$ & $\ll 0.001 \;\text{fb}$ \\
  \midrule
  $\chi_1^+ \, \chi_2^- \, jj$      & no b & $0.016 \;\text{fb}$ & $<0.001 \;\text{fb}$ & $< 0.001 \;\text{fb}$ \\
  $\chi_1^- \, \chi_2^+ \, jj$      & no b & $0.015 \;\text{fb}$ & $<0.001 \;\text{fb}$ & $< 0.001 \;\text{fb}$ \\
  $\chi_4^0 \, \chi_2^0 \, jj$      & no b & $0.013 \;\text{fb}$ & $<0.001 \;\text{fb}$ & $< 0.001 \;\text{fb}$ \\
  $\chi_2^+ \, \chi_2^0 \, jj$      & no b & $0.012 \;\text{fb}$ & $<0.001 \;\text{fb}$ & $\ll 0.001 \;\text{fb}$ \\
  $\chi_2^- \, \chi_2^0 \, jj$      & no b & $0.007 \;\text{fb}$ & $<0.001 \;\text{fb}$ & $\ll 0.001 \;\text{fb}$ \\
  \midrule
  \multicolumn{2}{l}{$\sum$ further processes} & $0.639 \;\text{fb}$ & $<0.024 \;\text{fb}$ & $< 0.021 \;\text{fb}$ \\
  \bottomrule
  \end{tabular}
  \end{center}
  \caption{Total cross sections of further SUSY backgrounds to
           the VBF $h\rightarrow WW$ channel for average squark masses of 1.1 TeV.}
  \label{tab_msq1tev_other}
\end{table}

As described in Section~\ref{scenarios}, we vary the first and second
generation squark masses and the gluino mass by scaling the low energy soft SUSY breaking terms
$M_{q_1L}$, $M_{q_2L}$, $M_{uR}$, $M_{dR}$, $M_{cR}$, $M_{sR}$ and $M_3$ by a factor
$(1+\xi)$, with $0 \leq \xi \leq 2$. Fig.~\ref{massdep_plot_squark} shows the cross
section as a function of the average squark mass for two different sets of cuts: The
left plot is for more inclusive cuts, taking only Eqs.~(\ref{cuts_W_min}) and
(\ref{cuts_W_deltaeta}) into account, while the right plot uses the full set of cuts
Eqs.~(\ref{cuts_W_min} -- \ref{cuts_W_nomass}), (\ref{cuts_W_paper}),
(\ref{cuts_W_paper_ptmiss}) and (\ref{cuts_jetveto}). 

We see a strong suppression with higher squark and gluino masses of the processes 
containing two neutralinos. The decrease of the $\chi_2^0\chi_1^0 jj$ 
cross section with increasing squark masses is even more pronounced as even at 
high masses the dominant contributions come from the squark production with a 
t-channel gluino exchange~\footnote{This holds for our approximation containing
only $\alpha_s^2\alpha^2$-contributions as well as for the full process with
all possible Feynman graphs.}, which also depends on the increased gluino mass. 
For the processes involving charginos the dominant contributions at low masses involve squarks,
but a noticeable fraction of the electroweak gaugino pairs originate from virtual W or Z boson decay. As these 
contributions are independent from the squark mass, there is a noticeable fraction
of the cross section which remains at higher squark masses. 
The slepton processes do not involve any squarks
or gluinos and therefore stay independent of the mass variation.\footnote{The 
slepton branching ratios change slightly due to higher-order corrected
$\chi_1^\pm$ and $\chi_2^0$ masses.}

For average squark masses of 1.1 TeV (see column 2 of Tables \ref{masstable} and \ref{brtable})
we checked the size of further SUSY contributions to see if the dominant processes
at low squark masses are still the largest ones at higher squark masses. For
the processes with not too tiny cross sections at low squark masses, the results are presented in
Tables~\ref{tab_msq1tev} and~\ref{tab_msq1tev_other}. Most of the other processes also have a reduced cross 
section at
higher squark masses, therefore the processes with the largest cross sections are again
within the list of the previously selected processes. One exception is the process
$\chi_1^\pm \, \chi_2^0 \, j$, which shows a slightly increased cross section.
This is due to interference effects between graphs with and without squarks.
But still the overall contribution of this process to the SUSY background is only a few
percent.

As the mass of the light neutral Higgs boson remains nearly unaffected by a variation of the
first and second generation squark masses and the gluino mass, the cross section
of the signal process $h \rightarrow WW$ remains unchanged. It is $\sigma = 2.91$ fb for the
cuts in the left plot and $\sigma = 1.33$ fb for the ones of the right plot of 
Fig.~\ref{massdep_plot_squark}.

For higher squark and gluino masses, the processes $\chi_1^\pm \, \chi_2^0 \, jj$
give the largest contributions to the SUSY backgrounds, at a total 
SUSY background level which is smaller by a factor of 
\begin{equation}
 S / B^{SUSY} = 6.2
\end{equation}
compared to the signal cross section.

\subsection{Scenario SPS1a-slope}

The scenario SPS1a-slope is characterized by SUSY particle masses which are
about 30\% higher compared to our default scenario SPS1amod (average 
$m_{\widetilde{q}}=703 \;\textrm{GeV}$, $m_{\chi_1^+} \approx m_{\chi_2^0} \approx 245 \;\textrm{GeV}$).
The cross sections at this scenario are shown in Table \ref{xs_W_table_sps-slope}.
Most noteworthy is the behavior of the slepton contributions.
They are not affected by the variation of gluino or squark mass, but here the
30\% higher slepton masses lead to a reduction of the slepton contribution by 65\% compared to
the scenario SPS1amod.

The higher Higgs boson mass results in a higher signal cross section.
Combined with the smaller backgrounds, the signal to 
background ratio increases strongly compared to SPS1amod.

\tabcolsep2.5mm
\begin{table}[p]
  \begin{center}
  \begin{tabular}{llccc}
  \toprule
  & Cuts & basic + rap. gap & + $m_{inv} + \slashed{p}_T + \phi_{\ell\ell} + \textrm{b-tag}$ & + CJV on $j_{decay}$ \\
  Processes && Eqs.~(\ref{cuts_W_min}) + (\ref{cuts_W_deltaeta}) & + \eqref{cuts_W_nomass}, \eqref{cuts_W_paper}, \eqref{cuts_W_paper_ptmiss}, \eqref{btag} & + \eqref{cuts_jetveto} \\
  \midrule
  \midrule
  $\chi_1^+ \, \chi_1^- \, jj$ & no b                   & $0.60 \;\text{fb}$ & $0.032 \;\text{fb}$ & $0.032 \;\text{fb}$ \\
  $\chi_1^+ \, \chi_1^- \, jj$ & b-contr.               & $0.03 \;\text{fb}$ & $0.002 \;\text{fb}$ & $0.002 \;\text{fb}$ \\
  $\chi_2^0 \, \chi_1^0 \, jj$ & no b                   & $0.25 \;\text{fb}$ & $0.013 \;\text{fb}$ & $0.013 \;\text{fb}$ \\
  $\chi_2^0 \, \chi_1^0 \, jj$ & b-contr.               & $0.03 \;\text{fb}$ & $0.001 \;\text{fb}$ & $0.001 \;\text{fb}$ \\
  $\widetilde{\ell}^+\, \widetilde{\ell}^-\, jj$ & no b & $0.81 \;\text{fb}$ & $0.010 \;\text{fb}$ & $0.010 \;\text{fb}$ \\
  \midrule
  $\chi_1^+ \, \chi_2^0 \, jj$ & no b                   & $1.12 \;\text{fb}$ & $0.069 \;\text{fb}$ & $0.039 \;\text{fb}$ \\
  $\chi_1^- \, \chi_2^0 \, jj$ & no b                   & $0.58 \;\text{fb}$ & $0.040 \;\text{fb}$ & $0.023 \;\text{fb}$ \\
  $\chi_2^0 \, \chi_2^0 \, jj$ & no b                   & $0.39 \;\text{fb}$ & $0.027 \;\text{fb}$ & $0.011 \;\text{fb}$ \\
  $\chi_2^0 \, \chi_2^0 \, jj$ & b-contr.               & $0.03 \;\text{fb}$ & $0.001 \;\text{fb}$ & $0.001 \;\text{fb}$ \\
  \midrule
  \multicolumn{2}{l}{$\sum B^{SUSY}$}                   & $3.84 \;\text{fb}$ & $0.195 \;\text{fb}$ & $0.132 \;\text{fb}$ \\
  \midrule
  \multicolumn{2}{l}{VBF $h\rightarrow WW$}             & $5.32 \;\text{fb}$ & $2.37 \;\text{fb}$ & $2.37 \;\text{fb}$ \\
  \midrule
  \multicolumn{2}{l}{$S / B^{SUSY}$}                    &      1.4       &    12        &            18  \\
  \bottomrule
  \end{tabular}
  \end{center}
  \caption{Total cross sections of SUSY backgrounds dominant at low squark masses 
           to the VBF $h\rightarrow WW$ channel for the scenario SPS1a-slope.}
  \label{xs_W_table_sps-slope}
\end{table}

\begin{figure}[p]
  \begin{center}
    \includegraphics[height=0.36\textwidth]{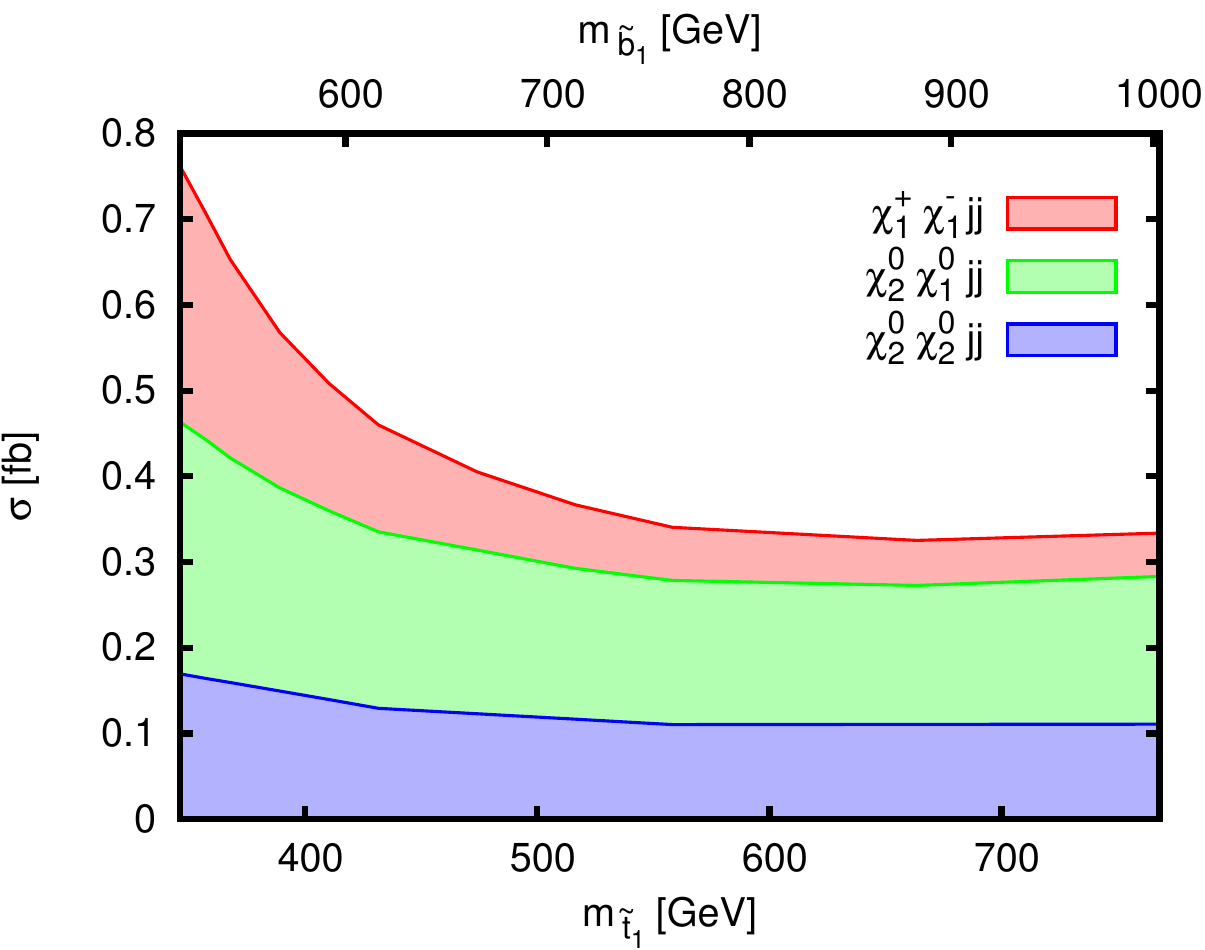}
    \hskip15pt
    \includegraphics[height=0.36\textwidth]{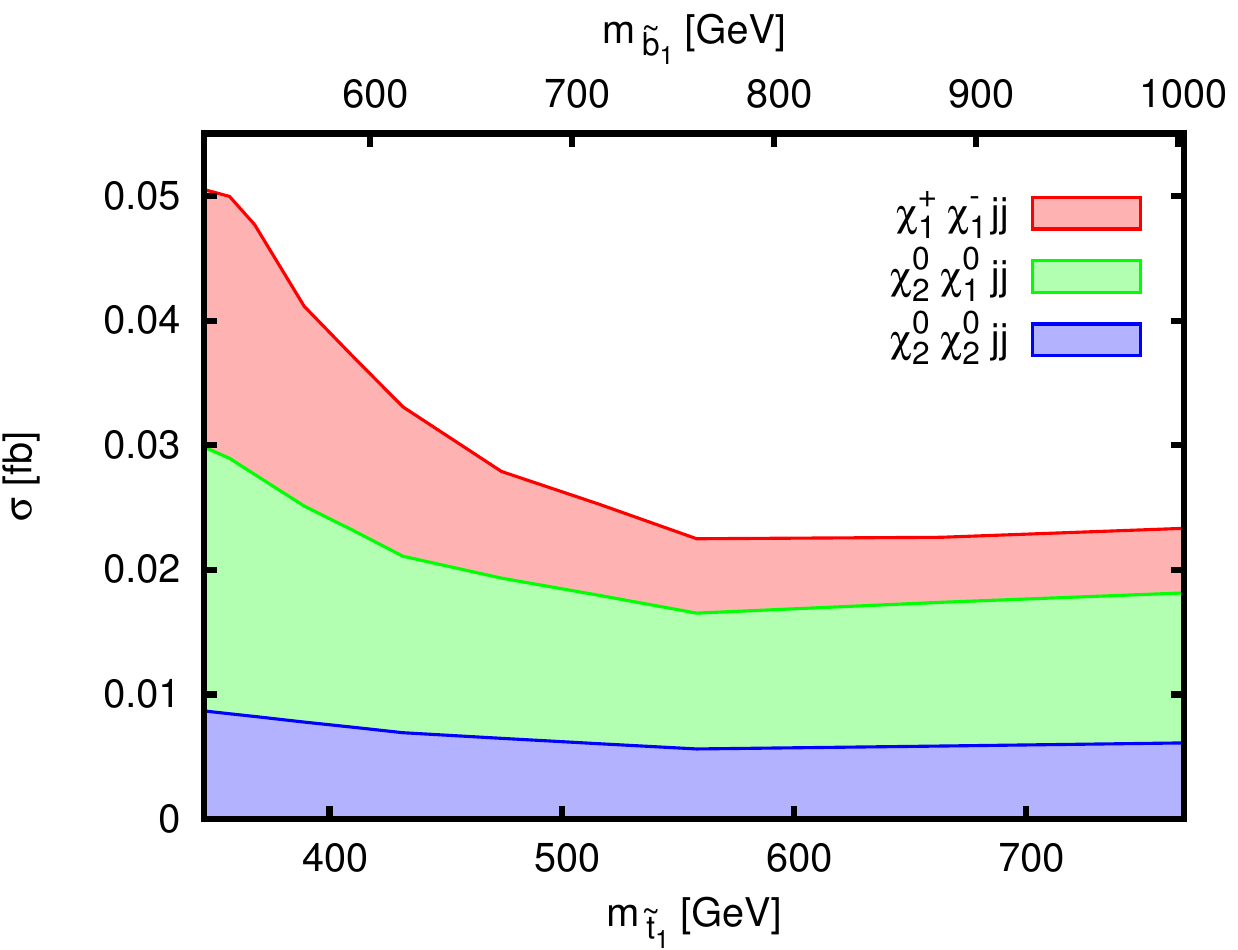}
  \end{center}
  \caption{Third generation squark mass dependencies for the b-quark
    contributions of the 
           SUSY-backgrounds as a function of the $\widetilde{t}_1$- mass. Left
           panel: with cuts from Eqs.~(\ref{cuts_W_min}) +
           (\ref{cuts_W_deltaeta}). 
           Right panel: with additional cuts from Eqs.~(\ref{cuts_W_nomass}),
           (\ref{cuts_W_paper}), (\ref{cuts_W_paper_ptmiss}) and (\ref{cuts_jetveto}).}
  \label{massdep_plot_bcontrib}
\end{figure}

\subsection{Stop/Sbottom Mass Dependence of b-Quark Contributions}

The b-quark contributions of the SUSY processes involve mostly third
generation squarks. We therefore check the dependence on their mass by varying $M_{q_3L}$,
$M_{tR}$, $M_{bR}$, $A_{t}$, $A_{b}$ and $M_3$ by a factor $(1+\rho)$ with 
$0 \leq \rho \leq 1$.

At low $\widetilde{t}_1$ masses, the chargino pair production is dominated by
$\widetilde{t}_1$ pair production. However, there are also small contributions
from the production of heavy Higgs bosons with a subsequent decay into a chargino
pair which remain as the only contributions for larger $\widetilde{t}_1$
masses. The cross section as a function of the $\widetilde{t}_1$ mass is shown in
Fig.~\ref{massdep_plot_bcontrib}. 
The slight increase of the cross section at low masses on the right plot is due to 
the cut on the invariant jet pair mass, which is less efficient for larger masses. 

For the next-to-lightest neutralino plus lightest neutralino production, the dependence on the 
$\widetilde{b}_1$ mass is smaller, for reasons described in Sect.~\ref{sps1amod-WW}.
At high $\widetilde{b}_1$ masses, only the contributions from the heavy Higgs 
bosons remain (see Fig.~\ref{massdep_plot_bcontrib}). The slight increase
above $\widetilde{b}_1$ masses of 750 GeV
is due to small changes in the mass of the heavy Higgs bosons and the next-to
lightest neutralino which lead to minor changes to branching ratios.
For the b-quark contributions of the production of a next-to lightest neutralino
pair plus two jets the dominant contributions come from the same Feynman Graph
topologies as in the $\chi_2^0\,\chi_1^0\,jj$ case. Therefore the mass dependence
of both processes is much alike.

In this scenario series, the Higgs boson mass slightly increases from 118 GeV
for $m_{\widetilde{t}_1}=345 \;\text{GeV}$ up to  
123 GeV for $m_{\widetilde{t}_1}=768 \;\text{GeV}$. This
leads to an increased cross section for 
the $h\rightarrow WW$ signal process ranging from $\sigma = 1.33$ fb up to
$\sigma = 1.98$ fb for the final set of cuts.

\section{Contributions to the Background in other Scenarios}
\label{otherscenarios}

In this section, we investigate effects due to different sparticle masses
which do change the characteristics of our scenario like branching
ratios. More specifically, we consider a scenario with light sleptons
at low and high squark masses for the dominant SUSY processes.
For the process $\chi_1^+\,\chi_1^-\,jj$ we also study the effect of 
a very light LSP as well as small mass differences of the particles 
within one cascade decay.
Again, we do not conserve the mSUGRA assumptions.

\subsection{Scenario with Light Selectrons and Smuons}

\subsubsection{Squark / Gluino Masses like in SPS1amod}

\begin{figure}[tb]
  \begin{center}
    \includegraphics[height=0.3\textwidth]{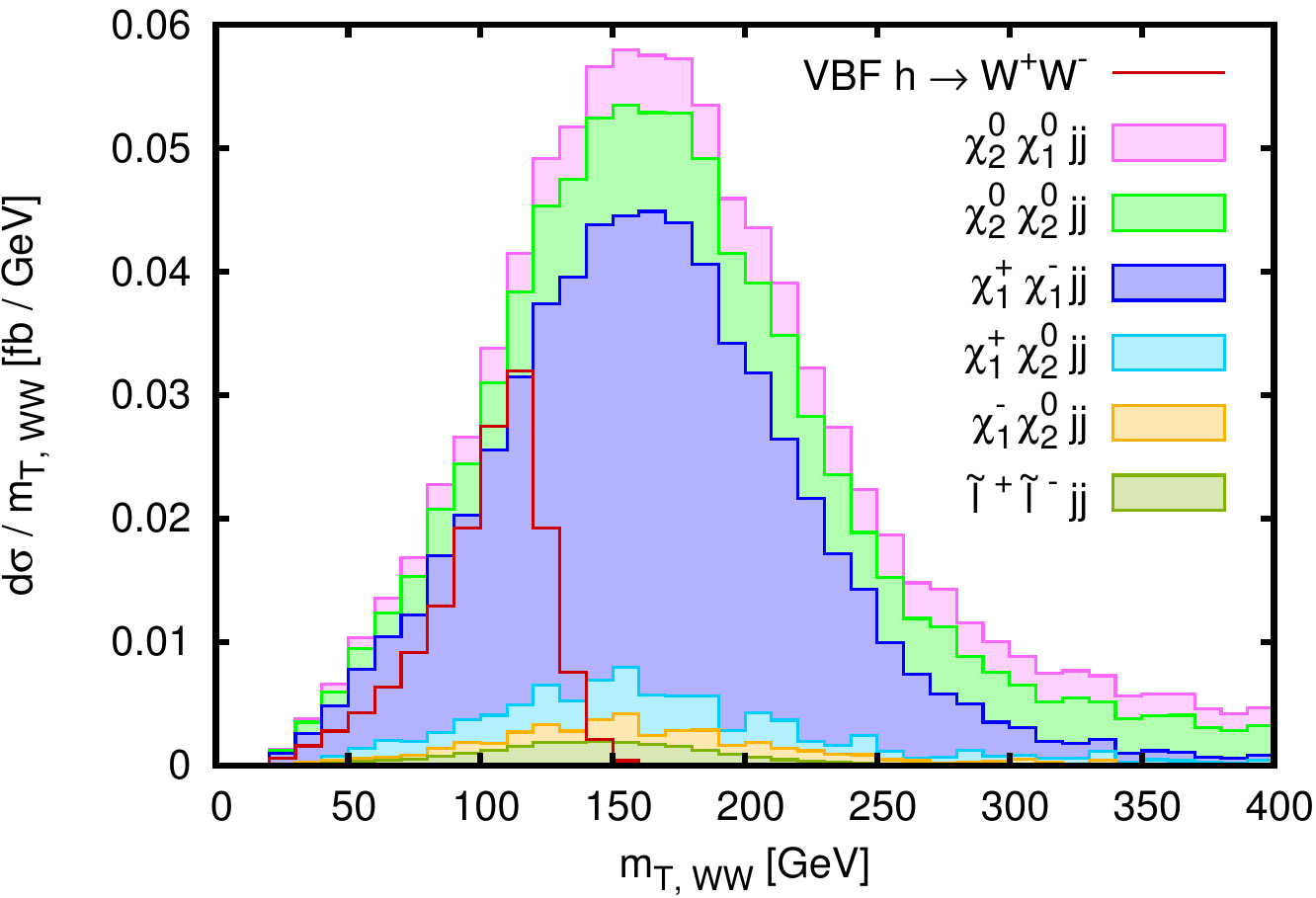}
    \hskip20pt
    \includegraphics[height=0.3\textwidth]{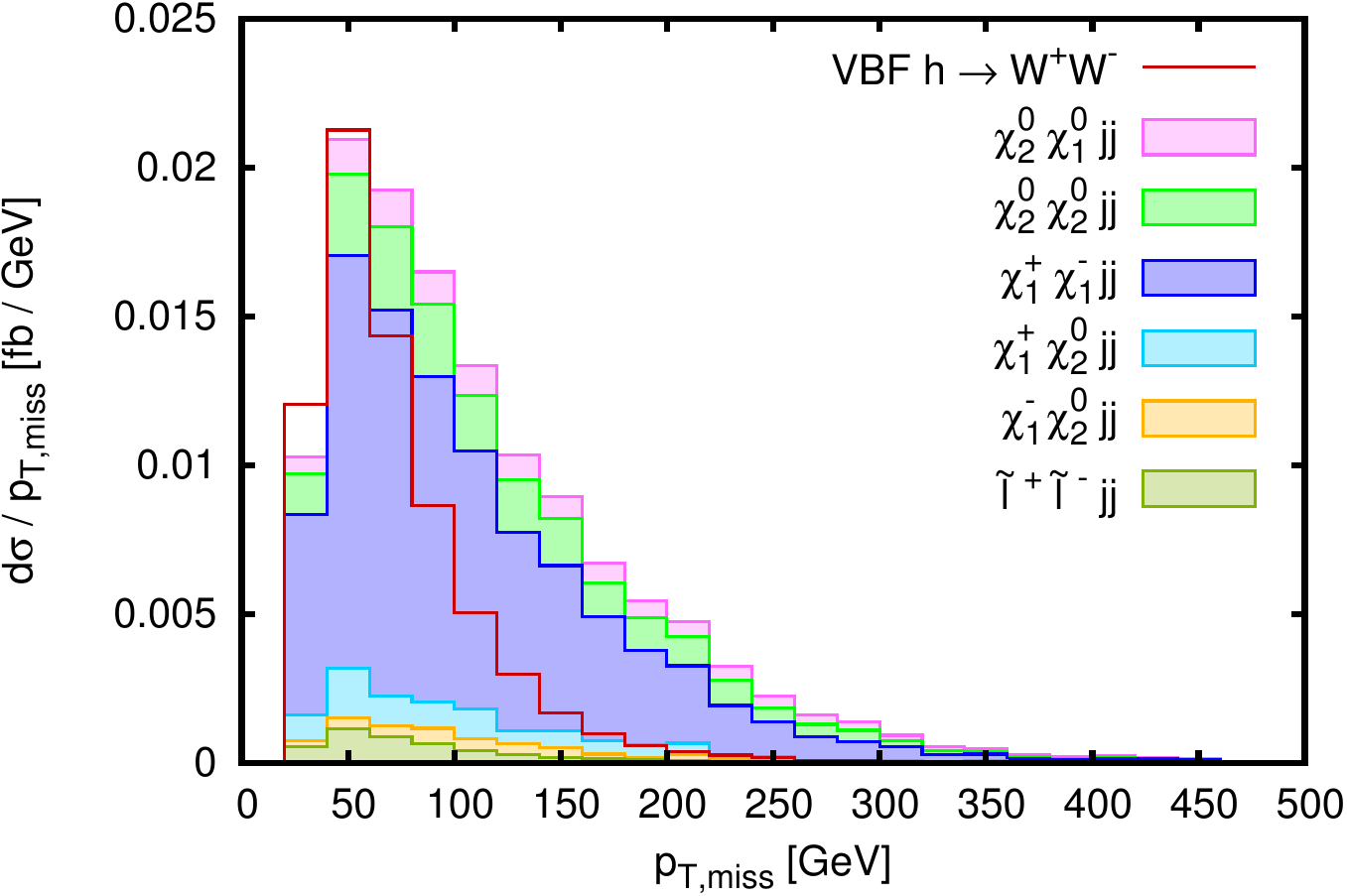}
  \end{center}
  \caption{Transverse WW mass and $\slashed{p}_T$ distributions analogous to Fig.~\ref{Wjj_plots},
           but for the scenario with light sleptons, including 
           the slepton channel and the reducible SUSY background processes. 
           B-quark contributions included where relevant.}
  \label{slepton_plots_W}
\end{figure}

\begin{figure}[tb]
  \begin{center}
    \includegraphics[height=0.3\textwidth]{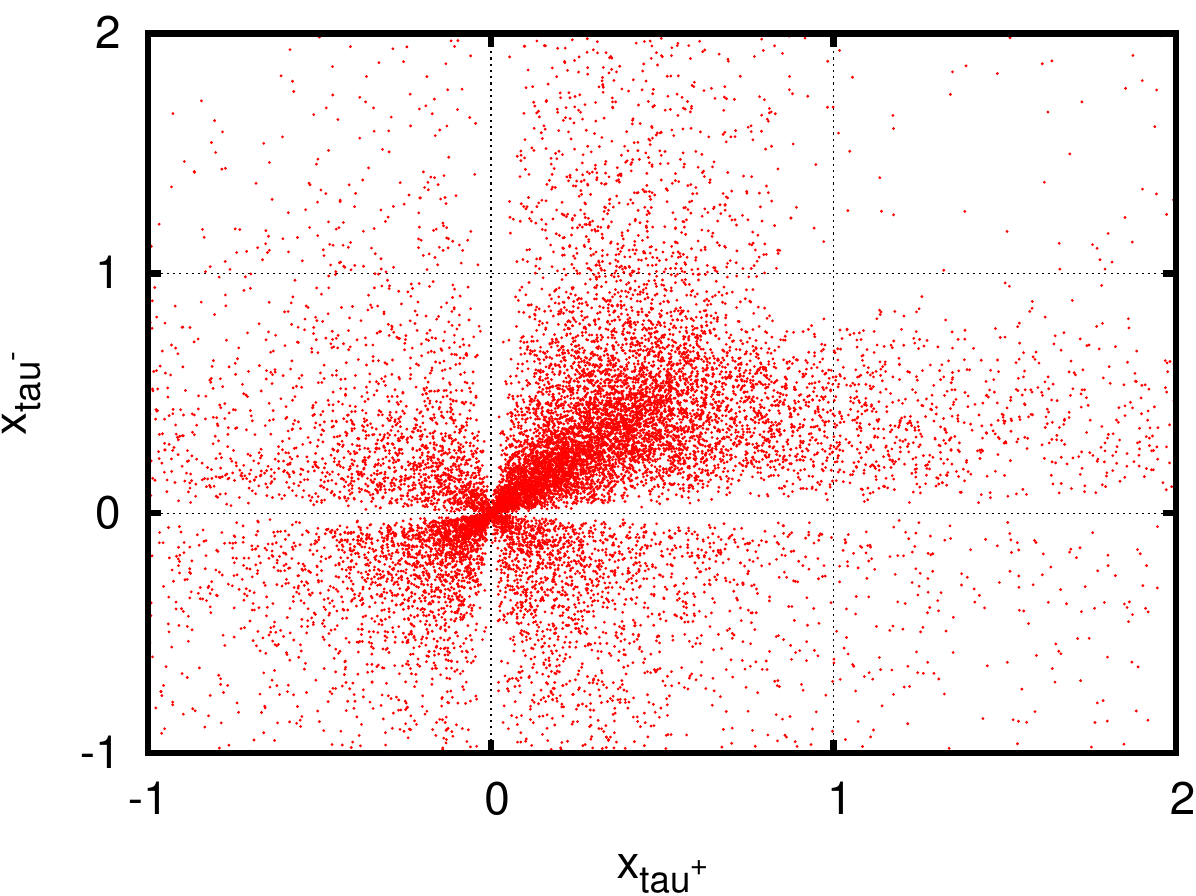}
    \hskip20pt
    \includegraphics[height=0.3\textwidth]{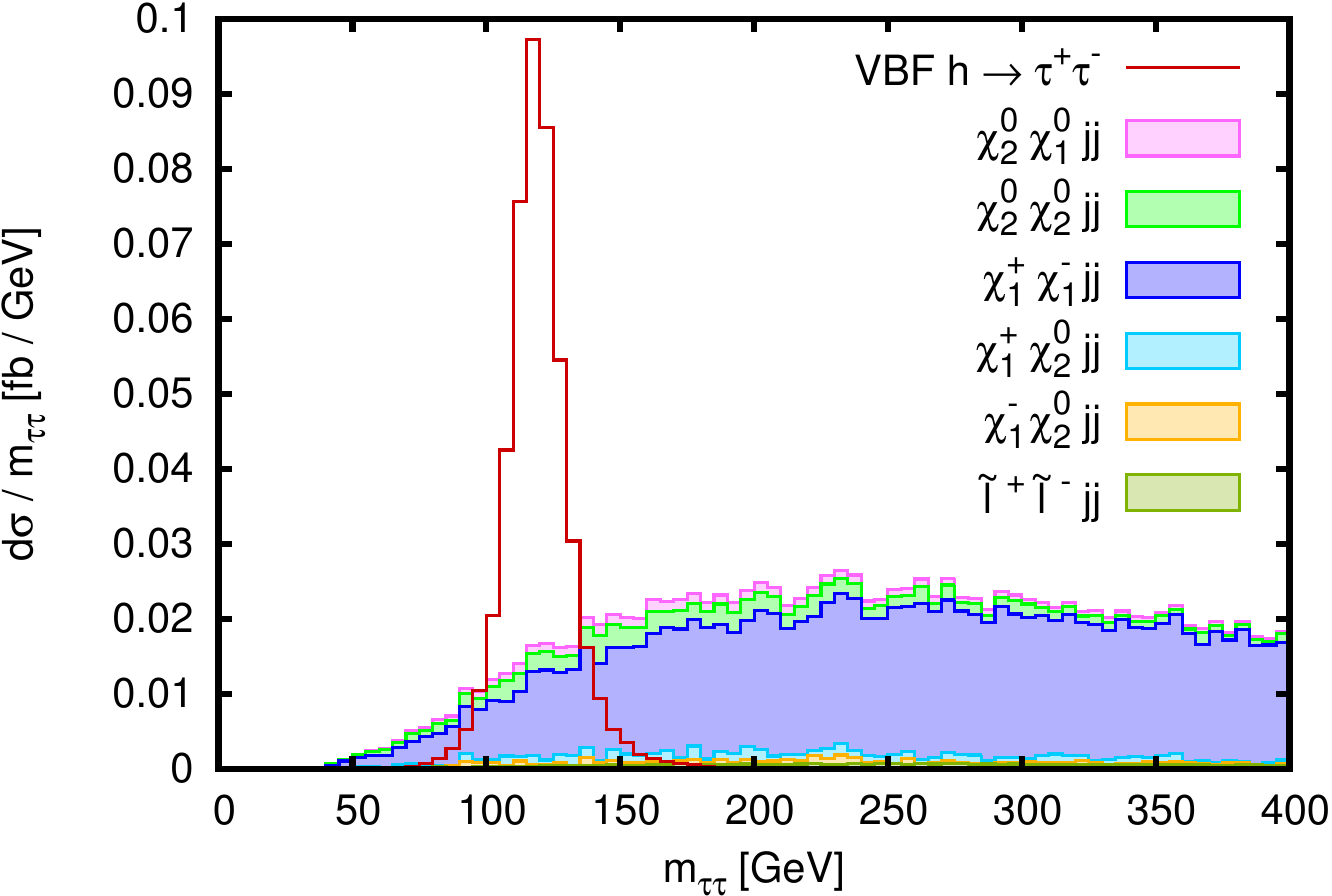}
  \end{center}
  \caption{Left panel: tau momentum fractions for the chargino process, analogous to Fig.~\ref{tau_masscut},
           but for the scenario with light sleptons. The right panel shows the reconstructed tau pair mass
           in this scenario with the same cuts as Fig.~\ref{tau_massrec}, including b-quark contributions,
           the slepton channel and the reducible SUSY background processes.
           The detector effects on the
           $\slashed{p}_T$ measurement from Eq.~(\ref{fakeptmiss}) are taken into account in both plots.}
  \label{slepton_plots_tau}
\end{figure}

In a scenario where $\chi_1^\pm$ and $\chi_2^0$ are heavier than
$\widetilde{e}_L$ and $\widetilde{\mu}_L$ as well as lighter than
$\widetilde{\tau}_{1/2}$, the chargino and next-to-lightest
neutralino will decay into $\widetilde{e}_L$ and $\widetilde{\mu}_L$ instead
of into $\widetilde{\tau}$ leptons; the 
subsequent slepton decay will not produce a tau lepton, but an electron or muon,
which can be detected directly.

These values for slepton masses (see Table~\ref{masstable}) are obtained by lowering the soft SUSY breaking
mass parameters $M_{e L}$ and $M_{\mu L}$ at low energies from 194.4 GeV in
SPS1amod to 134.4 GeV 
and increasing $M_{\tau L}$ from 193.6 GeV to 393.6 GeV as well as $M_{\tau R}$
from 133.4 GeV to 333.4 GeV. The other SUSY particle masses and parameters 
remain the same, leaving the signal processes $h \rightarrow \tau\tau$ and 
$h \rightarrow WW$ in vector boson fusion
unaffected.

\tabcolsep2.5mm
\begin{table}[b]
  \begin{center}
  \begin{tabular}{llccc}
  \toprule
  & Cuts & basic + rap. gap & + $m_{inv} + \slashed{p}_T + \phi_{\ell\ell} + \textrm{b-tag}$ & + CJV on $j_{decay}$ \\
  Processes && Eqs.~(\ref{cuts_W_min}) + (\ref{cuts_W_deltaeta}) & + \eqref{cuts_W_nomass}, \eqref{cuts_W_paper}, \eqref{cuts_W_paper_ptmiss}, \eqref{btag} & + \eqref{cuts_jetveto} \\
  \midrule
  \midrule
  $\chi_1^+ \, \chi_1^- \, jj$ & no b                   & $36.0 \;\text{fb}$ & $1.09 \;\text{fb}$ & $1.09 \;\text{fb}$ \\
  $\chi_1^+ \, \chi_1^- \, jj$ & b-contr.               & $12.5 \;\text{fb}$ & $0.14 \;\text{fb}$ & $0.14 \;\text{fb}$ \\
  $\chi_2^0 \, \chi_1^0 \, jj$ & no b                   & $2.38 \;\text{fb}$ & $0.103 \;\text{fb}$ & $0.103 \;\text{fb}$ \\
  $\chi_2^0 \, \chi_1^0 \, jj$ & b-contr.               & $0.72 \;\text{fb}$ & $0.020 \;\text{fb}$ & $0.020 \;\text{fb}$ \\
  $\widetilde{\ell}^+\, \widetilde{\ell}^-\, jj$ & no b & $2.77 \;\text{fb}$ & $0.082 \;\text{fb}$ & $0.082 \;\text{fb}$ \\
  \midrule
  $\chi_1^+ \, \chi_2^0 \, jj$ & no b                   & $2.00 \;\text{fb}$ & $0.133 \;\text{fb}$ & $0.133 \;\text{fb}$ \\
  $\chi_1^- \, \chi_2^0 \, jj$ & no b                   & $0.97 \;\text{fb}$ & $0.052 \;\text{fb}$ & $0.051 \;\text{fb}$ \\
  $\chi_2^0 \, \chi_2^0 \, jj$ & no b                   & $4.92 \;\text{fb}$ & $0.267 \;\text{fb}$ & $0.267 \;\text{fb}$ \\
  $\chi_2^0 \, \chi_2^0 \, jj$ & b-contr.               & $0.59 \;\text{fb}$ & $0.023 \;\text{fb}$ & $0.023 \;\text{fb}$ \\
  \midrule
  \multicolumn{2}{l}{$\sum B^{SUSY}$}                   & $62.9 \;\text{fb}$ & $1.91 \;\text{fb}$ & $1.91 \;\text{fb}$ \\
  \midrule
  \multicolumn{2}{l}{VBF $h\rightarrow WW$}             & $2.91 \;\text{fb}$ & $1.32 \;\text{fb}$ & $1.32 \;\text{fb}$ \\
  \midrule
  \multicolumn{2}{l}{$S / B^{SUSY}$}                    &      0.046      &     0.69        &            0.69  \\
  \bottomrule
  \end{tabular}
  \end{center}
  \caption{Total cross sections of SUSY backgrounds dominant at low squark masses
           to the VBF $h\rightarrow WW$ channel for the scenario with light sleptons.}
  \label{xs_W_table_sleptons}
\end{table}

The production cross sections for the processes involving electroweak gaugino pairs
remain unchanged; the important effects show up in
the decay chains. Looking at the electron and muon final states, we get an
increased cross section by a factor of 8.3 before lepton cuts for the chargino
process, which equals roughly $BR(\tau \rightarrow \ell \bar{\nu}_{\ell}
\nu_\tau)^{-2}$, and accounts for the missing suppression factor corresponding 
to the leptonic $\tau$ decays. The effect on the decay chain of the
next-to-lightest neutralino is quite different: As the masses of the sneutrinos
are linked to the masses of the sleptons, they become light as well. Therefore
a large fraction of the next-to-lightest neutralinos now decays completely
invisibly, leading to a  cross section only enhanced by a small factor of
1.6 for the $\chi_2^0\chi_1^0jj$ channel (including decays) compared to the 
SPS1amod scenario, again before lepton cuts. 

The missing tau decay has further impact on distributions and cut efficiencies:
Without the tau decay the charged final state
leptons have larger transverse momenta and many more events pass 
the initial lepton cuts. 
The missing transverse momentum cut is not very efficient as well (Fig.~\ref{slepton_plots_W}),
because there are less invisible particles in the final state. On the other
hand, the cut on the transverse $WW$ mass is more efficient now, as can be
seen in Fig.~\ref{slepton_plots_W}. So after all cuts, the cross section
for the chargino process is increased by a factor of 15, while the next-to
lightest neutralino contributions remain the same. 

The $\widetilde{\ell}^+\, \widetilde{\ell}^- jj$ channel is increased by
about a factor of two to three with respect to the SPS1amod scenario due
to the lighter $\widetilde{e}_L$ and $\widetilde{\mu}_L$ and due to the fact
that these sleptons now always decay to an electron or muon and the LSP.
The minor $\widetilde{\tau}$ contributions become even smaller than
in the SPS1amod scenario.

The situation for the reducible background is completely different for
the light sleptons of the first and second generation.
Beforehand, we had a large amount of soft jets and leptons from tau
decays as additional particles which could easily be missed by the detector.
Now, as the stau leptons are heavy, we have additional hard electrons and
muons and no tau jets instead of the tau lepton decay products.
As a first consequence, all processes that required a tau jet as tagging
jet ($\chi_1^\pm \, \chi_2^0 \, j$ and $\chi_2^0 \, \chi_2^0 \, (j)$) give
no contribution. Also the veto on additional hard jets from the
SUSY particle decays has no effect.
The process $\chi_1^\pm \, \chi_2^0 \, jj$ shows a reduced contribution due
to the additional hard leptons.
The only process that is enhanced in this scenario is $\chi_2^0 \, \chi_2^0 \, jj$.
Here the invisible decay modes of $\chi_2^0$ play an important role in mimicking
 the signature of the signal process.

Altogether, the total signal to background ratio for the $h \rightarrow WW$ 
signal reduces by more than a factor of two to
\begin{equation}
 S / B^{SUSY} = 0.69
\end{equation}
compared to $S / B^{SUSY} = 1.9$ in the SPS1amod scenario.
Detailed numbers are given in Table \ref{xs_W_table_sleptons}.

For the $h \rightarrow \tau\tau$ signal process, in this scenario, the background also increases
significantly, especially the $\chi_1^+ \, \chi_1^- \, j j$ contribution due to 
the previously mentioned reasons.
However, the tau pair mass reconstruction is still very efficient in reducing the SUSY background
processes. The signal to background ratio with all cuts and detector effects
on the missing transverse momentum measurement is $S / B^{SUSY} = 4.3$, compared to 
$S / B^{SUSY} = 12$ for SPS1amod, including b-quark contributions. The background is 
quite flat in the reconstructed tau pair mass, allowing for an effective 
subtraction from a sideband analysis, see Fig.~\ref{slepton_plots_tau}.

\subsubsection{Higher Squark and Gluino Masses}
\label{light_sleptons_msq}

\tabcolsep2.5mm
\begin{table}[b]
  \begin{center}
  \begin{tabular}{llccc}
  \toprule
  & Cuts & basic + rap. gap & + $m_{inv} + \slashed{p}_T + \phi_{\ell\ell} + \textrm{b-tag}$ & + CJV on $j_{decay}$ \\
  Processes && Eqs.~(\ref{cuts_W_min}) + (\ref{cuts_W_deltaeta}) & + \eqref{cuts_W_nomass}, \eqref{cuts_W_paper}, \eqref{cuts_W_paper_ptmiss}, \eqref{btag} & + \eqref{cuts_jetveto} \\
  \midrule
  \midrule
  $\chi_1^+ \, \chi_1^- \, jj$ & no b                   & $11.79 \;\text{fb}$ & $0.266 \;\text{fb}$ & $0.266 \;\text{fb}$ \\
  $\chi_1^+ \, \chi_1^- \, jj$ & b-contr.               & $ 2.43 \;\text{fb}$ & $0.040 \;\text{fb}$ & $0.040 \;\text{fb}$ \\
  $\chi_2^0 \, \chi_1^0 \, jj$ & no b                   & $0.069 \;\text{fb}$ & $0.001 \;\text{fb}$ & $0.001 \;\text{fb}$ \\
  $\chi_2^0 \, \chi_1^0 \, jj$ & b-contr.               & $0.473 \;\text{fb}$ & $0.011 \;\text{fb}$ & $0.011 \;\text{fb}$ \\
  $\widetilde{\ell}^+\, \widetilde{\ell}^-\, jj$ & no b & $ 2.78 \;\text{fb}$ & $0.086 \;\text{fb}$ & $0.086 \;\text{fb}$ \\
  \midrule
  $\chi_1^+ \, \chi_2^0 \, jj$ & no b                   & $1.08 \;\text{fb}$ & $0.058 \;\text{fb}$ & $0.058 \;\text{fb}$ \\
  $\chi_1^- \, \chi_2^0 \, jj$ & no b                   & $0.47 \;\text{fb}$ & $0.019 \;\text{fb}$ & $0.019 \;\text{fb}$ \\
  $\chi_2^0 \, \chi_2^0 \, jj$ & no b                   & $0.48 \;\text{fb}$ & $0.018 \;\text{fb}$ & $0.018 \;\text{fb}$ \\
  $\chi_2^0 \, \chi_2^0 \, jj$ & b-contr.               & $0.40 \;\text{fb}$ & $0.014 \;\text{fb}$ & $0.014 \;\text{fb}$ \\
  \midrule
  \multicolumn{2}{l}{$\sum B^{SUSY}$}                   & $19.97 \;\text{fb}$ & $0.513 \;\text{fb}$ & $0.513 \;\text{fb}$ \\
  \midrule
  \multicolumn{2}{l}{VBF $h\rightarrow WW$}             & $4.50 \;\text{fb}$ & $2.00 \;\text{fb}$ & $2.00 \;\text{fb}$ \\
  \midrule
  \multicolumn{2}{l}{$S / B^{SUSY}$}                    &      0.23      &         3.9      &            3.9  \\
  \bottomrule
  \end{tabular}
  \end{center}
  \caption{Total cross sections of SUSY backgrounds dominant at low squark masses to the
           VBF $h\rightarrow WW$ channel for the scenario with light sleptons but higher 
           squark and gluino masses.}
  \label{xs_W_table_sleptons_highmass}
\end{table}

We have seen in the last subsection that light sleptons can increase the SUSY
background cross sections substantially. As the numbers above were produced for
relatively light squark masses, we now want to check 
the size of the SUSY background processes with higher squark and gluino
masses, combined with these 
light selectrons and smuons. Therefore we take the SUSY breaking parameters
from the scenario discussed 
in the last subsection, and increase the values for $M_{q_1L}$, $M_{q_2L}$, $M_{uR}$, $M_{dR}$, $M_{cR}$,
$M_{sR}$, $M_{q_3L}$, $M_{tR}$, $M_{bR}$, $A_{t}$, $A_{b}$ and $M_3$ by a factor of $2$. This leads
to squark and gluino masses like in the second column of Table \ref{masstable}, and stop and sbottom mass
values like in the third column.
As the stop masses increase, the Higgs boson mass is shifted to a value of 123
GeV, leading to 
higher values for the $h \rightarrow WW$ signal process. As expected from the
mass behavior 
discussed in Section~\ref{massdep}, the processes involving charginos and neutralinos are reduced
by at least a factor of two, leading to a signal to background ratio of
\begin{equation}
 S / B^{SUSY} = 3.9
\end{equation}
after all cuts. 
Numbers at different cut levels are given in Table \ref{xs_W_table_sleptons_highmass}.

\subsection{Scenario with Light LSP}
\label{lightLSP}

Next we address the question whether the backgrounds increase
substantially when increasing the available
phase space for the decay leptons, especially in context of the tau pair mass reconstruction. 
We modify the mass values for some decay products
in the output from the spectrum generator. This is then again fed into
SDECAY~\cite{SDECAY} for the
branching ratio calculation. While we do not have a genuine MSSM model anymore, we still can use it
to investigate some characteristics.
We want to focus on the chargino decay, starting from the scenario SPS1amod.
The chargino can decay in two possible ways:
\begin{eqnarray}
\chi_1^\pm \rightarrow \widetilde{\tau}^\pm \,\nu \rightarrow \tau^\pm\,\chi_1^0\,\nu \;,\nonumber \\
\chi_1^\pm \rightarrow W^\pm \, \chi_1^0 \rightarrow \ell^\pm\,\chi_1^0\,\nu \;.
\end{eqnarray}
When we change $m_{\widetilde{\tau}}$ from 133 GeV to 100 GeV and $m_{\chi_1^0}$ from 98 GeV to 1 GeV, 
there is roughly 80 GeV available for kinematics at each step.

\begin{figure}[htb]
  \begin{center}
    \includegraphics[height=0.3\textwidth]{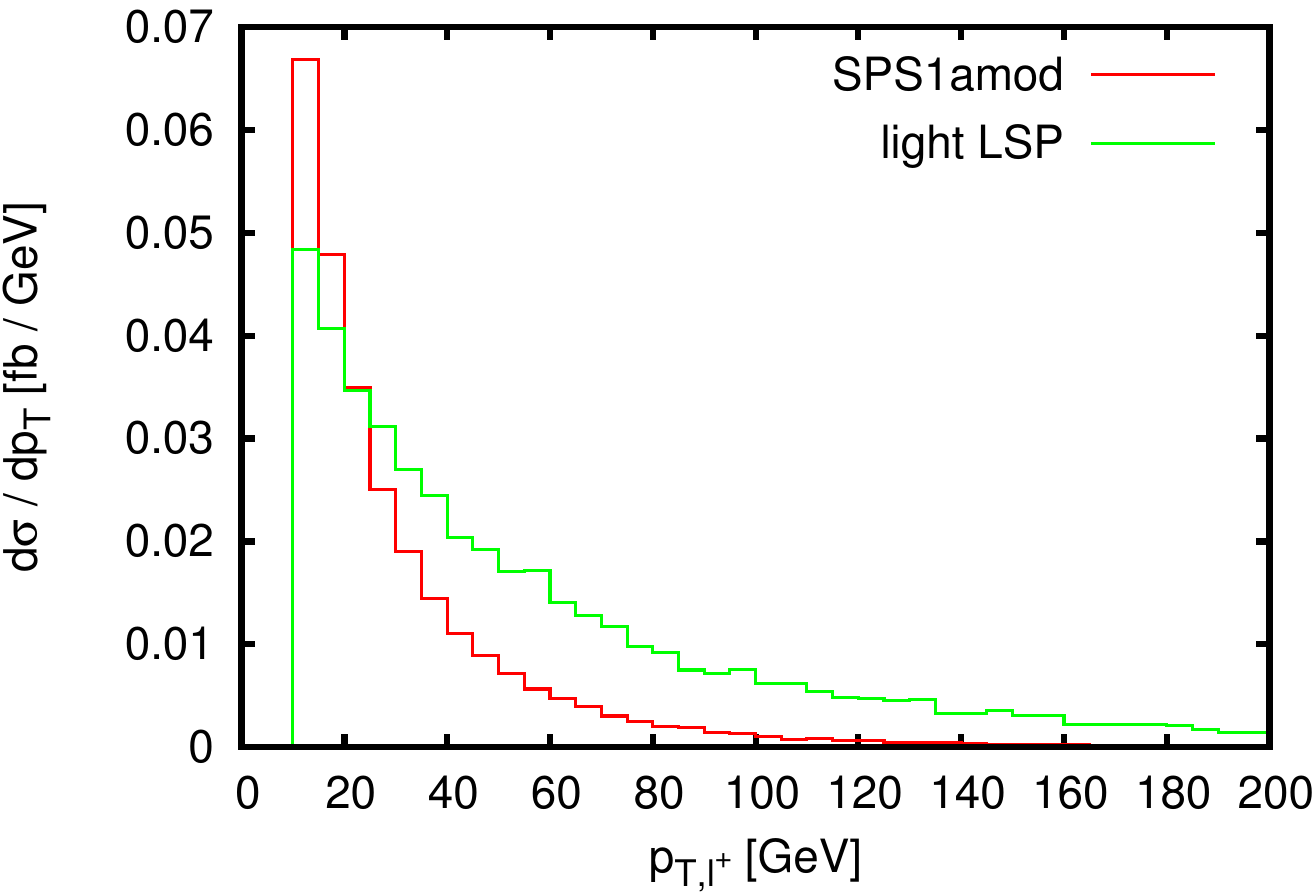}
    \hskip20pt
    \includegraphics[height=0.3\textwidth]{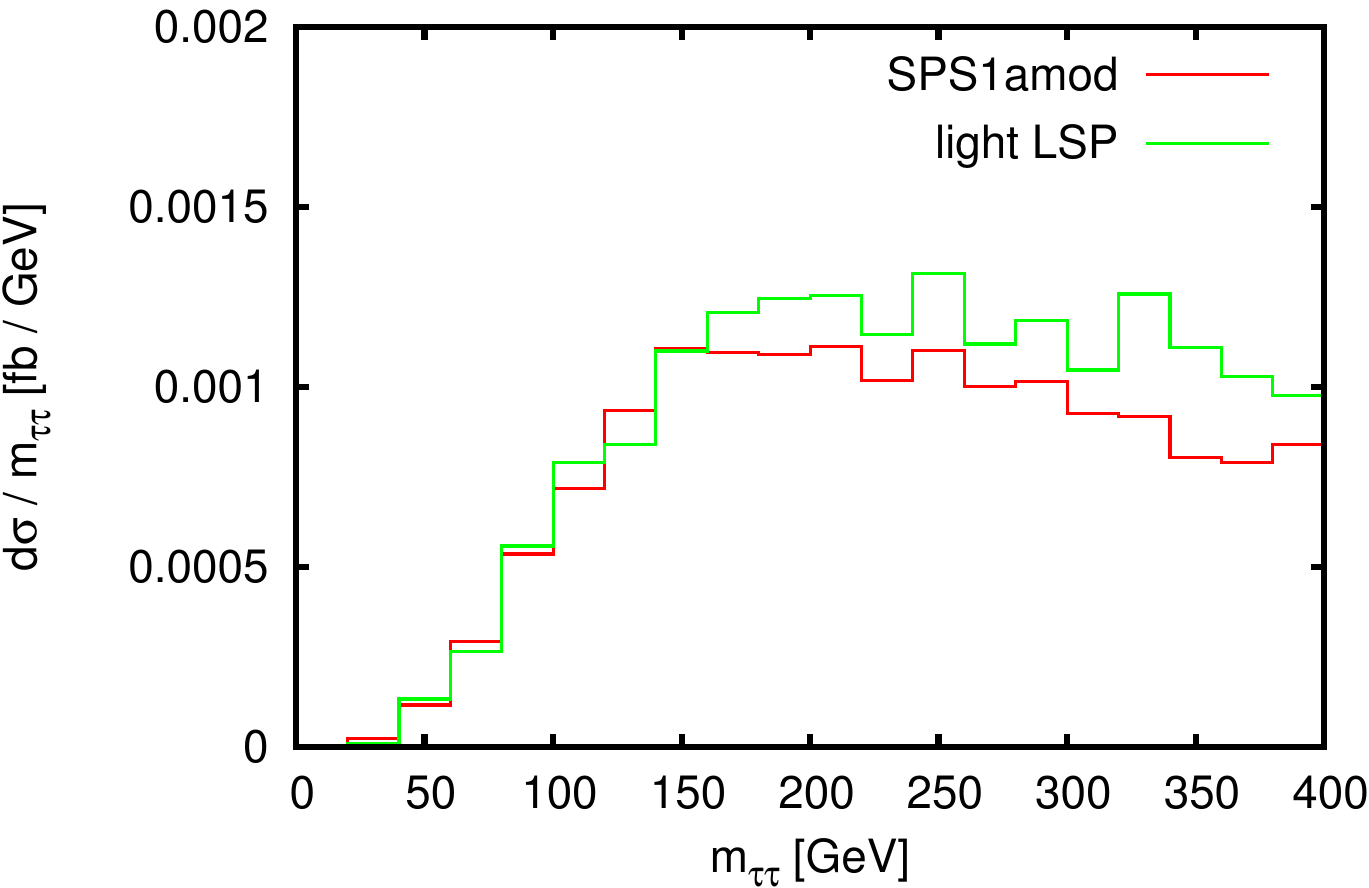}
  \end{center}
  \caption{Left panel: $p_{T,\ell^+}$ distribution of $\chi_1^+\chi_1^-jj$
    with cuts (\ref{cuts_tau_min}) - (\ref{cuts_tau_atlas}). 
           Right panel: reconstructed invariant tau pair mass with the
           additional cuts on $x_i$ and  
           $\text{cos}\, \phi_{\ell\ell}$ from Eq.~(\ref{cuts_tau_massrec}).}
  \label{lightlsp_plots}
\end{figure}

As shown in Fig.~\ref{lightlsp_plots}, we get harder leptons as expected, and thus more events pass the
transverse momentum cut on the leptons. Depending on the cuts, this results in
factors up to 1.8 on the cross section.
For the final cuts of the $h\rightarrow WW$ analysis, this factor is 1.4. 
There is no strong enhancement of the background when 
the cuts of $h\rightarrow \tau\tau$ after mass reconstruction are applied, as
can be seen on the right 
plot of Fig.~\ref{lightlsp_plots}. For the $h \rightarrow \tau\tau$ channel, the ratio of
lepton and missing  
transverse momentum in the scenario with light sleptons is more signal like
than in the scenario of this section.

\subsection{Scenario with Small Mass Differences in the Decay Chain}

In this SPS1a-like scenario with the unified trilinear coupling $A_0$ shifted from
-100 GeV to -750 GeV, we can see the effect of a very small mass difference of
only 9 GeV between the masses of the tau  
slepton and the LSP on the chargino decay chain. As, in this scenario, the
decay of the chargino into a tau slepton is dominant, we actually probe the effects of
this mass difference on the distributions of the $\chi^+_1\chi^-_1jj$ channel.

\begin{figure}[tb]
  \begin{center}
    \includegraphics[height=0.3\textwidth]{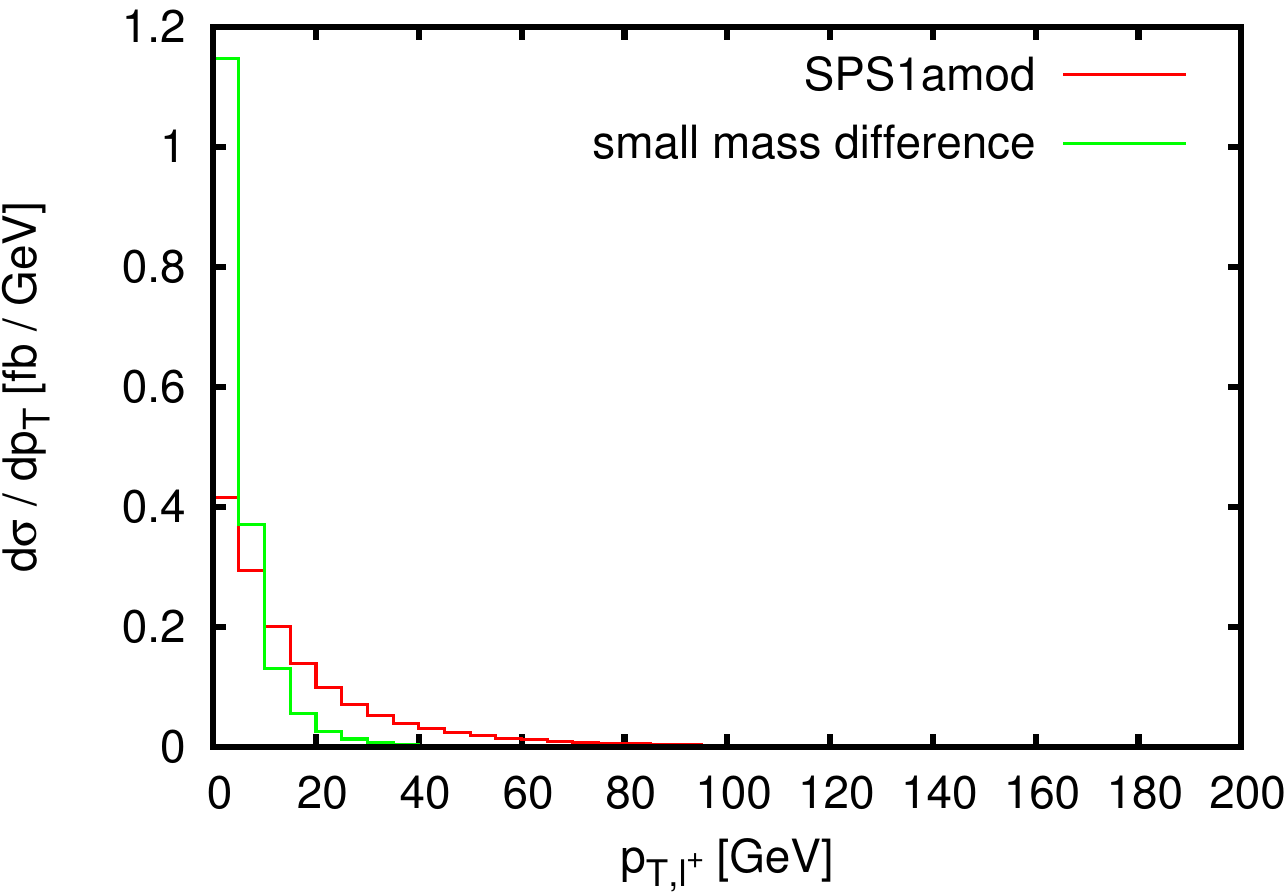}
    \hskip20pt
    \includegraphics[height=0.3\textwidth]{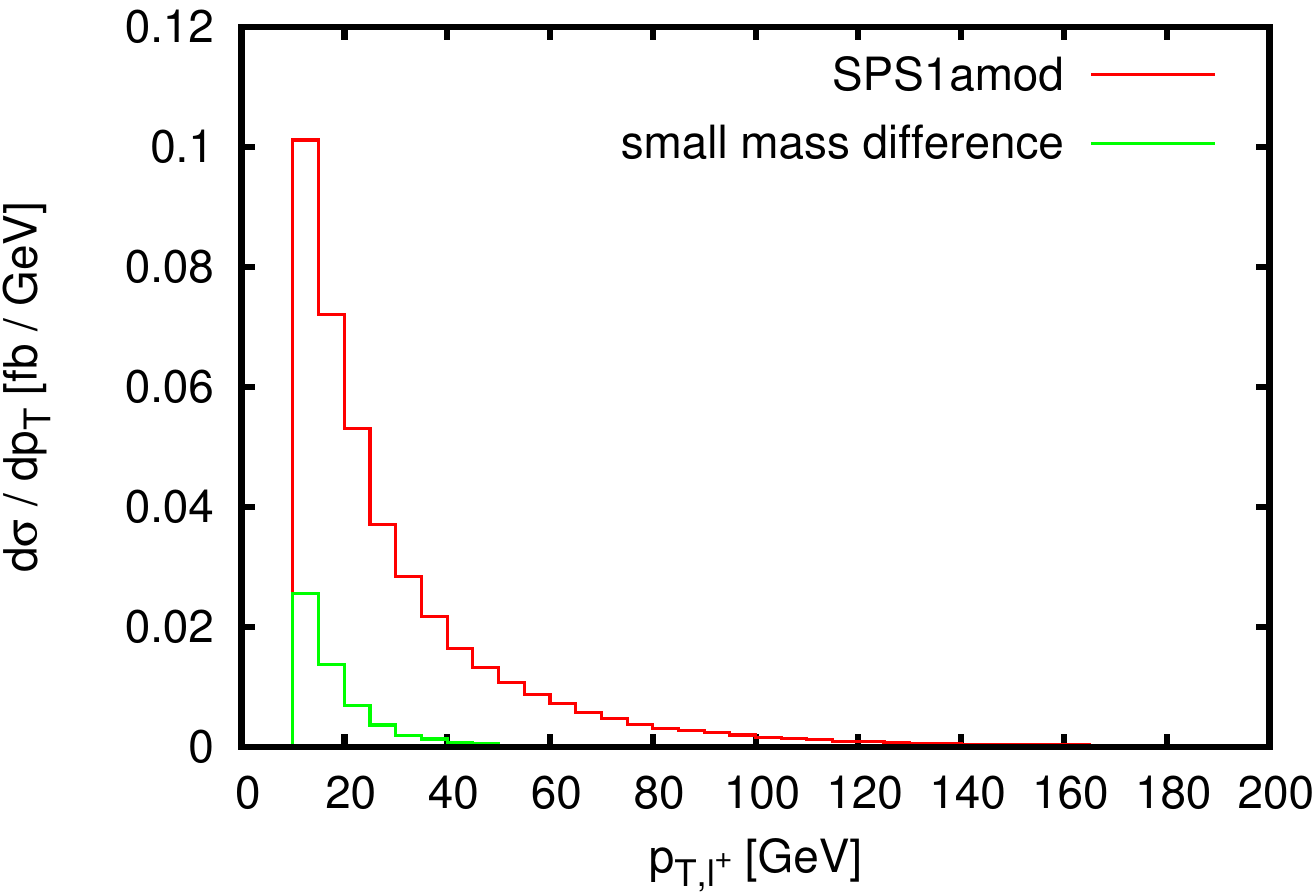}
  \end{center}
  \caption{$p_T$ distribution for positively charged lepton in a scenario with
    a small difference between 
           the $\widetilde{\tau}_1$ and $\chi_1^0$ mass. In the left plot,
           otherwise generated
           with the cuts from 
           Eqs.~\eqref{cuts_tau_min}+\eqref{cuts_tau_deltaeta}, also the
           region $p_{T,\ell} < 10\;\text{GeV}$ is shown.}
  \label{small_mass_diff_plots}
\end{figure}

As depicted in Fig.~\ref{small_mass_diff_plots}, the transverse
momentum of the tau lepton is very small, and the cut on the lepton
transverse momentum  of 10 GeV removes most of the chargino contributions. 
If we compare this with the results for SPS1amod and the scenario with a light
LSP, we find that a $\widetilde{\tau}_1-\chi_1^0$ mass difference of about 10 GeV
leads to negligible cross sections while the 36 GeV from SPS1amod results in
cross section values where further cuts are necessary for the suppression of
the background. If we increase this difference
even more as in the light LSP scenario of Sect.~\ref{lightLSP}, the additional effect is much smaller.

\section{LHC with 7 TeV Center of Mass Energy}
\label{7tev}

So far we have shown results for the LHC at 14 TeV center of mass energy, as,
due to the small signal cross sections, the Higgs boson searches via the VBF
topology are more relevant for higher center of mass energies;
a detailed analysis of Higgs boson production in vector boson fusion will
need of order $30~\text{fb}^{-1}$ or more of integrated luminosity at 14 TeV
center of mass energy.
But as the LHC performs very well with 7 TeV so far, we give some 
results on SUSY backgrounds to VBF Higgs production at 7 TeV as well.

\subsection{Scenario SPS1amod}

\begin{figure}[p]
  \begin{center}
    \includegraphics[height=0.3\textwidth]{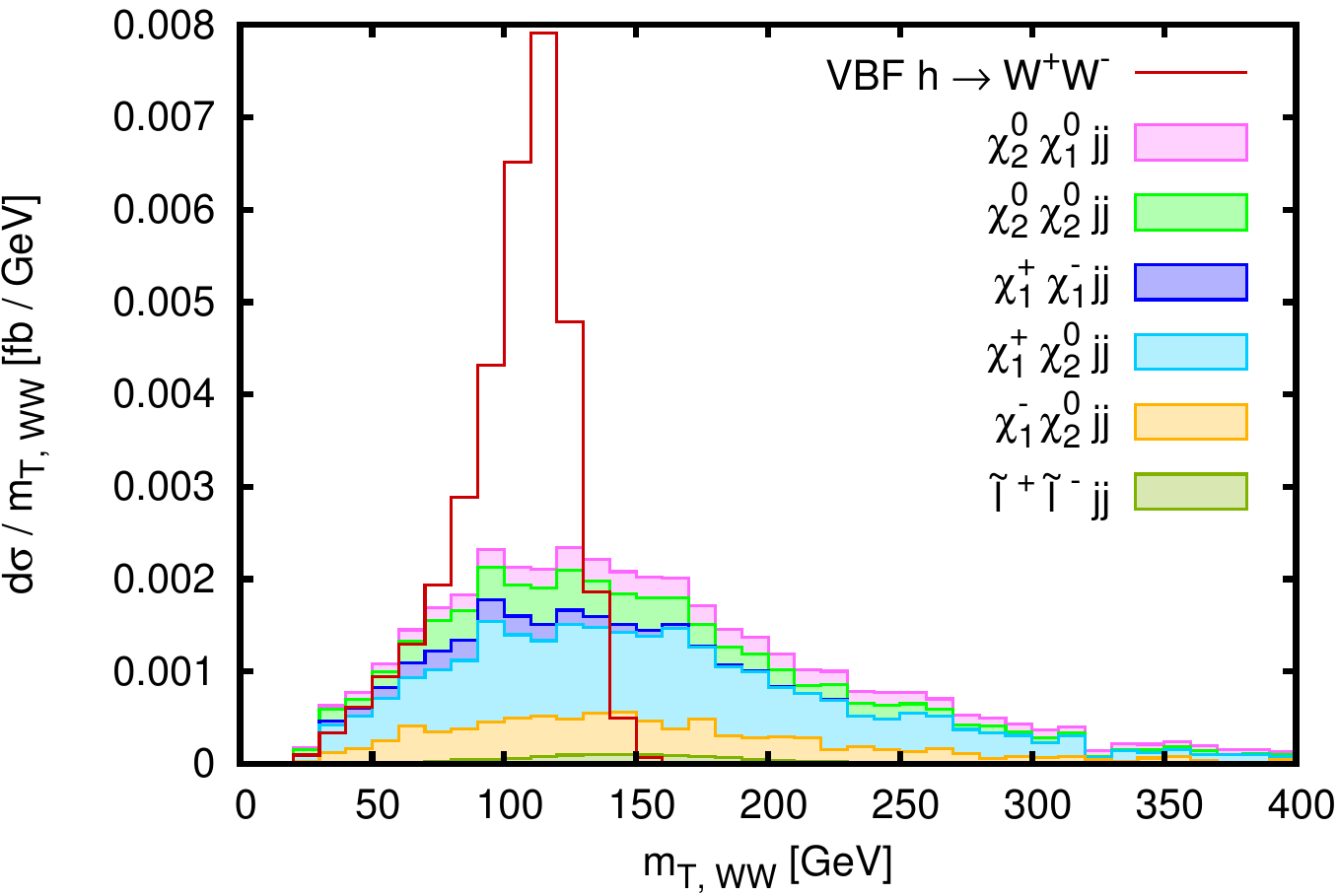}
    \hskip20pt
    \includegraphics[height=0.3\textwidth]{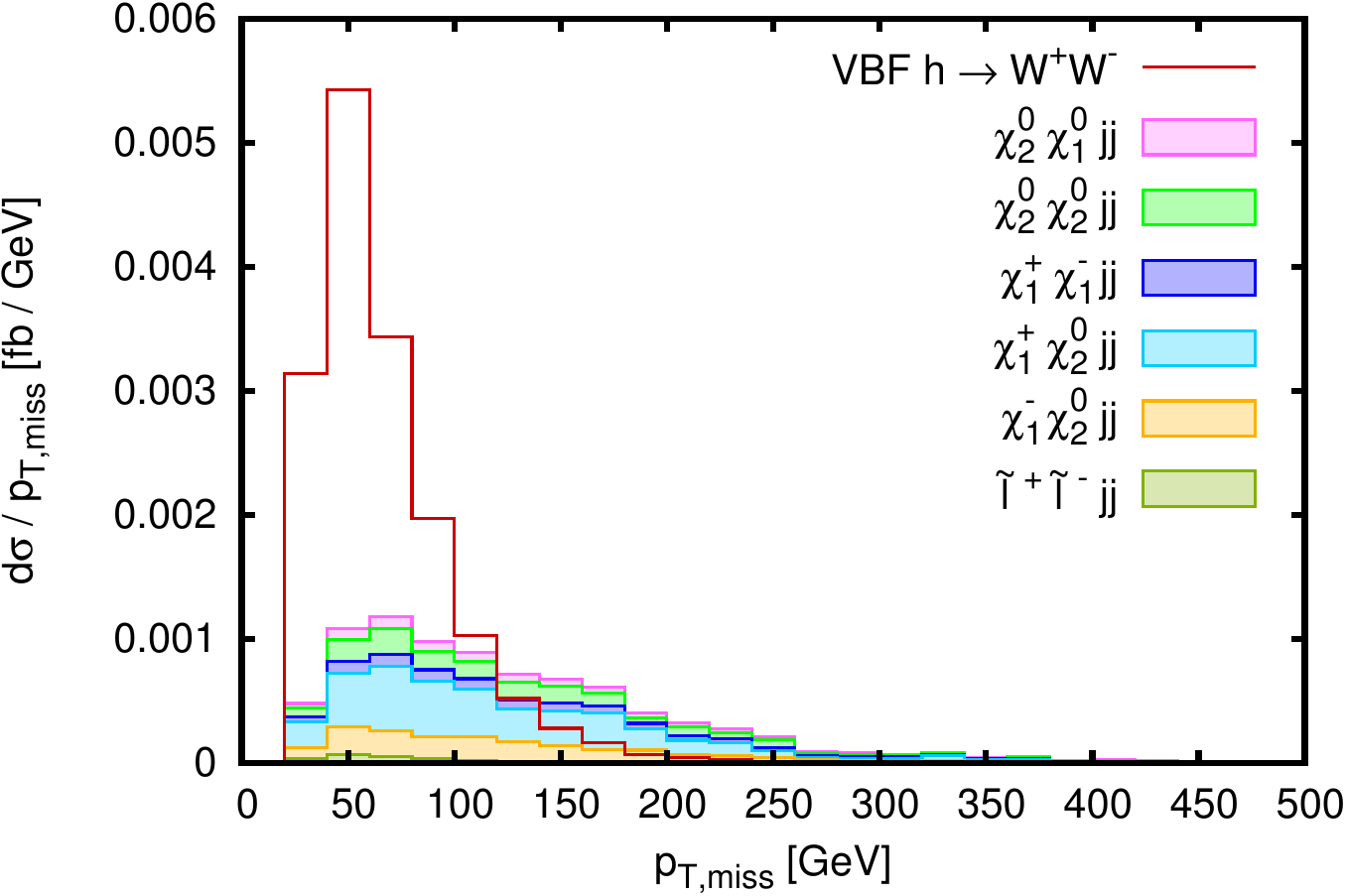}
  \end{center}
  \caption{Left panel: Transverse $WW$ mass distribution with cuts (\ref{cuts_W_min}) - (\ref{cuts_W_nomass}).
           Right panel: $\slashed{p}_T$ distribution with cuts (\ref{cuts_W_min}) - (\ref{cuts_W_nomass}), 
           (\ref{cuts_W_paper}). For both plots the scenario SPS1amod at 7 TeV
           center of mass energy is assumed and
           b-quark contributions are included.}
  \label{Wjj_plots_7TeV}
\end{figure}

\tabcolsep2.5mm
\begin{table}[p]
  \begin{center}
  \begin{tabular}{llccc}
  \toprule
  & Cuts & basic + rap. gap & + $m_{inv} + \slashed{p}_T + \phi_{\ell\ell} + \textrm{b-tag}$ & + CJV on $j_{decay}$ \\
  Processes && Eqs.~(\ref{cuts_W_min}) + (\ref{cuts_W_deltaeta}) & + \eqref{cuts_W_nomass}, \eqref{cuts_W_paper}, \eqref{cuts_W_paper_ptmiss}, \eqref{btag} & + \eqref{cuts_jetveto} \\
  \midrule
  \midrule
  $\chi_1^+ \, \chi_1^- \, jj$ & no b                   & $145 \;\text{ab}$ & $9.9 \;\text{ab}$ & $9.9 \;\text{ab}$ \\
  $\chi_1^+ \, \chi_1^- \, jj$ & b-contr.               & $ 26 \;\text{ab}$ & $0.6 \;\text{ab}$ & $0.6 \;\text{ab}$ \\
  $\chi_2^0 \, \chi_1^0 \, jj$ & no b                   & $ 99 \;\text{ab}$ & $8.9 \;\text{ab}$ & $8.9 \;\text{ab}$ \\
  $\chi_2^0 \, \chi_1^0 \, jj$ & b-contr.               & $ 24 \;\text{ab}$ & $0.9 \;\text{ab}$ & $0.9 \;\text{ab}$ \\
  $\widetilde{\ell}^+\, \widetilde{\ell}^-\, jj$ & no b & $206 \;\text{ab}$ & $4.4 \;\text{ab}$ & $4.4 \;\text{ab}$ \\
  \midrule
  $\chi_1^+ \, \chi_2^0 \, jj$ & no b                   & $588 \;\text{ab}$ & $53.7 \;\text{ab}$ & $36.5 \;\text{ab}$ \\
  $\chi_1^- \, \chi_2^0 \, jj$ & no b                   & $246 \;\text{ab}$ & $24.8 \;\text{ab}$ & $16.5 \;\text{ab}$ \\
  $\chi_2^0 \, \chi_2^0 \, jj$ & no b                   & $158 \;\text{ab}$ & $20.2 \;\text{ab}$ & $ 8.8 \;\text{ab}$ \\
  $\chi_2^0 \, \chi_2^0 \, jj$ & b-contr.               & $ 13 \;\text{ab}$ & $ 0.6 \;\text{ab}$ & $ 0.3 \;\text{ab}$ \\
  \midrule
  \multicolumn{2}{l}{$\sum B^{SUSY}$}                   & $1505 \;\text{ab}$ & $124 \;\text{ab}$ & $86.8 \;\text{ab}$ \\
  \midrule
  \multicolumn{2}{l}{VBF $h\rightarrow WW$}             & $777 \;\text{ab}$ & $316 \;\text{ab}$ & $316 \;\text{ab}$ \\
  \midrule
  \multicolumn{2}{l}{$S / B^{SUSY}$}                    &   0.52         &       2.5    &            3.6  \\
  \bottomrule
  \end{tabular}
  \end{center}
  \caption{Total cross sections of SUSY backgrounds
           to the VBF $h\rightarrow WW$ channel for the scenario SPS1amod at 7 TeV center of
           mass energy.}
  \label{xs_W_table_7TeV}
\end{table}

At first, the results for our base scenario SPS1amod are shown. As one can see
in  
Fig.~\ref{Wjj_plots_7TeV} and Table \ref{xs_W_table_7TeV}, the SUSY background
processes have a much stronger dependence on the center of mass energy than
the Higgs boson production in vector boson fusion, as they involve much heavier 
particles. Hence the signal to background ratio for the signal channel $h\rightarrow WW$ 
improves by about a factor of 2, yielding
\begin{equation}
 S / B^{SUSY} = 3.6 \,.
\end{equation}

This signal to background ratio is an upper limit calculated in a scenario with very light squark 
and gluino masses. For a center of mass energy of 14 TeV we found a reduction by a factor
of 3.3 when considering more realistic values of squark and gluino masses of the first
two generations. Assuming a reduction of the same order at 7 TeV, the SUSY backgrounds
would mean no harm. Furthermore the central jet veto used to suppress additional jets
from decayed SUSY particles also vetoes additional jets from QCD radiation which occur
more frequently in the SUSY processes than in the signal process (see Sect.~\ref{qcdjetveto}).

\subsection{\texorpdfstring{Scenario with Heavier {\boldmath $\widetilde{q}$} / {\boldmath $\widetilde{g}$}
           and Light {\boldmath $\widetilde{e}$} / {\boldmath $\widetilde{\mu}$}}
           {Scenario with Heavier Squarks / Gluinos and Light Selectrons / Muons}}

\begin{figure}[p]
  \begin{center}
    \includegraphics[height=0.3\textwidth]{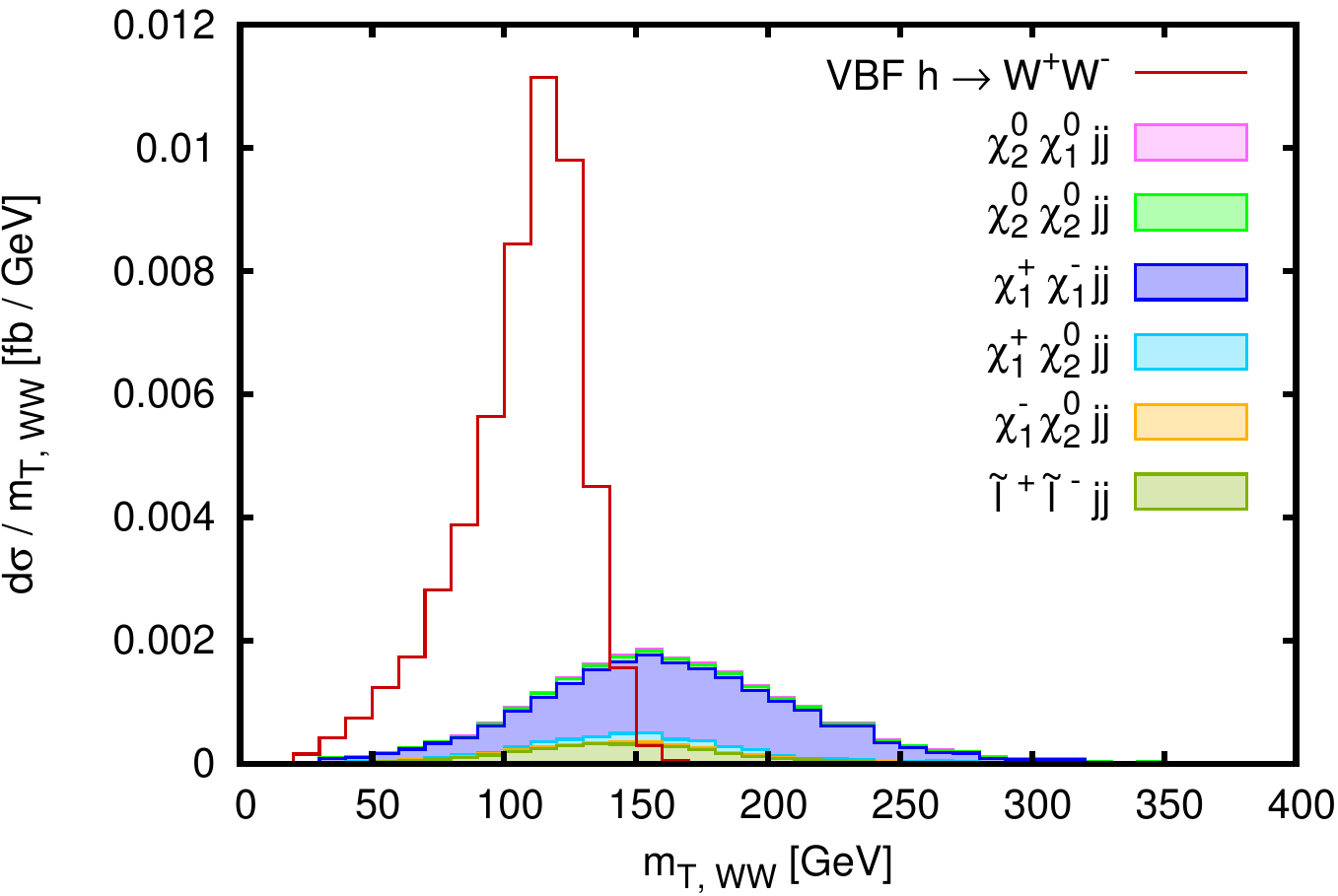}
    \hskip20pt
    \includegraphics[height=0.3\textwidth]{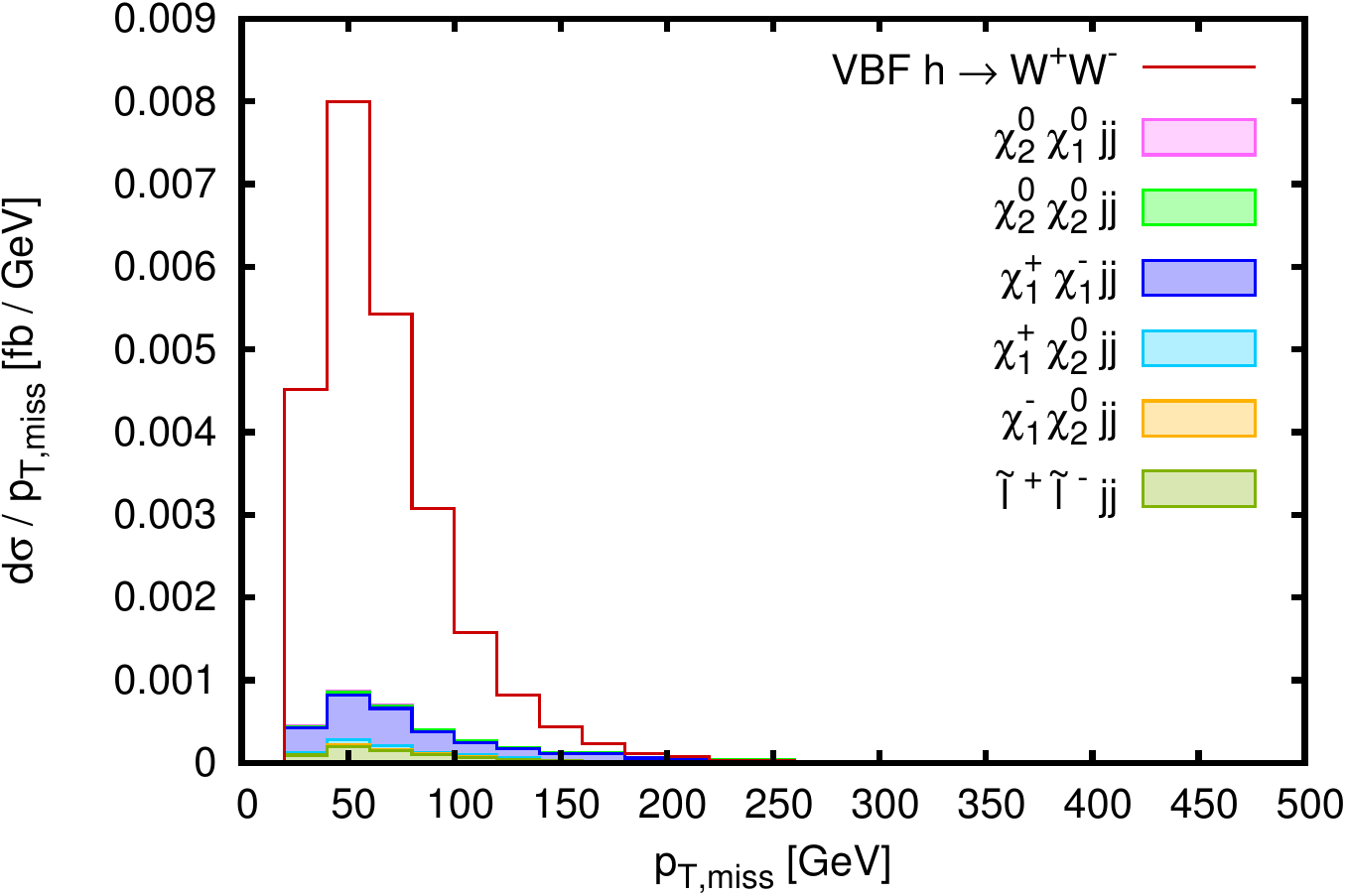}
  \end{center}
  \caption{Left panel: Transverse $WW$ mass distribution with cuts (\ref{cuts_W_min}) - (\ref{cuts_W_nomass}).
           Right panel: $\slashed{p}_T$ distribution with cuts (\ref{cuts_W_min}) - (\ref{cuts_W_nomass}), 
           (\ref{cuts_W_paper}). For both plots, the scenario with light
           sleptons and higher squark/gluino  
           masses at 7 TeV center of mass energy is assumed and b-quark
           contributions included.} 
  \label{Wjj_plots_sleptons_7TeV}
\end{figure}

Finally we show results for the scenario from Section~\ref{light_sleptons_msq},
which incorporates the higher cross sections of the SUSY background processes
together with squark masses that fit better to current exclusion limits of the
CMSSM parameter space by ATLAS~\cite{atlassusy1, *atlassusy2, *atlassusy3} and 
CMS~\cite{cmssusy1, *cmssusy2, *cmssusy3}.
As depicted in Fig.~\ref{Wjj_plots_sleptons_7TeV} and Table \ref{xs_W_table_sleptons_7TeV},
we get an improvement in the signal to background ratio by roughly a factor of 2 compared to
14 TeV center of mass energy, like for the scenario SPS1amod.
Therefore the SUSY background processes account for 12\% of the signal
cross section for $h \rightarrow WW$ in this optimized scenario.
As the dominant SUSY processes in this scenario do not lead to additional jets from
SUSY particle decays, a central jet veto would only reduce jets originating
from additional QCD radiation.

In this scenario the Standard Model like Higgs boson mass is assumed to be 123 GeV and therefore close
to the value where ATLAS and CMS have seen an excess in their Standard Model Higgs boson searches
\cite{atlas125, cms125}.
If we set up a model like SPS1amod2 introduced in Section~\ref{scenarios}, but increase the
SUSY breaking terms for the first and second generation squarks according to $\xi=1.5$
(corresponds to an average squark mass of 1.3 TeV) we increase the Higgs boson mass to
$m_h=124.3 \;\textrm{GeV}$. This raises the signal cross section with all cuts to
$\sigma^{h\rightarrow WW}=527 \;\textrm{ab}$.
With light sleptons of the first and second generation the background processes
account for $\sigma^{SUSY}=63 \;\textrm{ab}$ leading to roughly the same 
signal to background ratio
\begin{equation}
 S / B^{SUSY} = 8.4 \,.
\end{equation}

The situation for the $h \rightarrow \tau\tau$ signal process is again very comfortable.
With all cuts discussed for this channel and taking detector effects on the missing transverse
momentum measurement into account, we get $S/B^{SUSY}=43$, with a signal cross section
of 340 ab for the scenario with a Higgs boson mass of 123~GeV.

\tabcolsep2mm
\begin{table}[p]
  \begin{center}
  \begin{tabular}{llccc}
  \toprule
  & Cuts & basic + rap. gap & + $m_{inv} + \slashed{p}_T + \phi_{\ell\ell} + \textrm{b-tag}$ & + CJV on $j_{decay}$ \\
  Processes && Eqs.~(\ref{cuts_W_min}) + (\ref{cuts_W_deltaeta}) & + \eqref{cuts_W_nomass}, \eqref{cuts_W_paper}, \eqref{cuts_W_paper_ptmiss}, \eqref{btag} & + \eqref{cuts_jetveto} \\
  \midrule
  \midrule
  $\chi_1^+ \, \chi_1^- \, jj$ & no b                   & $1756 \;\text{ab}$ & $35.0 \;\text{ab}$ & $35.0 \;\text{ab}$ \\
  $\chi_1^+ \, \chi_1^- \, jj$ & b-contr.               & $ 216 \;\text{ab}$ & $ 2.2 \;\text{ab}$ & $ 2.2 \;\text{ab}$ \\
  $\chi_2^0 \, \chi_1^0 \, jj$ & no b                   & $ 1.6 \;\text{ab}$ & $0.03 \;\text{ab}$ & $0.03 \;\text{ab}$ \\
  $\chi_2^0 \, \chi_1^0 \, jj$ & b-contr.               & $  43 \;\text{ab}$ & $ 0.7 \;\text{ab}$ & $ 0.7\;\text{ab}$ \\
  $\widetilde{\ell}^+\, \widetilde{\ell}^-\, jj$ & no b & $ 482 \;\text{ab}$ & $13.4 \;\text{ab}$ & $13.4\;\text{ab}$ \\
  \midrule
  $\chi_1^+ \, \chi_2^0 \, jj$ & no b                   & $132 \;\text{ab}$ & $3.8 \;\text{ab}$ & $3.8 \;\text{ab}$ \\
  $\chi_1^- \, \chi_2^0 \, jj$ & no b                   & $ 48 \;\text{ab}$ & $1.3 \;\text{ab}$ & $1.3 \;\text{ab}$ \\
  $\chi_2^0 \, \chi_2^0 \, jj$ & no b                   & $ 32 \;\text{ab}$ & $1.7 \;\text{ab}$ & $1.7 \;\text{ab}$ \\
  $\chi_2^0 \, \chi_2^0 \, jj$ & b-contr.               & $ 35 \;\text{ab}$ & $0.8 \;\text{ab}$ & $0.8 \;\text{ab}$ \\
  \midrule
  \multicolumn{2}{l}{$\sum B^{SUSY}$}                   & $2746 \;\text{ab}$ & $58.9 \;\text{ab}$ & $58.9 \;\text{ab}$ \\
  \midrule
  \multicolumn{2}{l}{VBF $h\rightarrow WW$}             & $1193 \;\text{ab}$ & $476 \;\text{ab}$ & $476 \;\text{ab}$ \\
  \midrule
  \multicolumn{2}{l}{$S / B^{SUSY}$}                    &        0.44    &        8.1     &            8.1 \\
  \bottomrule
  \end{tabular}
  \end{center}
  \caption{Total cross sections of SUSY backgrounds
           to the VBF $h\rightarrow WW$ channel for the scenario with light sleptons and 
           higher squark/gluino masses at 7 TeV center of mass energy.}
  \label{xs_W_table_sleptons_7TeV}
\end{table}

\section{Summary and Conclusions}
\label{conclusion}

We have investigated SUSY induced background processes to the production of the 
light Higgs boson in the MSSM via vector boson fusion  at leading order
accuracy. Particularly, we discussed dominant background contributions 
to the Higgs boson decays into tau leptons and W bosons for fully leptonic
final states for a SPS1a-like scenario and scenarios with partly changed
characteristics with respect to the SPS1a-like one. 

Our analysis was split into two parts: First we searched for processes
that match the signal signature exactly and therefore contribute to
the irreducible background.
In this class the two production
processes $\chi_1^+\,\chi_1^-\,jj$ and $\chi_2^0\,\chi_1^0\,jj$ with subsequent
decays of $\chi_1^\pm$ and $\chi_2^0$ have been found to give the largest SUSY
backgrounds to the mentioned leptonic signatures of the Higgs boson.
The contributions with b-quarks in the final state have been checked
separately; though sizable if only basic cuts are applied, taking into account
the complete set of cuts, including a b-jet veto, reduces these b-quark
related background processes sufficiently. We estimated the effect of
a central jet veto on these processes and showed that the veto can lead to a further 
reduction of the SUSY background. Smaller irreducible background contributions come 
from the $\widetilde{\ell}^+\,\widetilde{\ell}^-\,jj$ production.

The processes $\chi_1^\pm\,\chi_2^0\,jj$ and $\chi_2^0\,\chi_2^0\,jj$ contribute
to the reducible background, as they lead to additional jets or leptons from the
decay of the SUSY particles. Even with a central jet veto these contributions
are twice as large as the irreducible SUSY background in SPS1a-like scenarios.

Starting from the SPS1a-like scenario, we have shown the SUSY background cross section  
dependence on the squark and gluino masses for mass values roughly between 550
GeV and 1.6 TeV. The SUSY background decreases with increasing squark and gluino
masses and is lowered by 70\% for squark masses of 1.1 TeV. For higher masses 
the 
SUSY contributions remain constant.
We also checked a scenario where the selectrons and smuons are the 
lightest sleptons, which leads to significantly larger backgrounds.

For the $h \rightarrow \tau\tau$ signal, we saw that the reconstruction of the
invariant tau pair mass,
which is possible in the collinear approximation of tau lepton decays, is very
efficient in reducing the cross sections of the
SUSY background processes. For the SPS1a-like scenario the backgrounds after
mass reconstruction 
and a mass window cut around the Higgs boson mass are more than one order 
of magnitude smaller than the signal. Even for the scenario with light
sleptons and squarks as light as in SPS1a, the SUSY background processes
account for not more than 23\% of the signal cross section.

SUSY backgrounds to the $h \rightarrow WW$ channel are a bit more
troublesome. They are under 
sufficient control for the SPS1a-like scenario with a ratio $S / B^{SUSY} = 1.9$
after passing all cuts. 
With larger squark masses the SUSY background rates decrease substantially.
However, for the scenario with light sleptons and SPS1a-like squarks, the
background can be larger than the signal.
With higher squark masses of around 1.1 TeV, which are still in agreement with
current 
measurements from ATLAS and CMS, a ratio $S / B^{SUSY} = 3.9$
remains.

These numbers have been calculated for the LHC operating at 14 TeV center of
mass  energy, as Higgs boson production in vector boson fusion is mainly
targeted at this energy. The analysis for the 7 TeV LHC shows that the cross
sections of our SUSY  background processes decrease about twice as much as the
ones from the signal processes. 

Overall, we could confirm that VBF Higgs boson production with a subsequent
decay into tau leptons is a very clean signal, not only in the Standard Model
but also in the MSSM. Considering the subsequent decay into W bosons,
particularly with an additional $\slashed{p}_T$-cut, the SUSY background
processes are mostly under control for squark and gluino masses
at or above the current exclusion limits from the LHC.

\section*{Acknowledgments}

We would like to thank Stefan Gieseke, Simon Pl\"atzer, Christian R\"ohr
and Peter Richardson for discussions and {\tt Herwig++} support and J\"urgen Reuter
and Christian Speckner for helping with the {\tt Whizard} comparison
and Margarete M\"uhlleitner for answering questions concerning
decays of SUSY particles and {\tt SUSYHIT}. We greatfully acknowledge helpful
discussions with Sophy Palmer, Eva Popenda and Michael Rauch. The Feynman diagrams
were done using JaxoDraw~\cite{jaxodraw1, *jaxodraw2, *axodraw}.
 This work was supported by the BMBF under Grant No.~05H09VKG 
(\textquotedblleft Verbundprojekt HEP-Theorie\textquotedblright) 
and by the Initiative and Networking Fund of the Helmholtz 
Association, contract HA-101 (\textquotedblleft Physics at the Terascale\textquotedblright) .

\bibliography{SUSYback}

\end{document}